\documentclass[12pt]{article}
\pdfoutput=1

\usepackage{putex}
\usepackage{amsmath,amssymb,amsfonts}
\usepackage{psfrag}
\usepackage{footmisc}
\usepackage{url}
\usepackage{icomma}
\usepackage{mathtools}
\usepackage{color}

\usepackage{tikz}
    \usepackage{amssymb,amsfonts,amsmath}
    \usepackage{tkz-euclide}
        \usetikzlibrary{arrows,calc,patterns}
\usepackage{pgfplots}

\definecolor{darkblue}{rgb}{0.1,0.1,.7}
\definecolor{purple}{rgb}{0.6,0,0.6}
\definecolor{orange}{rgb}{0.9,0.6,0}
\usepackage[colorlinks, linkcolor=darkblue, citecolor=darkblue, urlcolor=darkblue, linktocpage]{hyperref} 
\usepackage[square, comma, compress,numbers]{natbib}
\usepackage[]{amsmath}
\usepackage[]{graphicx}
\usepackage[]{latexsym}
\usepackage[utf8]{inputenc}
\usepackage{geometry}
\usepackage{amscd}
\usepackage[all,cmtip]{xy}
\usepackage{mathrsfs}

\usepackage[margin=10pt,font=small,labelfont=bf]{caption}
\usepackage{dsdshorthand}
\usepackage{changepage}
\usepackage{setspace}
\usepackage{tikz}
\usepackage{subcaption}

\usepackage[titles]{tocloft}
\usetikzlibrary{arrows,positioning} 
\newcommand\textred[1]{\text{{\color{red} #1}}}

\def\SL2{\widetilde{SL}(2,\mathbb R)}

\newcommand\mR{\mathbb{R}}
\usepackage{enumitem}
\newcommand\mZ{\mathbb{Z}}

\numberwithin{equation}{section}

\newcommand{\grp}[1]{\mathrm{#1}}

\newcommand {\bes} {\begin {equation*}}
\newcommand {\ees} {\end {equation*}}
\newcommand {\beq} {\begin {equation}}
\newcommand {\eeq} {\end {equation}}
\newcommand {\bea} {\begin {eqnarray}}
\newcommand {\ea} {\end {eqnarray}}
\newcommand {\eea} {\end {eqnarray}}

\newcommand{\Sch}{\text{Sch}}

\newcommand{\ESchw}{{E_\text{gap}}}

 \newcommand\bck[1]{{\bar{#1}}}

\numberwithin{equation}{section}

\def\<{\langle}
\def\>{\rangle}

\tikzset{
    >=stealth',
    punkt/.style={
           rectangle,
           rounded corners,
           draw=black, very thick,
           text width=15em,
           minimum height=2em,
           text centered},
    pil/.style={
           ->,
           thick,
           shorten <=2pt,
           shorten >=2pt,}
}

 \def\ie{\begin{equation}\begin{aligned}}
\def\fe{\end{aligned}\end{equation}}

\newcommand{\CN}{\mathcal{N}} 
\newcommand{\CQ}{\mathcal{Q}}      
\newcommand{\ndt}{\noindent} 
\renewcommand{\=}{\; = \;} 
\newcommand{\wt}{\widetilde}       
\newcommand{\IR}{\mathbb{R}} 
\newcommand{\IZ}{\mathbb{Z}} 
\newcommand{\IC}{\mathbb{C}} 
\newcommand{\IH}{\mathbb{H}} 
\renewcommand{\v}{\varphi}         
\newcommand{\vth}{\vartheta}         
\renewcommand{\t}{\tau}         
\newcommand{\rme}{{\rm e}}         
\newcommand{\ii}{{\rm i}}  
\newcommand{\bdy}{\text{bdry}}
\newcommand{\orbiso}{H_{\rm orb.}}

\newcommand{\CV}{\mathcal{V}}         
\newcommand{\CK}{\mathcal{K}}

\newcommand{\nv}{n_\text{v}}

\newcommand{\qeq}{Q_\text{eq}}

\newcommand{\dd}{{\rm d}}
\newcommand{\ve}{\varepsilon}

\begin{document}

    \institution{SU}{${}^1$ Stanford Institute for Theoretical Physics, Stanford University, Stanford, CA 94305, USA}
    \institution{OX}{${}^2$ Department of Mathematics, King’s College London, The Strand, London WC2R 2LS, UK}
\institution{IAS}{${}^3$  School of Natural Sciences, Institute for Advanced Study, Princeton, NJ, USA}

\institution{UW}{${}^4$ Physics Department, University of Washington, Seattle, WA, USA}

\title{
Black hole microstate counting from the gravitational path integral 
}

\authors{Luca V. Iliesiu${}^1$, Sameer Murthy$^2$, and Gustavo J. Turiaci${}^{3,4}$ }

\abstract{
Reproducing the integer count of black hole micro-states 
from the gravitational path integral is an important problem in quantum gravity. In this paper, we show that, by using supersymmetric localization, the gravitational path integral for $\frac{1}8$-BPS black holes in $\cN=8$ supergravity reproduces the index obtained in the string theory construction of such black holes, including all non-perturbatively suppressed geometries. A more refined argument then shows that, not only the black hole index, but also the total number of black hole microstates within an energy window above extremality that is polynomially suppressed in the charges, also matches this string theory index.

To achieve such a match we compute the one-loop determinant arising in the localization calculation for all $\cN=2$ supergravity supermultiplets in the $\cN=8$ gravity supermultiplet. Furthermore, we carefully account for the contribution of boundary zero-modes, that can be seen as arising from the zero-temperature limit of the $\cN=4$ super-Schwarzian, 
and show that performing the exact path integral over such modes provides a critical contribution needed for the match to be achieved. A discussion about the importance of such zero-modes in the wider context of all extremal black holes is presented in a companion paper. 
  }

\date{}

\maketitle
\tableofcontents

\newpage

\section{Introduction}

In the past few years a tremendous amount of progress has been made in understanding the extent to which the gravitational path integral reproduces the various features of a conventional quantum mechanical system \cite{Saad:2019lba, Almheiri:2019qdq,Penington:2019kki,Almheiri:2020cfm,Saad:2019pqd,Marolf:2020xie,Giddings:2020yes, Stanford:2020wkf, Maxfield:2020ale,Blommaert:2019wfy,Blommaert:2020seb,Pollack:2020gfa,Cotler:2020ugk,Chen:2020tes,Iliesiu:2021ari,Meruliya:2021utr,Stanford:2019vob,Verlinde:2021jwu,Saad:2021uzi, Blommaert:2021fob, Blommaert:2022ucs}. One such characteristic that is notoriously difficult to detect by using the path integral description is the discreteness of the spectrum of black hole microstates or, equivalently, the integrality of the total number of states within a given energy window. In the concrete context of extremal supersymmetric black holes in sectors of fixed charge, a similar question arises \cite{Sen:2008vm}: can the gravitational path integral reproduce the integer degeneracy of such black holes which was computed in string theory \cite{Strominger:1996sh,Maldacena:1999bp,Gaiotto:2005gf, Pioline:2005vi,Shih:2005uc,Shih:2005qf,David:2006yn}? There has been much progress in computing the index of such black holes by using supersymmetric localization for the gravitational path integral \cite{Dabholkar:2010rm, Dabholkar:2010uh, Dabholkar:2011ec, Gupta:2012cy, Dabholkar:2014ema, Gupta:2015gga, Murthy:2015yfa,deWit:2018dix,Jeon:2018kec}. Nevertheless, a complete match with the string theory prediction has not been achieved until now. 

In this paper, we address the
problem of reproducing the exact (integer valued) index of  $\frac{1}{8}$-BPS black holes in $4$d $\cN=8$ supergravity, by using  the supersymmetric localization of the gravitational path integral. For the gravity picture we work in Type IIA gravity compactified on $T^6$ with D0-D2-D4 charges. The microscopic evaluation of the index ({i.e.}~the difference between the number of degenerate bosonic and fermionic  BPS states) using string theory was originally obtained in Type IIB string theory compactified on~$T^6$ \cite{Maldacena:1999bp}. This 
index admits a Hardy-Ramanujan-Rademacher expansion, which is 
a convergent sum
that can be expressed as
\be 
\label{eq:Rademacher-intro}
C(\Delta) \=  \sum_{c \geq 1} K_c(\Delta) \int \overbrace{\frac{d\phi^0}{(\phi^0)^{3/2}}}^{\substack{\text{{\color{red}Induced}}\\\textred{integration measure} \\ \textred{on localization locus}}}\, \underbrace{(\phi^0)^5}_{\substack{\textred{Bulk modes}\\ \textred{one-loop determinant}}} \overbrace{\frac{1}{c \phi^0}}^{\substack{\textred{Boundary modes}\\ \textred{one-loop determinant}}}\exp \underbrace{\left[{-\frac{\pi \,\Delta \,\phi^0}{4c} -\frac{ \pi}{c \,\phi^0}}\right]}_{\substack{\text{Supergravity action}\\\text{along locus}}}\,,
\ee 
up to an overall constant that is independent of the charges of the microstate. Here, the charges of the black hole microstate enter through $\Delta$, a duality invariant combination of the charges 
that in the classical limit determines the area of the black hole, i.e.~$\text{Area}/4G_N  = \pi \sqrt{\Delta}$.  
The sum over~$c$ is identified as a sum of  geometries with different topologies all of which preserve some amount of supersymmetry~\cite{Dabholkar:2014ema}.
The~$c=1$ term corresponds to the usual extremal black hole geometry, while the $c>1$ terms come from specific orbifolds of this geometry~\cite{Banerjee:2009af,Murthy:2009dq}.\footnote{These orbifolds are closely related to (but different from) the set of orbifolds that contribute to the  expansion of the AdS$_3$ partition function~\cite{Maldacena:1998bw,Dijkgraaf:2000fq}.}
For all geometries,  the integration space in \eqref{eq:Rademacher-intro} was identified as a direction within a localization locus, along which the supergravity action can be found to exactly reproduce the exponent inside the integral \cite{Dabholkar:2010uh, Dabholkar:2011ec}. 
The term $K_c(\Delta)$ is usually referred to as the Kloosterman sum and does not grow with $\Delta$ -- rather, it involves an intricate sum of number theoretic phases that from the path integral perspective were found by evaluating various topological terms in the supergravity action~\cite{Dabholkar:2014ema}. 

Not all terms in \eqref{eq:Rademacher-intro} have been previously matched by a gravitational (macroscopic) computation, and the main goal in this paper is to fill these gaps. First of all, previous attempts worked within an $\mathcal{N}=2$ truncation of $\mathcal{N}=8$ supergravity instead of the full theory, something we correct in this paper. Secondly, the one-loop determinant around the leading extremal black hole saddle or around the subleading geometries have not fully been computed until now (terms labeled with red text in \eqref{eq:Rademacher-intro}). Progress has however been made in determining the one-loop determinant around the leading saddle of some (but not all) of the supermultiplets in the supergravity theory \cite{Gupta:2015gga, Murthy:2015yfa, Jeon:2018kec}. Thirdly, before this paper, to our knowledge the computation of the one-loop determinant around the subleading saddles with $c\neq 1$ in \eqref{eq:Rademacher-intro} had not previously been attempted in the context of localization. 
Finally, a crucial ingredient in reproducing the integer degeneracies through \eqref{eq:Rademacher-intro} is the proper treatment of super-diffeomorphisms in the gravitational path integral that had not been correctly addressed in the past.  

A complete calculation of the one-loop factors is critical in obtaining the integer degeneracies found in string theory. 
We start by computing the Euclidean path integral of such black holes in the low-temperature limit and in a sector of fixed charge for all vector fields in the theory, but with a specific angular velocity that corresponds to computing the index of such black holes.\footnote{The Euclidean angular velocity necessary when computing an index is $\Omega_E=i\Omega = \pi/\beta$, where $\beta$ is the inverse temperature in the theory. 
The gravitational saddles of the flat space supergravity found with these boundary conditions were shown to all be smooth \cite{Iliesiu:2021are} and we will show the same is true in the localization computation. 
See~\cite{Cabo-Bizet:2018ehj} for a similar discussion about the meaning of the index for supersymmetric black holes in~$AdS_5$.  } 
Performing the supergravity localization procedure with such boundary conditions for all fields, including the metric and its superpartners, the one-loop determinant is seen to have two contributions: that of bulk and boundary modes that are not zero-modes of the localizing deformation, and that of boundary modes that are zero-modes of both the localizing deformation and of the undeformed supergravity action. Using the ideas of \cite{Murthy:2015yfa, Jeon:2018kec}, we explain that the one-loop determinant around the localization locus of all $\cN=2$ supermultiplets can be obtained by classifying all bulk fields into a cohomology complex, from which the determinant can be written as an index within the cohomology.\footnote{While some the bulk modes for some $\cN=2$ supermultiplets (the Weyl, vector and hypermultiplets) have been arranged in cohomology complexes \cite{Murthy:2015yfa, Jeon:2018kec}, other  $\cN=2$ supermultiplets that are part of the $\cN=8$ Weyl multiplet have not (in particular, the spin 3/2 and chiral supermultiplets). In Section~\ref{sec:one-loop-det-results} we will show how to arrange all  these latter multiplets into such complexes.  } We then explain how the Atiyah-Bott fixed point formula can be used to easily evaluate such an index both around the AdS$_2 \times S^2$ localization saddle and around the non-perturbatively subleading geometries in \eqref{eq:Rademacher-intro}. For each individual $\cN=2$ supermultiplet the dependence on the orbifold parameter $c$ in the one-loop determinant is highly intricate. However, when putting all such multiplets together within an $\cN=8$ theory of supergravity, this dependence simplifies drastically leading to the simple result for bulk modes shown in \eqref{eq:Rademacher-intro}.

The integral over boundary modes is even more subtle. For the ones that are zero-modes, the path integral over the full moduli space of such modes needs to be performed.
We show that such a functional integral can be obtained from the zero temperature limit of the $\cN=4$ super-Schwarzian partition function which has the non-trivial yet simple $c$-dependence shown in \eqref{eq:Rademacher-intro}. 
Putting these results together in the integral over the entire localization locus, and integrating out all other directions with the exception of the variation $\phi^0$ of the scalar in the graviphoton supermultiplet, we thus find that the gravitational path integral with index boundary conditions fully reproduce \eqref{eq:Rademacher-intro}.

In addition to reproducing the index using the gravitational path integral, we provide strong evidence that the computation of the index also reproduces the bosonic degeneracy at extremality and we explain why, even at a non-perturbative level, the gravitational path integral suggests that there are no other black hole microstates in a large energy window above extremality (large with respect to $\rme^{-\pi \sqrt{\Delta}}$). Explicitly, by performing the gravitational path integral with boundary conditions that no longer correspond to an index but instead to a 
regular partition function, we show that the contribution of both bulk and boundary modes remains unchanged.\footnote{A similar statement was made in \cite{Sen:2009vz} by imagining that the ``quantum mechanical dual'' of such extremal and near-extremal states has an  $PSU(1,1|2)$ symmetry. Using representation of  $PSU(1,1|2)$, \cite{Sen:2009vz} concluded that BPS states are entirely bosonic. Our analysis of the near zero-modes makes this statement more precise since we explicitly compute the path integral with different boundary conditions that correspond to either an index or a degeneracy and find explicit agreement without the need to imagine the existence of a conformal quantum mechanical dual. } Additionally, while in principle other geometries that  preserve no
supersymmetries may contribute to the degeneracy but not to the index,\footnote{Here we discuss geometries that can locally be described as AdS$_2\times S^2$. These do not include the geometries that are explicitly summed over in \eqref{eq:Rademacher-intro} since those preserve no supersymmetries. Examples of such geometries include near-horizon regions that can have a higher-genus or generalizations of Seifert manifolds.   } the integral over boundary modes in a large class of geometries excludes this possibility. Finally, we show that the integral over the nearly zero-modes on all geometries, including those that contributed to both the index and to the degeneracy, guarantees that the density of states has a large gap above extremality that scales like $1/\Delta^{3/2}$, at sufficiently large $\Delta$.

 While deriving the full one-loop factors is the main technical goal of this paper, along the way we also clarify numerous other aspects in the localization computation. For instance, we shall show how to perform the localization procedure when keeping track of all the scalars from the $\cN=2$ vector supermultiplets within the $\cN=8$ theory, rather than performing a truncation which turns out to be inconsistent if carefully accounting for all one-loop factors.\footnote{This inconsistency was also pointed out, though not fully addressed in \cite{Murthy:2015yfa, Jeon:2018kec}.} We also explain why the integration measure on the localization locus obtained from the natural ultralocal measure on the space of scalar fluctuations is necessary in order to obtain a duality invariant index and partition function from the path integral perspective.

The rest of this paper is organized as follows. In Section~\ref{sec:BPS-BHs-in-N=8}, we review the Rademacher expansion of the microscopic count for $\frac{1}{8}$-BPS black hole microstates (Section~\ref{sec:microscopic-index}), and then present a rough outline for the macroscopic interpretation of this expansion (Section~\ref{sec:macroscopic-interpretation-of-BH-degeneracy}). In Section~\ref{sec:localization-in-supergravity} we present the basics of the localization formalism in $\cN=2$ supergravity (Section~\ref{sec:locformalism}) and then give a careful treatment of all the aspects that we shall encounter in $\cN=8$ supergravity (Section~\ref{sec:black-holes-in-N=8}). Next, in Section~\ref{sec:one-loop-det-theoretical-prelims} we compute the one-loop determinant using index theorem around both the AdS$_2\times S^2$ geometry (Section~\ref{sec:one-loop-determinant-AdS2-S2}) and the orbifold geometries (Section~\ref{sec:one-loop-det-orbifold}). To exemplify the difference when computing the one-loop determinant on each of the two geometry classes, we present a pedagogical derivation of the index associated to the $\cN=2$ vector supermultiplet (Section~\ref{sec:N=2-vector-supermultiplet-example}).  We then move on to computing the  one-loop determinant of all bulk modes which are part of the $\cN=2$ supermultiplets that form the $\cN=8$ Weyl supermultiplet in Section~\ref{sec:one-loop-det-results}. For readers that are interested in the actual value of the determinants for bulk modes rather, than their derivation, readers can skip Section~\ref{sec:one-loop-determinant-AdS2-S2}--\ref{sec:one-loop-det-results} and directly read the summary in section   \ref{sec:summary-results-bulk-one-loop-det}.  We then address the contribution of the boundary modes to the one-loop determinant in Section~\ref{sec:role-of-large-gauge-transformations}, explaining (in Section~\ref{sec:effect-of-large-gauge-transf}) how such modes affect the index computation discussed in the prior section, how their integration measure is determined (Section~\ref{sec:integration-measure-large-gauge-transf}), and how their path integral is related to that of the $\cN=4$ super-Schwarzian (Section~\ref{sec:connection-to-N=4-super-Schw}). In Section~\ref{sec:putting-it-all-together}, we put all the pieces together and verify the matching of the path integral result to the microscopic index. Next, in Section~\ref{sec:index-vs-partition-function} we explain why the partition function is the same as the index for all geometries, including those that preserve some number of supersymmetries and those that do not, yielding a vanishing contribution. Finally we discuss the meaning of our results in Section~\ref{sec:discussion}, presenting a comprehensive picture about the spectrum of $\frac{1}{8}$-BPS and near-BPS black holes.

\section{$\frac18$-BPS black holes in~$\CN=8$ string theory} 
\label{sec:BPS-BHs-in-N=8}

Consider Type-II string theory compactified on~$T^{6}$. This theory has 32 supercharges, 
and has~$E_{7, 7} (\mathbb{Z})$ U-duality symmetry \cite{Cremmer:1979up}. 
At low energies the theory is described by (the unique) two-derivative~$\CN=8$ four-dimensional 
supergravity, whose fields are summarized by the~$\CN=8$ graviton multiplet. 
The bosonic field content includes the graviton, 28 $U(1)$ gauge fields, and 70 massless scalar fields. 

The supergravity theory admits dyonic black hole solutions preserving 4 supercharges~\cite{Cvetic:1995uj,Cvetic:1995bj,Cvetic:1996zq}.
They carry electric and magnetic charges under the~$U(1)^{28}$, which together transform 
in the~$\mathbf{56}$ of the U-duality group. In terms of these charges~$\CQ_a$, $a=1,\dots, 56$,
there is a unique quartic invariant of the U-duality group
given by 
\be \label{quarticinv}
\Delta(\CQ) \= C^{abcd} \, \CQ_a \,  \CQ_b \,  \CQ_c \,  \CQ_d \,,
\ee
where~$C^{abcd}$ is the rank-4 invariant tensor of~$E_{7, 7} (\mathbb{Z})$ \cite{Cremmer:1979up, Kallosh:1996uy}.
The Bekenstein-Hawking entropy of these black holes is given in terms of the charge invariant as 
\be \label{BHentropy}
S_\text{BH}(\CQ) \= \pi \sqrt{\Delta(\CQ)} \,.
\ee

On the microscopic side, we are interested in $\frac18$-BPS dyonic states carrying the same 
charges as the black holes. We now proceed to discuss the supersymmetric index that receives 
contributions from these states.

\subsection{Microscopic index of $\frac18$-BPS states}
\label{sec:microscopic-index}

In order to count the microscopic degeneracy of states we choose a particular duality frame,
and use a duality rotation to map the general charge configuration~$\CQ_a$ to a specific 
charge configuration where one can carry out a counting of states. 
The microscopic counting of BPS states is best performed in the duality frame described by  
Type IIB string theory compactified on $T^6= T^4 \times S^1 \times \wt{S}^1$. 
We consider a charge configuration consisting of one D1-brane wrapping~$S^{1}$, one 
D5-brane wrapping $T^{4}\times S^{1}$, $n$ units of momentum around~$S^1$, $\wt K =1$ unit of 
Kaluza-Klein charge around~$\wt S^1$, and~$\ell$ units of momentum around~$\wt S^1$. 
This configuration of charges has~$E_{7, 7} (\mathbb{Z})$ charge invariant 
\be \label{defDeltanell}
\Delta \= 4  \, n - \ell^2 \,. 
\ee

Having picked a particular 
five-charge configuration as above, one can further ask if the most general charge configuration 
in the~$\CN=8$ theory can be mapped to it by a duality rotation. The analysis of this question 
involves the classification of all charge invariants of~$E_{7,7}(\mathbb{Z})$. 
It is known that the quartic invariant~\eqref{quarticinv} is the unique continuous invariant of the duality 
group~$E_{7,7}(\mathbb{Z})$, but there can also be other discrete invariants
depending on e.g.~common divisors of the charges which have not been completely classified (see~\cite{Banerjee:2008ri,Banerjee:2008pu,Dabholkar:2008zy,Sen:2009gy} for a discussion of the 
discrete invariants in the above context.). 
If we choose the charges $n$ and $\ell$ to be relatively prime to each other then 
all the discrete invariants are trivial. In this case, the above five-charge configuration
is in the duality orbit of any given choice of charges for an appropriate choice of~$n$ and~$\ell$.

The calculation of the index of BPS states with the above five charges proceeds by first considering the 
compactification~IIB/$T^4 \times S^1$, in which one calculates the degeneracies of the five-dimensional 
D1-D5 system with momentum~$n$ along~$S^1$ and angular momentum~$\ell$~\cite{Strominger:1996sh,Maldacena:1999bp}. 
One then reaches a four-dimensional configuration by placing the D1-D5 system at the tip of a KK monopole.
The degeneracies of the four-dimensional system are calculated using the 
4d/5d lift~\cite{Gaiotto:2005gf,Shih:2005uc,Shih:2005qf,David:2006yn}.  
The answer is stated in terms of the function (with~$\t \in \IH$, $z \in \IC $)
\be \label{phi21}
\v_{-2,1}(\t,z) \= \frac{\vth_1(\t, z)^2}{\eta(\t)^6} \,,
\ee
where~$\vth_{1}$ is the odd Jacobi theta function and  $\eta$ is the Dedekind function, 
given explicitly in~\eqref{theta1def}, \eqref{etadef}.

The function~$\v_{-2,1}(\t,z)$ is a weak Jacobi form of weight~$k=-2$ and index~$m=1$. 
In Appendix~\ref{app:Jacobi} we recall the definitions and some basic properties of Jacobi forms.
From the general properties of Jacobi forms, we have  
the double Fourier expansion 
\be \label{phim21}
\v_{-2,1}(\t,z) \= \sum_{n, \ell \, \in \, \IZ} c_{-2,1}(n, \ell)\,q^n\,\z^\ell\,, \qquad q \= \rme^{2\pi \ii \tau}\,, \quad \zeta \= \rme^{2\pi \ii z}\,.
\ee
As described in Appendix~\ref{app:Jacobi}, the Fourier coefficients~$c_{-2,1}(n,\ell)$
are completely captured by one function~$C$ of one variable,
\be
c_{-2,1}(n,\ell) \= C(\Delta)\,, \qquad \Delta = 4 \, n - \ell^2\,.
\ee

Now, it is a remarkable fact of analytic number theory that the Fourier coefficients of a Jacobi form,  
under suitable restrictions, is given by an exact analytic formula. This formula, known 
as the Hardy-Ramanujan-Rademacher expansion, takes the form of a convergent series of successively 
exponentially smaller terms. We present the general form of this expansion in~\eqref{RademacherGeneral}.
For the Jacobi form~\eqref{phi21}, the Rademacher expansion simplifies from this general form,  
and the coefficients~$C(\Delta)$ discussed in the previous paragraph
are given by the following formula,  
\be\label{RadexpC} 
 C(\Delta) \=   2{\pi} \, \Bigl( \frac{\pi}{2} \Bigr)^{7/2} \, \sum_{c=1}^\infty 
  c^{-9/2} \, K_{c}(\Delta) \; \wt I_{7/2} \Bigl(\frac{\pi \sqrt{\Delta}}{c} \Bigr)  \,.
\ee
Here~$\wt{I}_{\rho}$, a modification of the standard $I$-Bessel function,
is given by the following integral formula, 
\begin{equation}\label{Besselintrep}
 \wt{I}_{\rho}(z)\=  \Bigl(\frac{z}{2} \Bigr)^{-\rho} I_{\rho}(z) 
 \=\frac{1}{2\pi \ii}\int_{\epsilon-\ii\infty}^{\epsilon+\ii\infty} \, 
 \frac{d\s}{\s^{\r +1}}\exp \Bigl(\s+\frac{z^2}{4\s} \Bigr) \, .
\end{equation}
The Kloosterman sum~$K_c$  in~\eqref{RadexpC} is a sum over phases, simplified from its general expression~\eqref{kloos0}. 
Given a pair of integers~$(c,d)$ with~$c>0$,~$-c \le d <0$, and~$(c,d)=1$, denote by~$\gamma_{c,d}$
an $SL_2(\IZ)$ matrix~$\bigl(\begin{smallmatrix} a & b \\ c & d \end{smallmatrix} \bigr)$.  
Then 
\bea
\label{kloos}
K_{c}(\Delta) \;\coloneqq \;
\rme^{5\pi \ii /4}
\sum_{-c \leq d< 0 \atop (d,c)=1}
\rme^{2\pi \ii \frac{d}{c} (\Delta/4)} \; M(\gamma_{c,d})^{-1}_{\nu 1} \; 
\rme^{2\pi \ii \frac{a}{c} (-1/4)} \,,  \qquad \nu = \Delta \; \text{mod} \; 2 \,,
\eea
where the multiplier matrix is  
\be \label{multiplier1}
\begin{split}
  M^{-1}(\gamma_{c,d})_{\nu 1} &\= 
 \rme^{\frac{\pi \ii}{4}}  \frac{1}{\sqrt{6c}} \exp \Bigl( -\frac{\ii\pi}{6}\Phi(\gamma) \Bigr) \times \cr
& \qquad   \sum_{\varepsilon\=\pm 1}\sum_{n=0}^{c-1}\varepsilon \, 
  \exp \Bigl(\, \frac{\ii\pi}{6c} \bigl( d (\nu +1)^2-4(\nu +1)(3n+\varepsilon )+4a(3n+ \varepsilon )^2 \bigr) \Bigr) \,,
\end{split}
\ee
where~$\Phi$ is the Rademacher function given in~\eqref{RadPhi}.
Notice that, as written above, we need to make a choice of the~$SL(2,\IZ)$ matrix~$\g_{c,d}$ in~\eqref{kloos} 
for given~$(c,d)$, but one can check from the above expressions that~$K_c(D)$ is independent of this choice. 
The more fundamental way to write the sum is as a sum over the elements of the 
coset~$\Gamma_{\infty}\backslash SL(2,\IZ) /\Gamma_{\infty}$.

\subsection{Macroscopic interpretation of the index}
\label{sec:macroscopic-interpretation-of-BH-degeneracy}

Now we compare the microscopic index of~$\frac18$-BPS states with its macroscopic interpretation. 
The contributions to the macroscopic index come from supergravity 
configurations in asymptotically flat space that preserve the same supercharges as 
the~$\frac18$-BPS states. 
Apart from the~$\frac18$-BPS black hole, one could, in principle have multi-black-hole or more exotic configurations.
Such configurations are known to contribute to the supersymmetric index in theories with lower supersymemtry, 
but it was shown in~\cite{Dabholkar:2009dq} by an analysis of fermion zero modes that the only macroscopic configurations that contribute to
the~$\frac18$-BPS helicity supertrace  in~$\CN=8$ string theory are the~$\frac18$-BPS black holes. 
The~$\frac18$-BPS black holes are labelled by the U-duality invariant~\eqref{defDeltanell} of supergravity, which is to be 
identified with the elliptic invariant~$\Delta=4n-\ell^2$ in the Rademacher expansion~\eqref{RadexpC}.

In the background of the~$\frac18$-BPS black hole, we need to worry about whether the index is counting fluctuations of supergravity that are localized inside the horizon as opposed to outside---these latter modes would constitute the so-called hair modes.
As explained in~\cite{Sen:2009vz,Dabholkar:2010rm},
in the duality frame with only D-brane charges, the only hair modes come from the 28 
fermionic Goldstino zero-modes associated 
to the supersymmetries broken by the~$\frac18$-BPS black hole, and contribute an overall sign~$(-1)^{\Delta+1}$ to the index. 
Upon factoring out this contribution, 
we obtain the Witten index of the black hole degrees of freedom only (see \textit{e.g.} equation (1.7) of \cite{Denef:2007vg}). 
In the rest of this paper we 
compare a macroscopic calculation of the index with
\be 
\label{dCrel}
W_\text{micro}(\Delta) \= (-1)^{\Delta+1} \, C(\Delta) \,,
\ee
which we call the microscopic Witten index of the black hole. 
Note that although it is difficult to directly calculate the microscopic index of BPS states that constitute a BPS black hole in general, we can, through the above set of arguments, do so very precisely for theories with extended supersymmetry.

The Rademacher expansion~\eqref{RadexpC} (combined with~\eqref{dCrel})
lends itself to a precise macroscopic interpretation. 
In the rest of this subsection we explain the ideas behind this interpretation, using 
increasingly accurate analytic approximations to the integer~$W_{\rm micro}(\Delta)$.

Recalling that~$I_\rho(x) \sim \rme^x$, $x \to \infty$, it is clear from the form of the Rademacher expansion~\eqref{RadexpC} that the~$c>1$ 
terms are all exponentially suppressed compared to the~$c=1$ term.
The leading asymptotic growth of~$W_{\rm micro}(\Delta)$ is controlled by the large-argument 
asymptotics of the first~$I$-Bessel function, and is given by  
\be \label{logDeltaBH}
\log \, W_\text{micro}(\Delta) \; \sim \; \pi \sqrt{\Delta} + \dots \,, \qquad \Delta \to \infty  \,.
\ee
The right-hand side of~\eqref{logDeltaBH} agrees precisely the Bekenstein-Hawking entropy 
formula~\eqref{BHentropy} for the BH entropy, which from the path integral perspective comes from the Euclidean action evaluated on the corresponding  geometry that solves the equations of motion. 

The subleading corrections in the microscopic formula (the dots in~\eqref{logDeltaBH})
can be systematically computed using the asymptotic development of the $I$-Bessel function
 as~$\Delta \to \infty$ (see Formula~\eqref{Besselasymp}). 
As we shall see, these corrections to the leading growth of states correspond to quantum corrections
in supergravity to the leading Bekenstein-Hawking entropy formula. 
The subleading corrections to $\log W_\text{micro}(\Delta)$ 
in decreasing 
magnitude as~$\Delta \to \infty$, consist of a logarithmic term~$\log \Delta$,
followed by power-law corrections in~$1/\sqrt{\Delta}$, and then finally non-perturbatively
small corrections of order~$\rme^{- \alpha \Delta}$, $\alpha >0$, which we now discuss in turn.

Including the first subleading correction to~\eqref{logDeltaBH}, we have 
\be \label{leadinglog}
\log \, W_\text{micro}(\Delta) \= \pi \sqrt{\Delta} - 2 \log \Delta + \text{O}(1/\sqrt{\Delta}) \,.
\ee
As was seen in \cite{Banerjee:2011jp}, the logarithmic correction in the above formula arises from the one-loop fluctuations of the massless fields 
of two-derivative supergravity around the near-horizon~AdS$_2 \times$~S$^2$ region of the black hole.
We review these calculations and rectify some previously incomplete treatment of boundary modes in the companion paper \cite{Iliesiu:companionPaper}. 
The power-law corrections in~\eqref{leadinglog} come from corrections to the local effective 
action of two-derivative supergravity in string theory~\cite{LopesCardoso:1998tkj} arising from integrating out the massive modes of the string theory in the background 
of the black hole. They affect the BH entropy in two ways: firstly, in the presence of 
higher-derivative operators in the gravitational effective action the Bekenstein-Hawking formula 
needs to be replaced by the Wald entropy formula~\cite{Wald:1993nt,Iyer:1994ys} and, 
secondly, the black hole solution itself gets corrected by the higher-order effects. 
We refer the reader to~\cite{Sen:2007qy} for a nice review of these ideas and calculations.

Continuing on to a better approximation 
to the integer degeneracy, 
we consider the first term ($c=1$) in the 
Rademacher expansion~\eqref{RadexpC}, i.e.,
\be \label{dFirstBessel}
W_{\rm micro}(\Delta)\= 
\underbrace{\frac{\pi^{9/2}}{8} \; 
\wt I_{7/2}(\pi \sqrt{\Delta}) }_{\equiv\,\, W^{(1)}(\Delta)  }+ \text{O} (\rme^{\sqrt{\Delta}/2}) \,.
\ee
This result is interpreted in gravity as the result of summing up the entire perturbation series for the quantum entropy of the~$\frac18$-BPS BH.
Such an interpretation is made possible in the framework of the quantum black hole 
entropy 
expressed as a functional integral over the fluctuating fields 
of supergravity around the full-BPS AdS$_2 \times$~S$^2$ background~\cite{Sen:2008yk,Sen:2008vm}. 
The calculation of this integral to all 
orders in perturbation theory 
is made possible by  using the technique of supersymmetric localization applied to the  
gravitational path integral. 
There are many subtleties underlying this calculation, both at the conceptual as well as the technical level,
that we shall discuss in Section~\ref{sec:localization-in-supergravity}.

As we explain there, at the end of these calculations, the full functional integral 
reduces to an integral over over a 16-dimensional space. 
15 of the 16 integrals are Gaussian, after performing which one reaches precisely the
leading ($c=1$) term of the Rademacher expansion~\eqref{RadexpC} with the integral
representation~\eqref{Besselintrep} of the 
Bessel function.

\begin{figure}[t!]
    \centering
    \includegraphics[width=0.49\textwidth]{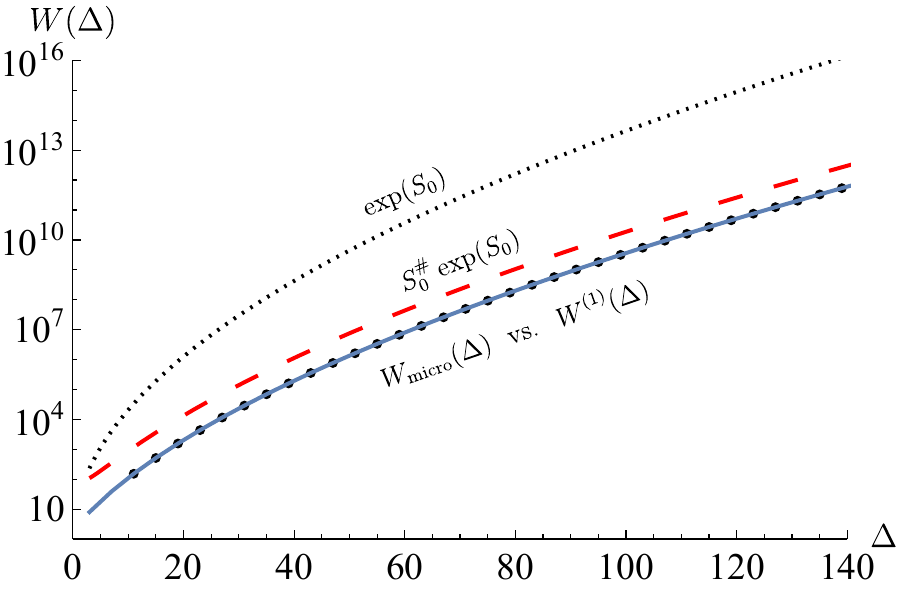}
     \includegraphics[width=0.49\textwidth]{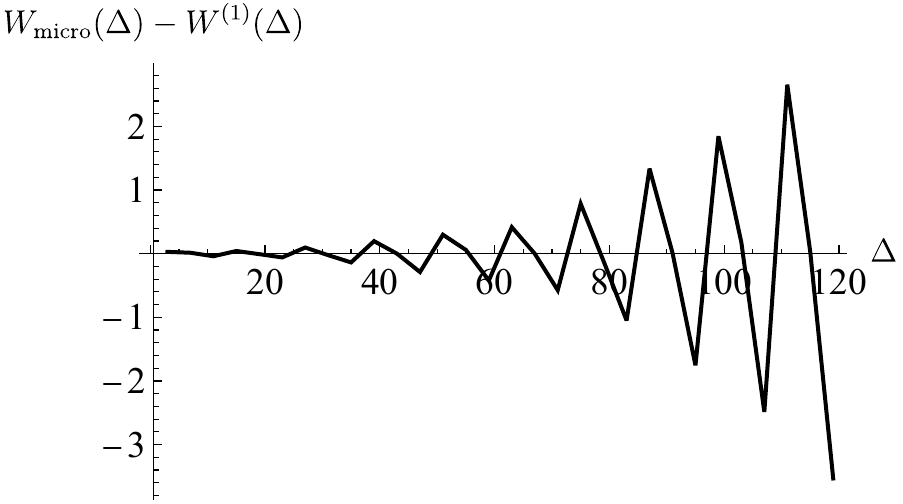}
    \caption{\textit{Left}: Indices of $1/8$-BPS black holes estimated to four different levels of precison: (i)~the dotted black curve represents the roughest Bekenstein-Hawking estimate given by $\exp(S_0)$; (ii)~the dashed red curve represents the degeneracy corrected at one-loop around the leading saddle; (iii)~the solid blue curve is the estimate from considering the entire asymptotic   perturbation theory series around the leading AdS$_2 \times S^2$ localizing saddle, neglecting non-perturbative corrections; (iv)~the dots represent the exact index as obtained from string theory. \textit{Right}: The difference between the string theory index and the entire contribution of the leading AdS$_2 \times S^2$ localizing saddle. Note that while the plot on the left shows the various degeneracies on a logarithmic scale, the plot on the right does not. This emphasizes the great accuracy of even the leading localization result before other non-perturbative corrections (the orbifold geometries) are included.  }
    \label{fig:degeneracy}
\end{figure}

This procedure should be understood as the result of summing up the entire perturbation series 
in quantum supergravity for the quantum entropy of the~$\frac18$-BPS BH. We emphasize the difference between the rough Bekenstein-Hawking approximation to the black hole degeneracy, the one-loop corrected degeneracy, the contribution of the leading localization saddle and the exact string theory result in the left panel of Figure \ref{fig:degeneracy}.  In the right panel, we show the difference between the exact string theory result and the all-order perturbation series around the leading AdS$_2 \times S^2$ saddle. 

It is striking that the macroscopic quantum entropy formula approximates the microscopic index in magnitude, as well as in the positive sign \cite{Sen:2011ktd,Bringmann:2012zr}, all the way to very small values of~$\Delta$ where, a priori, the gravitational approximation breaks down.\footnote{In fact, $W_{\rm micro}$ is also defined for~$\Delta \ge -1$, with $W_\text{micro}(-1)=1$, $W_\text{micro}(0)=2$, where there is no single-center black hole interpretation. Instead, these states are interpreted as two-center black hole bound states~\cite{Murthy:2015yfa,Chowdhury:2019mnb}. The macroscopic entropy formula also works remarkably well for these values, with~$\exp\bigl(S^\text{quant}(\Delta) \bigr)_\text{pert}(-1)=1.04\dots$ and~$\exp\bigl(S^\text{quant}(\Delta) \bigr)_\text{pert}(0)=1.86\dots$.}

Now that the asymptotic series coming from perturbation theory has been summed up into a function
in~\eqref{dFirstBessel}, we are now in a position to rigorously discuss the exponentially suppressed 
corrections~\cite{Dabholkar:2014ema}. 
Using the Rademacher expansion~\eqref{RadexpC}, \eqref{kloos}, we write
\be \label{Radcterm}
\frac{1}{2{\pi}} \, \Bigl( \frac{2}{\pi} \Bigr)^{7/2} \, 
W_\text{micro}(\Delta)  \= \sum_{c=1}^\infty \, A_c(\Delta) \,, \qquad \quad
A_c (\Delta)  \= 
  c^{-9/2} \, K_{c}(\Delta) \; \wt I_{7/2} \Bigl(\frac{\pi \sqrt{\Delta}}{c} \Bigr) \,.
\ee
We see that the exponentially suppressed corrections to the perturbatively exact~\eqref{dFirstBessel} are precisely the $c>1$ terms in the above expansion. 
To get a sense of how the non-perturbative corrections correct the leading saddle to give an integer degeneracy, consider the example string theory index for $\Delta = 159$ for which $W_\text{micro}(159) = 5,429,391,518,385$. Consider the sum of the first six saddles (together with all the perturbative corrections around those saddles) in the non-perturbative expansion, 
\begin{table}[h!]
    \centering
    \begin{tabular}{c|l c}
    Orbifold & Localization contribution \,\,\,\,\,$\text{ vs. }$\,\,\, String theory index \\ \hline
       $c=1$ &+5,429,391,518,372.2546  \\
       $c=3$ & \hspace{2.75cm}+13.4240 \\
       $c=4$ & \hspace{3.15cm}-0.5948\\ 
       $c=5$ & \hspace{3.15cm}-0.0797\\ 
       $c=7$ & \hspace{3.15cm}-0.0016\\ 
       $c=8$ & \hspace{3.15cm}-0.0028\\ \hline
       & \hspace{0.38cm}5,429,391,518,384.9997$\,\,\,\,\,$\text{ vs. }$\,\,\, $ $W_\text{micro}(159) = 5,429,391,518,385$
    \end{tabular}
    \caption{An example of the Rademacher expansion when summing the first six non-trivial saddles (and the entire perturbative expansion around each saddle) corresponding to orbifold geometries labeled by $c$ for $\Delta = 159$. The orbifold with $c=2$ and $c=6$ happen to have vanishing Kloosterman sums for this specific value of $\Delta$. }
    \label{tab:my_label}
\end{table}\\
which we see quickly converges to the answer in string theory.\footnote{In fact we only need $\sim \pi \sqrt{\Delta}/\log \Delta$ terms in the expansion to see what integer the expansion will be converging towards.}

These terms should be interpreted as the contribution of saddles of the bulk string theory which are not 
smooth configurations in 4d supergravity but are smooth when analyzed in higher dimensions. Explicitly, the Rademacher expansion~\eqref{RadexpC}, \eqref{kloos} is a sum over two coprime integers~$(c,d)$ 
with~$c\ge 1$,~$-c<d<0$. In~\cite{Dabholkar:2014ema}, a family of orbifolds of the near-horizon BH geometry
with the same labels was identified in the string theory. 
Writing the IIA compactification as AdS$_2 \times$~S$^2 \times$~T$^6$,  
the orbifold acts as a~$\IZ_c$ quotient of the AdS$_2 \times$~S$^2$ which preserves the 
original boundary conditions of the near-horizon BH, and acts locally as a~$\IC^2/\IZ_c$ 
quotient near the center of~AdS$_2$. On top of this orbifold we set a non-trivial holonomy around the thermal circle (proportional to $d/c$) for the D0 RR gauge field. When lifted to M-theory this holonomy is realized geometrically as a shift along the M-theory circle and the 11D geometry is smooth since there are no fixed points. 

We should now sum up the perturbative quantum fluctuations around each of these saddles
labelled by~$(c,d)$. 
The argument~$\pi\sqrt{\Delta}/c$ of the~$I$-Bessel function in \eqref{Radcterm} 
is interpreted as the real part of the action on the orbifold saddle, which is reduced by a factor of~$c$ 
compared to the~$c=1$ AdS$_2$ saddle, as consistent with the~$\IZ_c$ action of the orbifold. Besides this, there are imaginary contributions to the action which are more subtle. These were computed in~\cite{Dabholkar:2014ema} by lifting to M-theory and reducing on $S^2 \times T^6$ to $AdS_2 \times S^1$ resulting in Chern-Simons terms in the action. It was shown that these terms give 
a sum over phases which are precisely identified with the phases in the expressions~\eqref{kloos}, \eqref{multiplier1},
so that they add up to~$\sqrt{c} \, K_c$.  
Finally, we have to perform the one-loop determinants. 
As we shall see, the Gaussian integrals, the topological terms, and the one-loop determinant of the non-zero modes together account 
for almost the full formula~\eqref{Radcterm} or, equivalently, ~\eqref{RadexpC} 
\emph{except for one factor of~$1/c$}, that we are left to explain. 
We show in Section~\ref{sec:role-of-large-gauge-transformations} that precisely this factor arises from a careful quantization of the zero-modes of the supergravity.

\section{Localization in  supergravity}
\label{sec:localization-in-supergravity}

In this section we start by reviewing the main points of the calculation of the black hole index using localization in 
supergravity following~\cite{Dabholkar:2010uh,Dabholkar:2011ec,Jeon:2018kec}. The black hole solutions 
that we discuss in this paper are solutions of four-dimensional ungauged~$\CN=2$ supergravity with a 
number of gauge fields, and we review the calculation of the quantum entropy of such black holes in 
Section~\ref{sec:locformalism}. The black holes discussed in Section~\ref{sec:BPS-BHs-in-N=8} are lifts 
of these~$\CN=2$ solutions to~$\CN=8$ string theory. 
As we shall explain in detail in Section~\ref{sec:black-holes-in-N=8}, the index calculation of these black holes
involves additional subtleties compared to the~$\CN=2$ theory.

\subsection{The formalism of localization in supergravity \label{sec:locformalism}}

\subsubsection{The framework}

Consider a theory of four-dimensional ungauged ~$\CN=2$ supergravity (8 supercharges) coupled to~$\nv$ 
vector multiplets labelled by~$I=1,\dots, \nv$. The on-shell graviton multiplet contains a vector field (the graviphoton), 
so that we have a total of~$\nv+1$ gauge fields in the theory, labelled by~$\Lambda=0, \dots, \nv$ (greek letter labels 
include 0 while italic letters labels do not). This theory has black hole solutions carrying electric and magnetic 
charges~$(q_\Lambda, p^\Lambda)$, which preserve 4 out of 8 supercharges~i.e.~they are~$\frac12$-BPS. 
These solutions can be lifted to~$\cN=8$
supergravity~\cite{Cvetic:1995uj,Cvetic:1995bj,Cvetic:1996zq}, where
they are~$\frac18$-BPS black holes preserving 4 out of 32 supercharges.

The near-horizon configuration is maximally supersymmetric~$AdS_{2} \times S^{2}$ with constant electric and 
magnetic fields. Our starting point is the gravitational path integral (with appropriate boundary conditions) for the near-horizon region of such 
black holes, which should reproduce the index of $\frac{1}{8}$-BPS black holes~\cite{Sen:2008yk,Sen:2008vm}:
\be \label{qef}
W (q, p) &= 
\Bigl\langle \exp\bigl(-\ii \, q_\Lambda \oint  A^\Lambda \bigr) \Bigr\rangle_{\rm{AdS}_2 \times S^2 \text{ with } T \to 0}^\text{reg} \nn \\ 
&= \int [Dg \,D {\Psi} \,D A \, D \Phi] \,\,e^{-S_\text{sugra}^\text{bulk} - S_\text{sugra}^\text{bdry}}\,.
\ee 
We now explain all the elements that go into this formula. 
\begin{itemize}
\item In the first line, the angular brackets indicate the formal Euclidean path integral over the infinite set of fields 
coming from string theory reduced on the compact space, weighted by the action of the theory. 
The lower right subscript means that we perform the path integral on spaces that are asymptotically~AdS$_2 \times S^2$. 
Imposing the boundary conditions on all the fields of the theory needs some care, this is indicated by the 
phrase~$T \to 0$. 
In particular, we first turn on a small temperature~$T$, by imposing Dirichlet boundary conditions on the metric and appropriately 
tuning the value of the proper length of the $AdS_2$ boundary, the size of the~$S^2$, and the boundary values of the 
electric and magnetic fields. We then take a~$T\to 0$ limit of our calculation in order to obtain the  
macroscopic index~$W (q, p)$ of the black hole.

\item In the second line, we have written the full functional integral as an integral over the fields of low-energy
supergravity, namely the metric, the gravitini, a number of gauge fields (schematically labeled by~$D A$) and 
a number of matter fields (schematically labeled by~$D \Phi$). 
Here the action is the effective action of the fields of supergravity obtained by integrating out all the massive fields. 
whose integration measure we choose to be given by the ultralocal measure in the space of fields that 
preserves the symmetries of the theory (e.g.~superdiffeomorphism invariance). 

\item The action of the theory is divergent due to the infinite volume of~AdS$_2$. We regulate this divergence 
by the standard procedure of holographic renormalization indicated by the superscript ``reg". 
The procedure requires us to include appropriate boundary counterterms in the gravitational variables, which
comprise one part of~$S^\text{bdry}_\text{sugra}$ in the second line. 

\item It is more natural in~AdS$_2$ to fix the charge of gauge fields instead of their chemical potential. 
To make the variational problem well-defined for fixed charges requires (in any dimensions) adding 
an electromagnetic boundary term 
$S^\text{bdry}_\text{sugra} \supset \int_{\partial(\text{AdS}_2\times S^2)} d^3 x \sqrt{h} A^\mu F_{\mu \nu} n^\nu$ 
\cite{Braden:1990hw, Hawking:1995ap}. After writing each field strength~$F^\Lambda$ in terms of the charges $q_\Lambda$, 
this boundary term becomes precisely the Wilson loop insertion in the expression above.

\item In order to calculate the functional integral~\eqref{qef} using supersymmetric localization, it is important to  
impose supersymmetric boundary conditions, {i.e.}~compute $\Tr \,(-1)^F \rme^{-\beta H}$ in the putative black hole Hilbert space.
This is equivalent to turning on 
an angular velocity $\Omega$ such that $\Tr\, \rme^{-\beta H + \beta \Omega J } \to \Tr\, \rme^{-\beta H +2\pi i J}=\Tr \,(-1)^F \rme^{-\beta H}$ by the spin-statistics theorem. In the dual 
interpretation this corresponds to studying black holes in the grand canonical ensemble (with respect to angular momentum) with a real Euclidean angular velocity $\Omega_E = i \Omega = 2\pi/\beta$  at the asymptotically flat boundary.\footnote{The answer for the path integral with supersymmetric boundary conditions should be independent 
of temperature so we can compute the gravitational path integral at any temperature as long 
as $e^{\beta \Omega J } = \rme^{2\pi i J}$, corresponding to an analytic continuation of Kerr-Newman solution. 
For simplicity we will work anyways at low temperatures so we can perform the calculation in $AdS_2\times S^2$. }. The gravitational path integral~$W(p, q)$
in~\eqref{qef} can thus be viewed as the ``Witten index'' associated to such black holes.  

Smoothness implies that the configurations that we consider in the path integral should have fermions anti-periodic and bosons periodic around all contractible cycles (which depending on the choice of gauge might \emph{not} be the thermal circle). 
From an AdS$_2$ perspective this requires fixing the chemical potential of the $SU(2)$ gauge field that arises from the isometries of $S^2$, see~\cite{Iliesiu:2021are}\footnote{More precisely, as explained in \cite{Iliesiu:2020qvm,Iliesiu:2021are} fixing any chemical potential at the asymptotically 4d flat boundary corresponds to mixed boundary conditions in the boundary of the AdS$_2$ throat, between the holonomy and the field strength. Whenever we say we fixed a chemical potential in AdS$_2$ we refer to these mixed boundary conditions, and \textit{not} fixed holonomy. }. This point is particularly relevant for the treatment of 
boundary zero-modes in Section~\ref{sec:effect-of-large-gauge-transf}. Nevertheless, choosing such boundary conditions does not affect the path integral over all other fields.

\item Finally, there are a number of scalar fields in the vector supermultiplets and hypermultiplets in $\CN=2$ supergravity. 
The value of the scalars in the vector multiplet are fixed to their attractor value in 
the entire near-horizon region of the classical full-BPS extremal black hole, independently of their values 
at asymptotic flat space. The hypermultiplets, on the other hand, do not couple to the black hole, 
and are flat directions. 
At the boundary of~AdS$_2$ we impose Dirichlet boundary conditions on all the scalars and, in particular, the constant modes for all scalars in~AdS$_2$ are not normalizable. 
The values of 
vector multiplet scalars are fixed in terms of the charges to the attractor values, while the hypermultiplet scalars are  arbitrary.

\end{itemize}

To apply localization to \eqref{qef} we need a formalism where supersymmetry is realized off-shell.
Such a formalism for~$\CN=2$ supergravity coupled to vector multiplets is given by the 
superconformal formalism~\cite{deWit:1979dzm,deWit:1980lyi}. 
In this off-shell formalism, the  theory is described by the Weyl multiplet coupled to~$\nv+1$ vector 
multiplets,  labelled by~$\Lambda=0,\ldots, \nv$, and an auxiliary hypermultiplet. 
Of these, one vector multiplet and the hypermultiplet 
play the role of compensating multiplets for some of the gauge 
symmetries of the superconformal theory~\cite{deWit:1980lyi}. 
These compensating fields are gauged-fixed to reach the super-Poincar\'e formulation.

There are many new issues that show up in  
the application of the localization technique to supergravity, compared to ordinary gauge theories.  
\begin{itemize}
\item The supergravity theory is not UV complete, and so 
the functional integral does not, a priori, make sense. The idea of~\cite{Dabholkar:2010uh,Dabholkar:2011ec} 
was to treat the path integral formally, maintaining consistency with supersymmetry, and reduce it to a sensible integral 
which can then be compared to microscopic string theory. 
\item Supergravity does not, a priori, have a rigid supercharge which is needed for 
localization. 
One can choose the attractor solution 
as a background and use one of its background supercharges as the localizing supercharge, as  in~\cite{Dabholkar:2010uh,Dabholkar:2011ec}, but 
there is an important 
assumption underlying this procedure, namely that all the gauge-invariances of supergravity  are 
consistently fixed in the quantum theory of the fluctuations of supergravity around AdS space. 
This assumption is non-trivial 
because the usual background field quantization of gauge theory does not hold for
supergravity (because its 
structure functions are not constants but field dependent, i.e.~it has a ``soft gauge algebra"). This problem was solved in~\cite{deWit:2018dix,Jeon:2018kec} in the context 
of a space with asymptotic boundary like AdS space. 
The solution involves 
requiring that the fields as well as background ghosts are BRST invariant,
which deforms the nilpotent BRST 
algebra to an equivariant algebra---that can then be used to carry out localization of supergravity on AdS space. 

\item Once we have set up the localization equations, we have to solve them to obtain
the localization manifold~\cite{Dabholkar:2010uh,Gupta:2012cy}. 
\item We have to deal with the fact that supergravity action is written
as a formal infinite series of terms with an arbitrary number of derivatives. 
The way forward is to separate the terms 
into chiral superspace integrals  (F-terms) which are controlled by the topological string, 
and full superspace integrals (D-terms) 
which turn out not to contribute to the localized integral~\cite{Murthy:2013xpa}. 
\item We have to calculate the one-loop determinants of the supergravity multiplets around each localization solution. 
Almost all fluctuations of supergravity are captured by the equivariant cohomology referred to above, and this can be used to 
calculate their quadratic fluctuation determinants. However, as mentioned in the introduction, one has to address the important 
subtlety of the zero-modes appearing from gauge fields, the gravitino, and the graviton.

\item We need to allow orbifold solutions of supergravity that contribute to the functional integral~\eqref{qef}.
\item Finally, we need to adapt the formalism of localization in~$\CN=2$ supergravity to~$\CN=8$ supergravity, where the moduli space of scalars mixes~$\CN=2$ vector and hyper multiplets, paying attention to the measure on the field space.
\end{itemize}

In the following subsections we expand on these points in some detail.

\subsubsection{The localization procedure}

The procedure for defining the localized 
functional integral uses a deformation of the
BRST technique in order to obtain a 
background field quantization of supergravity~\cite{deWit:2018dix}. 
We summarize some of the details in Appendix \ref{sec:BRST-variation-equiv-variation}. 
At a practical level, one begins by 
choosing a Killing spinor~$\varepsilon$ in the background 
attractor geometry, that 
generates a fermionic symmetry obeying the algebra
\be \label{specificQ}
\delta_{\varepsilon}^{2} \=  L_{0} - J_{0} \, , 
\ee
where~$L_{0}$ and~$J_{0}$ are the Cartan generators of the~$SL(2)$ and the~$SU(2)$ algebras of the near-horizon regions, 
respectively. One then promotes $\delta_\varepsilon$ to a covariant operator~$\qeq$ in the full quantum supergravity theory, including the ghosts 
for all the gauge symmetries.  
Here,~$\qeq$ is defined to act on arbitrarily large fluctuations around the attractor background. The details of this procedure for~$\CN=2$ supergravity are presented in~\cite{Jeon:2018kec}. 
We then consider the gauge-fixed functional integral using the action
\begin{equation} \label{eq:BRST-action}
S_\text{sugra} \=  \int  \mathrm{d}^{4} x \,  \Bigl( \mathcal{L}^\text{phys}_\mathrm{sugra} 
- \qeq \bigl( b_\alpha    \, F^\alpha \bigr) \Bigr) \,,
\end{equation}
where we introduce the ghosts/anti-ghost/Lagrange multiplier system $(c_\alpha, b_\alpha, B_\alpha)$ (with $\alpha$ labeling all gauge symmetries) and the gauge-fixing conditions~$F^\a$ are assumed to completely fix all the gauge invariances of the 
theory. $\mathcal{L}^\text{phys}_\mathrm{sugra}$ denote the unfixed supergravity Lagrangian. Next, we deform the action as
\be \label{deformact}
S_\text{sugra} = S(0) \to S(t) = S(0) + \lambda \, \qeq \, \CV \,, 
\ee 
with
\be \label{Vact}
\qquad \CV \= \int \, d^{4} x \, 
\sqrt{\mathring{g}}\, \sum_\psi \, \overline{\psi} \, \qeq \psi  
\ee
summed over all the physical fermions of the theory. The circle on top of quantities indicate it takes values in the supersymmetric background for the metric and matter fields (we will describe what they are below). Since~$L_0-J_0$ is a compact~$U(1)$ isometry,
this deformation obeys the condition~$\qeq^2 \CV=0$. This leads to the result that the functional integral 
reduces to an integral over the critical points of the deformation term, weighted by the original action 
times a one-loop determinant of the deformation action~$\qeq \, \CV$.
The critical points are given by the localization equations
\be
\label{eq:localization-equation}
\qeq \,\psi \= 0  \,, \qquad \text{for all physical fermions} \; \psi \, ,
\ee
to be solved along with the gauge conditions~$F^\a =0$.

\subsubsection{Localization configurations \label{sec:locsolns}}

We now discuss the solution of the equations~\eqref{eq:localization-equation}
with~AdS$_2 \times S^2$ boundary conditions that is relevant for the 
functional integral~\eqref{qef}.
In the superconformal formalism of~$\CN=2$ supergravity, the supersymmetry algebra closes off-shell and we can therefore discuss the
localization equations separately for each supermultiplet.

We start with the Weyl multiplet. As described above, we will impose Dirichlet boundary conditions for the metric at the edge of AdS$_2$.
One class of solutions to the localization equations is 
\be 
\label{eq:AdS2-S2-metric}
ds^2 =  d\rho^2 + \sinh^2 \rho \, d\tau^2 + d\theta^2 + \sin^2\theta d\psi^2 \,,
\ee
written in a gauge where $\sqrt{g}$ is chosen to be set to its asymptotic value (which we choose to be independent of all black hole charges) where we impose that the boundary is at some cut-off value of $\rho = \rho_{\rm bdy}$. Above, $\theta \in [0,\pi]$,
and the angular coordinate of $S^2$ and Euclidean time are identified as 
 \be 
    \label{eq:standard-periodic-ident}
(\tau, \psi) \sim (\tau+2\pi, \psi)\sim  (\tau, \psi+2\pi)   \,.
\ee
Besides \eqref{eq:AdS2-S2-metric},  
when imposing Dirichlet boundary conditions for the metric and gravitino at the edge of AdS$_2$, the metric \eqref{eq:AdS2-S2-metric} can be modified by acting with large diffeomorphisms to generate new solutions of \eqref{eq:localization-equation}. These diffeomorphisms can change the location of the boundary in a physically observable way (for instance the extrinsic curvature of the boundary is affected) and, consequently, the space of large diffeomorphisms should be part of our functional integral.
An exception is given by the isometry group of the background which is $H_{\rm Disk} = PSU(1,1|2)$,\footnote{See~\cite{Banerjee:2009af}, for conventions and a more detailed discussion of this point.} where $SU(2)\subset H_{\rm Disk}$ corresponds to rotations along $S^2$ while $SL(2,\mathbb{R})\subset H_{\rm Disk} $ corresponds to conformal transformations within $AdS_2$. Since large diffeomorphisms do not change the action for any $\lambda$, we will formally include the path integral over large diffeomorphisms in the one-loop determinant around the localization locus of the matter fields which we shall describe below. In the gauge where $\sqrt{g}$ is fixed to its asymptotic value, \eqref{eq:AdS2-S2-metric} and the solutions generated by these large diffeomorphisms are the only supersymmetric configurations in the
Weyl multiplet that are smoothly connected to the full-BPS 
attractor configuration.

In the vector multiplet sector,
the vector potentials  
are given by their on-shell values with field strength
\be 
\label{eq:onshellgaugefield}
F^\Lambda_{\rho \tau} \=-i e_\star^\Lambda~\sinh \rho\,,\qquad 
F^\Lambda_{\theta\psi}\=p^\Lambda~ \sin \theta,
\ee
where $e_\star^\Lambda$ are the electric fields (given by known functions of the charges $q_\Lambda$) of the $n_V+1$ gauge fields near the horizon and $p^\Lambda$ their magnetic charge. The complex scalar fields in the vector multiplets take the form
\be 
\label{eq:vector-mult-scalar-profiles}
X^\Lambda \= \frac{1}2 (e_\star^\Lambda + \ii p^\Lambda) + \frac{\phi^\Lambda - e_{\star}^\Lambda}{2\cosh\rho}\,, \qquad \bar X^\Lambda \= \frac{1}2 (e_\star^\Lambda -\ii p^\Lambda)  + \frac{\phi^\Lambda - e_{\star}^\Lambda}{2\cosh \rho}\,.
\ee
for some real free parameters $\phi^\Lambda$.  

The off-shell~$\mathcal{N}=2$ vector multiplets also contain auxiliary scalars~$Y^\Lambda_{ij}$, transforming as a triplet of the $SU(2)$ R-symmetry,
which take the form 
\begin{equation}
 Y^{\Lambda,1}_1 = -Y^{\Lambda,2}_2 = \frac{\phi^\Lambda - e_\star^\Lambda}{\cosh^2 \rho}\,.
\end{equation}
In contrast to the full-BPS attractor solution, the auxiliary fields $Y^\Lambda_{ij}$ have a nontrivial profile indicating that the localizing saddle point does not extremize the supergravity action, {i.e.}~it is an off-shell configuration.
 The boundary conditions of $X^\Lambda$ at the boundary of AdS$_2$ are fixed their attractor value $X^\Lambda(\rho\to\infty) = \frac{1}2 (e_\star^\Lambda + \ii p^\Lambda)\,$ and  the real parameters $\phi^\Lambda$ give the value of the scalars evaluated at the horizon (i.e. the center of AdS$_2$) $X^\Lambda(\rho=0)= \frac{\phi^\Lambda+\ii p^\Lambda}{2}$ and  we use them as coordinates on the localizing manifold.  

The physical size of  AdS$_2 \times S^2$
in~\eqref{eq:AdS2-S2-metric} at the center is set by 
\begin{eqnarray}
\ell^2 = \rme^{-\CK} \equiv -\ii(X^\Lambda \,\bar{F}_\Lambda   - \bar{X}^\Lambda F_\Lambda)\Big|_{\rho = 0},
\end{eqnarray}
where the scalars~$X^\Lambda, \,\overline{X}^\Lambda$ take the values in equation~\eqref{eq:vector-mult-scalar-profiles}. $\CK$ is called the K\"ahler potential and  $F$ is the holomorphic prepotential. Thus, for the vectors, the most general such smooth solution 
is parameterized by one real parameter~$\phi^\Lambda$ 
in each vector multiplet~$\Lambda=0\,,\,1\,,\dots, \nv\,$.

In the hypermultiplet sector, the scalar fields take constant values which are fixed by boundary conditions. 
No calculation below depends on the value of these constants. However, as we see below, it is important to include the hypermultiplet fluctuations to calculate the correct the one-loop determinant.

We shall see in Section \ref{sec:putting-it-all-together} that the localizing configurations determined so far reproduce the $c=1$ contribution of the Rademacher expansion described in Section~\ref{sec:macroscopic-interpretation-of-BH-degeneracy}. 
With AdS$_2 \times S^2$ asymptotics, and the same canonical boundary conditions for the gauge field and index boundary conditions for the angular velocity,\footnote{See section 3.8 and, in particular, Footnote 4 in the companion paper~\cite{Iliesiu:companionPaper}  for more details about the boundary conditions for the angular velocity.}
there are also other solutions to the localization equations. These are generated by orbifolds of~AdS$_2 \times S^2$~\cite{Dabholkar:2014ema}. 
This can be obtained by considering a quotient of AdS$_2 \times S^2$ by the $\mathbb Z_c$ action generated $\rme^{\frac{2\pi i}c(L_0-J_0)}$.
This is equivalent to taking the metric \eqref{eq:AdS2-S2-metric} and imposing the identification
\be 
\label{eq:orbifold-ident}
(\tau, \phi) \sim \left(\tau+\frac{2\pi}c, \psi+\frac{2\pi}c\right)\sim (\tau, \psi+2\pi)\,.
\ee
This solution also has large diffeomorphisms that are zero-modes of the action. In this case the isometry group is given by
the group $\orbiso $ whose bosonic subgroup corresponds to $U(1)$ rotations around the Euclidean horizon  in $AdS_2$ and  $U(1)$ rotations around the axis $\psi$ on $S^2$ (see Section~2 of~\cite{Banerjee:2009af}). This solution as described so far is singular. 
In the M-theory embedding of the black hole, this 
can be resolved in the following way. Take the RR gauge field associated to D0-charge $A^{0}$, which has a vanishing holonomy around the thermal circle. On the orbifold geometry keep all fields unchanged except for
\begin{eqnarray}
A^{0}_{(c,d)} = A^{0} + \frac{d}{c} d\tau,~~~A^{0} = -i e_\star^0 (\cosh \rho -1) ~d\tau,
\end{eqnarray}
with $d$ an integer defined mod $c$ and $A^{0}_{(1,0)}$ is its value on the original background. This makes the configuration smooth in the following sense: D0-charge in 10d can be geometrized as momentum in 11d along an extra circle. The nontrivial holonomy around thermal circle is equivalent to a rotation proportional to $d/c$ along the new circle. This 11d orbifold is completely smooth as long as $c$ and $d$ are relative prime. 
This is explained in detail in~\cite{Banerjee:2008ky,Murthy:2009dq,Sen:2009gy}.\footnote{After adding the extra M-theory circle $\xi \sim \xi + 2\pi$ the geometry is $AdS_2\times S^2 \times S^1$ with momentum along $S^1$ and the identification $(\tau,\psi,\xi) \sim (\tau+2\pi/c, \psi+2\pi/c, \xi + 2\pi d/c)$. 
The coordinates used in this paper correspond to shifting $\xi \to \xi' + \tau d/c$ such that the orbifold does not act on $\xi'$ and then dimensionally reducing on $\xi'$. 
This coordinate system is more convenient from the 4d perspective. } This motivates including these orbifolds as an allowed localization configuration and this is supported by the index describe in \ref{sec:macroscopic-interpretation-of-BH-degeneracy} since they correspond to each term in the Rademacher expansion\footnote{We could also shift the holonomies of the other gauge fields in the 
black hole charge configuration. 
Indeed, such a shift in a gauge field coupling to a particular D2-brane (determined by the background charges) is important to reproduce 
the multiplier matrix~\cite{Dabholkar:2014ema}. 
}

\subsubsection{The holomorphic prepotential}

The next step is to evaluate the action of the supergravity theory on the localization locus, first for the regular black hole solution \eqref{eq:standard-periodic-ident}, and then for the orbifold solutions \eqref{eq:orbifold-ident}. 
A special subset of terms in the~$\CN=2$ supergravity action comes from the so-called holomorphic prepotential~$F$, which 
is a degree-2 homogeneous function of the 
vector superfield (each of weight~1) and the 
Weyl-squared superfield (of weight~2). (More precisely, it is a function of the associated ``reduced chiral superfields".)
This action is a chiral superspace integral of the prepotential function~$F$ 
and can produce terms with an arbitrary number of derivatives. 
The holomorphic prepotential generically has the form
\be \label{prepF}
F(X^I) \= C_{IJK} \frac{X^I X^J X^K}{X^0} \;+\; \dots \,,
\ee
where the dots correspond to the higher-derivative terms. The leading-order cubic term is governed 
by the completely symmetric three-tensor which, for a CY$_3$ compactification of string theory,
is the intersection form of the CY$_3$. 

At two-derivative level, the complete action of~$\CN=2$ supergravity is given by the action arising from the 
prepotential,   
while at higher-derivative level the action can also contain other terms coming from full superspace integrals, so called D-terms. 
It was shown in~\cite{Murthy:2013xpa} that all known full superspace integrals vanish on the localization locus. 

\subsubsection{The integral over the localization locus}

The action of the holomorphic prepotential on the localization solutions
takes a very simple form in the gauge where~$\sqrt{g}$
is fixed to its asymptotic value, leading to the following result~\cite{Dabholkar:2010uh}
for the functional integral~\eqref{qef}, localized around the leading black hole saddle
\begin{equation} \label{LocIntegral}
 W^\text{(1)} (q, p) \= \int_{\mathcal{M}_{Q}}  \,[\dd^{n_v+1}\phi] \, \exp\Big(- \pi  \, q_\Lambda  \, \phi^\Lambda 
 + 4 \pi \, \Im{F \Big(\frac{\phi^\Lambda+\ii p^\Lambda}{2} \Big)} \Big)
 \, Z^{\qeq\CV}_\text{1-loop}(\phi^{\Lambda})\,.
\end{equation}
Here~$[\dd^{n_V+1}\phi]$ is the measure in field space over the coordinates of the localization manifold, 
and~$Z^{\qeq\CV}_\text{1-loop}(\phi^{\Lambda})$ is the one-loop determinant of the~$\qeq \CV$ action over all 
the non-BPS directions orthogonal to the localization locus. $Z^{\qeq\CV}_\text{1-loop}(\phi^{\Lambda})$ receives two contributions: that of the bulk modes (which vanish sufficiently quickly at the boundary of AdS$_2$), and that of the path integral over the localization locus of large gauge transformations which as mentioned earlier. 
The formula~\eqref{LocIntegral} is an all-order perturbation theory result around the attractor configuration 
(for this reason it is sometimes called~$W^\text{pert}$), and the superscript indicates that this 
the first of an infinite series of non-perturbative gravitational saddles, as we discuss below. Similarly, performing the path integral around the orbifolded solutions and evaluating the action in terms of the holomorphic prepotential yields a contribution
\begin{equation} \label{LocIntegral-orbifolds}
 W^\text{(c)} (q, p) \= \int_{\mathcal{M}_{Q}}  \, [\dd^{n_v+1}\phi]_c \, \exp\Big(- \frac{\pi  \, q_\Lambda  \, \phi^\Lambda}c 
 + \frac{4 \pi}c \, \Im{F \Big(\frac{\phi^\Lambda+\ii p^\Lambda}{2} \Big)} \Big)
 \, Z^{\qeq\CV}_{\text{1-loop},\,c}(\phi^{\Lambda}) Z^\text{top}_{c}(q, \, p)\,,
\end{equation}
that turns out to be non-perturbatively suppressed compared to the contribution of the standard black hole geometry in \eqref{LocIntegral}. The measure $[\dd^{n_v+1}\phi]_c$ originates from the same ultra-local measure that we consider in the  supergravity problem for the unorbifolded solutions (and will shortly be evaluated explicitly), 
and~$Z^{\qeq\CV}_{\text{1-loop},\,c}(\phi^{\Lambda})$ is the one-loop determinant of the~$\qeq \CV$ action around the orbifolded solution. Finally, as we shall review in more detail shortly, in addition to the contributions that were present on the regular black hole background, there is an additional  contribution on orbifolds, $ Z^\text{top}_{c}(q, \, p)$, which comes from evaluating the contribution of topological terms of the action on such geometries and includes a dependence on $d$ as well as other topological data.\footnote{On the regular black hole geometry, such terms evaluate to trivial, charge-independent, terms. }

The formalism of this subsection applies directly to~$\CN=2$ supergravity coupled to a number of vector multiplets. We now extend this formalism to black holes in~$\CN=8$ supergravity.

\subsection{Black holes in~$\CN=8$ supergravity}
\label{sec:black-holes-in-N=8}

The field content of~$\CN=8$ supergravity
in the~$\CN=2$ formalism consists of a graviton multiplet, 6 spin-$\frac32$ multiplets, 15 vector multiplets, 
and 10 hyper multiplets, with a total of~$1+6 \times 2 + 15 \times 1 + 10 \times 0 = 28$ gauge fields 
and $15 \times 2 + 10 \times 4=70$ scalars. 
Consider the lift of the black hole solution in the~$\CN=2$ theory discussed above to~$\CN=8$ supergravity. 
The extra~$\CN=2$ hypermultiplets and~$\CN=2$ spin-$\frac32$ multiplets that 
occur in the field content of~$\CN=8$ supergravity do not couple to the black hole classically. 
We can, therefore, calculate the classical action for the~$\CN=2$ Weyl and vector multiplets in the 
black hole background, and then take into all the multiplets---including the hypers and spin~$\frac32$---at the 
quantum level by calculating their one-loop determinant.

Let us first discuss the scalar fields. 
The separation of the~70 scalars into~$\CN=2$ vector multiplets and hyper multiplets is not immediately obvious, 
because the scalar manifold~$SU_c(8)\backslash E_{7(7)}$ \cite{Cremmer:1979up,deWit:1986oxb} mixes the vector and hyper 
scalars. (At the level of the~$\CN=8$ Lagrangian, this is seen through the kinetic mixing matrix.)
At a given point in field space we can  decompose the 
scalars into~$\CN=2$ vectors and hypers, but this decomposition changes as we move in field space and we cannot do a global reduction to say a theory with only vectors. 
The fact that allows us to proceed with an~$\CN=2$ reduction is that the attractor mechanism 
picks out a special choice of directions.  
Recall that the coupling of the vector multiplet scalars to gravity 
is governed by the prepotential. In background flat space these scalars do not experience a potential,
but near the horizon of the extremal black hole they do experience a locally quadratic potential, with the values of the scalars at the minimum determined completely by the black hole charges. 
The hyper multiplet scalars, on the other hand, are flat directions of the potential,
and form the coset~$(SU(2) \times SU(6))\backslash E_{6(2)}$~\cite{Bossard:2009we}.

Now we can perform the localization calculation of the path integral. For the vector multiplets
we use the~$\CN=2$ formalism of~\cite{deWit:1979dzm,deWit:1980lyi} as before. Although there is no Lorentz covariant 
off-shell formalism for~$\CN=2$ hypermultiplets with a finite number of auxiliary fields, 
we can, nevertheless, close the supersymmetry algebra off-shell on hypermultiplets for one complex 
supercharge~\cite{Hama:2012bg,Murthy:2015yfa}.
In the off-shell theory, the vector and hyper multiplets are decoupled.
The supersymmetric configurations consist of a one-parameter family of configurations
in each vector multiplet (which vary inside~AdS$_2$ as in Section~\ref{sec:locsolns}), while the hyper multiplets are 
completely fixed to their classical value by supersymmetry~\cite{Murthy:2015yfa}. 
At that fixed value of the hypers, we can go beyond the local quadratic fluctuations
for the vector scalars, and use the full~$\CN=2$ prepotential for the supersymmetric configurations.
For all the other (non-BPS) field modes (including the hyper-multipelts), it suffices to consider only quadratic fluctuations around each BPS configuration. 
In this manner we reach a consistent truncation to a theory of~$\CN=2$ supergravity coupled 
to~15 vector multiplets, whose scalars form the coset manifold~$SO^*(12)/(SU(6) \times U(1))$~\cite{Pioline:2005vi, Pioline:2006ni}.

Finally we consider the~$\CN=2$ spin-$\frac32$ multiplets. These multiplets do not have scalar fields and so
they do not affect the preceding discussion, but there is the ``opposite" problem to consider, namely they 
contain gauge fields under which the black hole can carry charges, and so they couple to the black hole. 
In order to disentangle these vector fields, 
recall that the~$\CN=2$ attractor mechanism singles out one gauge field
(the on-shell graviphoton) which carries all the charge of the black hole. In other words, one can make a 
symplectic transformation on the~$\nv+1$ vector multiplets so that the charges of the black hole under 
all gauge fields, except the graviphoton, vanish. Relatedly, the horizon area is given by the simple 
formula~$A= \pi \rme^{\CK} |Z|^2$, where~$Z = \rme^{\CK/2}(p^\Lambda F_\Lambda - q_\Lambda X^\Lambda)$ is the~$\CN=2$ central charge, and~$\rme^{-\CK}= \ii (F_\Lambda \overline{X}^\Lambda - \overline{F}_\Lambda {X}^\Lambda)$ defines the K\"ahler potential.
Now, a similar statement is true in the~$\CN=8$ theory: now we have a central charge antisymmetric matrix~$Z_{ij}$, $i,j=1,\dots, 8$, with four skew eigenvalues. 
The largest skew eigenvalue~$Z$ determines the potential and the horizon area
as above. This central charge effectively picks out the graviphoton and therefore 
the~$\CN=2$ subalgebra underlying the black hole. 

The bottom line is that, by a suitable choice of coordinates in the space of charges, we can 
view the classical black hole in the~$\CN=8$ theory as a black hole in an~$\CN=2$ theory of supergravity
coupled to 15 vector multiplets.
The full path integral now factors into a finite-dimensional integral over the~$\CN=2$ vector 
multiplet off-shell BPS configurations whose action is determined by the prepotential, and the one-loop
determinant over all the non-BPS configurations. The latter space consists of the rest of the modes of the 
Weyl and vector multiplets, and all the modes of the hyper multiplets and spin-$\frac32$ multiplets.
The next step is to determine the prepotential of the vector multiplet theory explicitly.

\subsubsection{The prepotential of the black hole in the~$\cN=8$ theory}\label{sec:N8prepot}

In order to calculate the prepotential, we use the string theory description of the system as 
Type IIA on~$T^6$, whose associated M-theory lift on an additional circle is the one described in~\cite{Maldacena:1997de}. 
In the M-theory description, each vector multiplet~$I=1,\ldots, 15$ is associated with one of the $({}^6_2)=15$ 2-cycles 
(or its Hodge-dual 4-cycle) inside $T^6$. For any Calabi-Yau 3-fold, the classical (tree-level) prepotential is given 
by the intersection numbers of these 4-cycles, i.e., 
\begin{eqnarray} \label{Fcub}
F^{(0)}(X) \= C_{IJK} \frac{X^I X^J X^K}{ X^0} \,, \qquad 
C_{IJK} \,\equiv \, \frac{1}{6} \int_{T^6} \alpha_I \wedge \alpha_J \wedge \alpha_K \,,
\end{eqnarray}
where~$C_{IJK}$ is the intersection matrix 
of~$T^6$ and $\alpha_I$ is an integral basis for $H^2(T^6;\mathbb{Z})$. 
Recall that the exact prepotential can be calculated as a genus expansion of the topological string 
on the CY$_3$~\cite{Bershadsky:1993cx, Antoniadis:1993ze}. In our case where~CY$_3=T^6$, the holomorphic prepotential 
is tree-level exact, i.e.~\eqref{prepF} does not receive any corrections.
This can be understood as due to the extra fermion zero-modes on $T^6$ which are the superpartners of 
the translation symmetries on the torus.

For concreteness, we use the identification between the 15 vector 
multiplet scalars and the corresponding 2-cycles described by equation \eqref{eqn:XvsSigma}. The precise choice will not be very important except when comparing with possible truncations of the 
theory, and therefore we leave it in Appendix~\ref{app:truncation}. The intersection matrix between cycles $ij$, $kl$ and $mn$ 
with $i,j,k,l,m=4,\ldots,9$ is given by $C_{ij;kl;mn}= \frac{1}{6}\varepsilon_{ijklmn}$ and using the relation 
\eqref{eqn:XvsSigma} we can obtain $C_{IJK}$ in the basis relevant for \eqref{Fcub}.

The electric and magnetic charge vectors $(q_\Lambda, p^\Lambda)=(q_0,q_I , p^0, p^I)$ of the black hole are given in terms of the number of D-branes wrapping cycles 
of~$T^6$ in the Type-IIA frame. The parameter~$q_0$ counts the number of D0-branes, $q_I$ the number of D2-branes wrapping 2-cycles in~$T^6$, $p^I$ the number of D4-branes wrapping the respective Hodge-dual 
4-cycles, and~$p^0$ the number of D6-branes wrapping~$T^6$. 
In the M-theory frame, the D0, D2, D4 branes 
lift to momentum around the M-theory circle, 
M2 branes transverse to the M-theory circle, and M5-branes wrapping the M-theory circle respectively. 
Note that D6-branes in string theory, as well as the KK-monopole manifestation in M-theory, 
are intrinsically heavy non-perturbative objects, and they have a very large backreaction even at small coupling. 
For this reason we restrict attention to~$p^0=0$, but otherwise keep an arbitrary number of D0-D2-D4 branes.

 $\mathcal{N}=8$ supergravity resulting from compactifying Type IIA on $T^6$ enjoys U-duality and the charge invariant \eqref{quarticinv} for arbitrary R-R charges and vanishing NS-NS charges (coming in $\mathcal{N}=2$ language from the $12$ vectors in the spin-3/2 multiplets) reduces to 
\begin{equation}
    \Delta = 4C_{IJK}p^Ip^Jp^K \Big(q_0+\frac{1}{12}  q_I C^{IJ} q_J\Big),\label{eq:DeltaRRcharges}
\end{equation}
where $C^{IJ}$ is the inverse of the matrix $C_{IJ}\equiv C_{IJK} p^K$. Setting the charges in the gravitino multiplets to zero preserves, $SO^*(12;\,\IZ)$ out of the full U-duality group. 
This~$SO^*(12;\,\IZ)$ naturally acts on the coset~$SO^*(12)/(SU(6) \times U(1))$
and the~16 remaining charges transform as a spinor under this~$SO^*(12;\,\IZ)$~\cite{Pioline:2006ni}.
Since we further set $p^0=0$, we only preserve $USp(6;\mathbb{Z})$ which acts on the remaining 15 charges and under which $\Delta$ in \eqref{eq:DeltaRRcharges} is invariant. 
 
The index calculation in the microscopic theory is done in a dual type IIB theory on $T^6$ with 
D1-D5-P-KK charges. The duality transformations between the localization calculation and the 
microscopic index calculation are described in Section~3.1 of~\cite{Dabholkar:2011ec}. 
In the IIA frame, a simple choice of charges with a non-zero invariant $\Delta$ involves the following charges. Denoting the six-torus by $T^6= T^4 \times S^1 \times \wt{S}^1$, we can choose~$q_0$ D0-branes, $q_1$ D2-branes wrapping 2-cycle $S^1 \times \wt{S}^1$, $p^{1}$ D4-branes wrapping 4-cycle $T^4$, $p^{2}$ D4-branes wrapping 4-cycle
$\Sigma_{67}\times S^1 \times \wt{S}^1$, and~$p^{3}$ D4-branes wrapping 
4-cycle $\Sigma_{89}\times S^1 \times \wt{S}^1$. 
The relation to the dual IIB frame microscopic result of Section~\ref{sec:BPS-BHs-in-N=8} is~$q_{0}=n$, $q_1=\ell$, and $(p^1,p^2,p^3) = (\wt K, Q_1, Q_5)$. For this configuration the U-duality invariant given in \eqref{eq:DeltaRRcharges} is given by $\Delta = 4 p^1 p^2 p^3 q_0 - (q_1 p^1)^2=4n-\ell^2$ for $(\wt K, Q_1, Q_5) = (1,1,1)$.

As we shall see in more detail in Section~\ref{sec:putting-it-all-together}, evaluating the prepotential $F(X)$ in the exponent of \eqref{LocIntegral} or \eqref{LocIntegral-orbifolds} and integrating-out the moduli $\phi^1,\,\dots,\phi^{15}$, which all have a Gaussian weight, yields a single integral over $\phi^0$. After the change of variable, the exponential in the integral over $\phi^0$ precisely matches that in the Bessel function that one expects from the microscopic description presented in Section~\ref{sec:microscopic-index}. What remains to be done however, is to match all one-loop factors of $c$ and $\phi^0$ in the Radamacher expansion of the microscopic degeneracy \eqref{RadexpC}. Before understanding how to compute this one-loop determinant 
we first discuss the integration measure in~\eqref{LocIntegral},~\eqref{LocIntegral-orbifolds} 
for the moduli $\phi^\Lambda$, and the topological term in~\eqref{LocIntegral-orbifolds}.  

\subsubsection{The integration measure along the localization locus}

In the localization calculation of the gravitational partition function computing the black hole index described before, the measure over the manifold $[\dd^{16}\phi]$ is not fixed by symmetries in an obvious way since for example the vector multiplets are abelian. If we had a path integral representation defining the theory this could be used as a starting point to derive the induced measure on the localization manifold. The situation in supergravity is less clear since the UV completion is not a quantum field theory but string theory. 

A natural candidate for $[\dd^{16} \phi]$ is the following. An ultralocal measure for a path integral is defined such that $\int [D\Phi^m] \rme^{- \int \sqrt{g} G_{mn}(\Phi)\delta \Phi^m \delta \Phi^n }=1$ where $\Phi^m$ denotes the set of fields in the theory and $G_{mn}$ defines an inner product. 
In the problem that we consider, the only dependence on the orbifold parameter $c$ appears, as a factor of~$1/c$ in the integral
of the exponent of the ultralocal measure. 
The dependence with the charges is still undetermined by this argument without knowing more details about the ultralocal measure. We propose to take $\partial_\Lambda \bar{\partial}_{\Sigma} \rme^{-K} = {\rm Im} F_{\Lambda \Sigma}$ as a natural metric on the localization manifold since this matrix appears in the quadratic action of the vector fields, where $F_{\Lambda\Sigma}$ is the $16\times 16$ matrix of second derivatives of the prepotential. This gives\footnote{In \cite{Dabholkar:2011ec} it was proposed to derive $[\dd \phi]$ from an ultralocal measure that diagonalizes kinetic mixing for the scalar fields in vector multiplets \cite{Dabholkar:2011ec}. It is important to note, however, that this ultralocal measure is separate from the one-loop determinant that we calculate in section \ref{sec:one-loop-det-theoretical-prelims} in contrast to what was initially suggested in \cite{Dabholkar:2011ec}. This distinction was further addressed in \cite{Murthy:2015yfa}.}
\begin{eqnarray}
[\dd^{16}\phi]_c \equiv  \dd^{16}\phi~ \sqrt{ 2~{\rm det}\Big( \frac{1}{2c} ~{\rm Im} ~F_{\Lambda\Sigma} \Big) },\label{eq:measure-over-MQ}
\end{eqnarray}
where $\dd^{16}\phi = \prod_{\Lambda=0}^{16} \dd \phi^\Lambda$. This choice remains to be derived from first principles but we will see it satisfies very interesting properties. The choice of numerical prefactor is arbitrary. 

Since we will restrict to $p^0=0$ we can compute this determinant in the following way. First separate the symmetric $16\times 16$ matrix into a $15\times 15$ matrix ${\rm Im}(F_{IJ}) = \frac{6C_{IJ}}{\phi^0}$, two 15-dimensional vector when one component is zero ${\rm Im}(F_{0I})={\rm Im}(F_{I0})= -\frac{6 C_{IJ}\phi^J  }{ (\phi^0)^2}$ and the 00-component ${\rm Im}(F_{00})=- \frac{2C_{IJ}p^Ip^J}{(\phi^0)^3}+\frac{6C_{IJ}\phi^I\phi^J}{(\phi^0)^3}$. The determinant can be computed in terms of these four blocks and the result simplifies considerably:
\begin{eqnarray}
[\dd^{16}\phi]_c =\dd^{16}\phi~\frac{1}{c^8}\sqrt{\frac{ -2C_{IJK}p^Ip^Jp^K ~{\rm det} \left(3C_{IJ} \right)}{(\phi^0)^{18}}}\,,\label{eq:measurelocN8}
\end{eqnarray}
where we defined the $15\times15$ matrix $C_{IJ} = C_{IJK} p^K$ whose determinant appears in the second line. This choice of measure is motivated by the way the scalars appear in the action, but it also has some further encouraging properties. First, it depends only on $\phi^0$, at least when $p^0=0$. Second, we will see the precise function of charges that appear in the numerator is necessary in order to obtain a T-duality invariant black hole index. Finally, the factor of $c$ will be necessary to reproduce the sum in the Rademacher expansion. Even though this measure works for the black holes in $\mathcal{N}=8$ supergravity, it is an open question whether it gets corrections in situation with less supersymmetry.

\subsubsection{The topological term and Kloosterman phases}

The term~$Z^\text{top}_{c}$ in~\eqref{LocIntegral-orbifolds} comes from terms in the action of the supergravity theory 
that are topological or non-local in nature, and account for the~$c$-dependence of the phases in the Kloosterman sum 
in the expression~\eqref{RadexpC},~\eqref{kloos}~\cite{Dabholkar:2014ema}. We now briefly summarize these 
observations. 
Recall that the black hole in embedded in the full string theory as a black string wrapping an~S$^1$. 
The holographically dual theory is a~$(4,4)$ SCFT$_2$, and the near-horizon geometry 
is~AdS$_3 \,\times$~S$^3$~($\times \, T^4$) with~$SU(2)_L \times SU(2)_R$ R-symmetry. 
There are two circle reductions which are important. 
Upon reducing~AdS$_3$ to~AdS$_2 \, \times$~S$^1$ one obtains the BMPV BH. Here 
the~$J_L$ rotation on the~S$^3$ breaks the symmetry to~$U(1)_L \times SU(2)_R$. 
On the other hand, we can reduce~S$^3$ to~S$^2 \, \times \, \wt S^1$ such that~$SU(2)_R$ is the 
rotational symmetry of~S$^2$ and~$U(1)_L$ rotates the~$\wt S^1$.  
The near complete-horizon geometry of the black hole is~AdS$_2 \,\times$~S$^2 \times S^1 \times \wt S^1$.

Now consider the configuration with one electric charge~$q^0=n$ as above. 
In the near-horizon geometry the electric charge~$n$ is the momentum 
around~$S^1$. The macroscopic observable~\eqref{qef} contains the Wilson line term. 
This Wilson line gives~$\exp(2 \pi \ii n) =1$ on the~AdS$_2$ geometry~($c=1$), but gives a 
non-trivial phase~$\exp(2 \pi \ii n d/c) =1$ on the orbifold~\eqref{eq:orbifold-ident} which includes 
a shift on the~S$^1$. Recalling that~$\Delta = 4n-\ell^2$, we see that this term precisely accounts for 
the first term in the sum~\eqref{kloos}.

Now, recall that the~$\IZ_c$ orbifold also acts on the~S$^2$ part of the geometry in order to preserve supersymmetry.
Thus it creates flux in the four-dimensional geometry which has an action associated to it. 
One way to calculate this action is to consider the lift to~AdS$_3$. As explained in~\cite{Murthy:2009dq},
the~$(c,d)$ orbifolds can be thought of as the near-horizon geometry of the 
Maldacena-Strominger family of~AdS$_3$ geometries with~$T^2$ boundary. 
On such a solid torus geometry, the~$SU(2)_R$ gauge field has a non-abelian Chern-Simons action, 
and the value of its bulk action precisely accounts for the last term~$\exp \bigl(-2\pi \ii a/4c \bigr)$ of the sum~\eqref{kloos}. 

As explained in~\cite{Dabholkar:2014ema}, the first Wilson line can also be understood
as arising from the abelian CS theory associated to~$U(1)_L$.  In this description the Wilson line arises 
as the boundary action needed to make the Chern-Simons theory 
well-defined, evaluated on the torus boundary. 

The final piece in~\eqref{kloos} is the multiplier system~$M(\gamma_{c,d})^{-1}_{\nu 1}$ given explicitly 
in~\eqref{multiplier1}. The phases in the exponent of~\eqref{multiplier1} arise from the evaluation of the bulk 
non-abelian CS term associated with the~$SU(2)_L$ gauge field on the~AdS$_3$. This calculation is quite subtle 
and uses the fact that the CS invariant on the~$(c,d)$ solid torus can be mapped to the 
Dehn surgery formula for the unknot on Lens spaces \cite{Witten:1988hf,Jeffrey:1992tk,Rozansky:1994qe,Beasley:2009mb}. 
The Rademacher~$\Phi$-function arises as the framing correction. 

Putting all these together we arrive at the~$(c,d)$ dependent part of the exponent of the Kloosterman sum~\eqref{kloos}.
The Chern-Simons partition functions do not have any additional~$(c,d)$-dependent pre-factor \cite{Beasley:2009mb,Rozansky:1994qe}.
Thus we are left with one overall constant factor which we determine by comparing the~$c=1$ term in~\eqref{RadexpC}. 
The final answer is 
\be 
Z^\text{top}_{c}\= \sqrt{c} \, K_c(\Delta) \,.
\ee
Here, we should note that the factor of $\sqrt{c}$ is meant to cancel the factor present in the definition \eqref{multiplier1} of the multiplier matrix, in order to reproduce the $(c,\,d)$-independent one-loop determinant of Chern-Simons theory.

\section{One-loop determinants around localizing saddles}
\label{sec:one-loop-det-theoretical-prelims}

The remaining bit of input needed in the localized integral~\eqref{LocIntegral} is the one-loop determinant. It was argued in~\cite{Murthy:2015yfa} that since there is only one scale 
set by~$\ell^2=\rme^{-\CK}$ in the localization background, the one loop determinant can only depend on $\ell$ and on the parameter $c$ of the orbifold. When restricting to $p^0=0$, the K\"ahler potential for~\eqref{Fcub} evaluated on the localization locus takes a simple form 
\be \label{KahlerVal}
\rme^{-\CK(\phi^\Lambda)} = \frac{-2C_{IJK}p^I p^J p^K}{\phi^0} \,,
\ee 
and thus the one-loop determinant can depend only on $\phi^0$, on the combination of charges $C_{IJK}p^I p^J p^K$ and on the orbifold parameter $c$. Computing the exact dependence of the one-loop determinant on these parameters is crucial in order to match the microscopic value of the degeneracy discussed in Section~\ref{sec:microscopic-index}.

As explained in the introduction, the one-loop determinant has two contributions, neither of which has so far been fully determined:
\begin{itemize}
    \item \textbf{Contribution captured by an index.} A formalism for calculating the one-loop determinant of such modes in generic~$\CN=2$ supergravity  theories in the off-shell formalism was developed in~\cite{Jeon:2018kec} by using the Atiyah-Bott fixed-point theorem. In order to apply this computation to the~$\CN=8$ theory we need to calculate the index associated to several ~$\CN=2$ supermultiplet: the graviton, the vector, the hyper, the chiral and the gravitino supermultiplets. While the index has been computed for the former three on AdS$_2 \times S^2$, the index of the latter two has not been computed. In the particular case of the gravitino supermultiplet, a difficulty in determining the index comes from the lack of an off-shell $\cN=2$ formulation. In this section, as well as in Section~\ref{sec:one-loop-det-results}, we will determine the index for all these multiplets by working in the off-shell $\cN=4$ formulation, which from the $\cN=2$ perspective includes both the gravitino and chiral supermultiplets.

    Additionally, in order to fully compute $W^{(c)}(q_\Lambda,p^\Lambda) $ in \eqref{LocIntegral-orbifolds} one has to also compute the index on the orbifold geometries which, from the four-dimensional perspective, are $\mathbb Z_c$ quotients of AdS$_2 \times S^2$. This calculation was not done previously in this context. In Section~\ref{sec:one-loop-det-orbifold}, we will adapt the index theorem to compute the one-loop determinant on the quotient geometry for all the above supermultiplets. While the $c$-dependence in the one-loop determinant of each of the previously mentioned $\cN=2$ supermultiplets turns out to be complicated, the one-loop determinant of the bulk modes in the $\cN=8$ theory turns out to be entirely $c$-independent.

    For readers solely interested in the actual value of the one-loop determinants (or at least their part captured by the index theorem) for the various supermultiplets in $\cN=2$, $4$ or $8$ supergravity theories, rather than their derivation, could skip to Section~\ref{sec:summary-results-bulk-one-loop-det}.

    \item \textbf{Contribution of zero modes.} As was already observed in \cite{Jeon:2018kec}, the contribution of the large gauge transformations (also called boundary modes)
    mentioned in Section~\ref{sec:locformalism} are not correctly captured in the index computation.
    Some boundary modes are not zero-modes of the action, and their contribution to the determinant can be computed from the index by carefully understanding their role in the $\qeq$-cohomology. Other boundary modes are zero-modes of both the deformed and undeformed supergravity action and, therefore, are technically part of the localization locus and cannot be treated only to quadratic order. To compute their contribution to  $Z_{\text{1-loop},\,c}^{Q_{\text{eq.}} \CV}(\phi^\Lambda)$ we will thus have to perform the explicit path integral over them. This is especially important if we compare their contribution on the AdS$_2 \times S^2$ geometry and on the orbifolded geometry, where the path integral could have a non-trivial $c$-dependence.
    In section  \ref{sec:integration-measure-large-gauge-transf} and \ref{sec:connection-to-N=4-super-Schw}, we will show that the path integral over the boundary zero-modes can be studied by taking the zero-temperature limit of the $\cN=4$ super-Schwarzian path integral with AdS$_2$ boundary conditions (in the case of the AdS$_2 \times S^2$ saddle) or with boundary conditions that correspond to the insertion of a supersymmetry preserving defect (in the case of the orbifolded saddles).    
\end{itemize}

Before we proceed with our one-loop determinant computation, it is worth noting some leading-order results for the one-loop determinant obtained by expanding the $\cN=8$ supergravity action to quadratic order around a classical solution (and therefore less generic than the localization manifold) and extracting the logarithmic area correction to the black hole entropy. 
As  mentioned above, the logarithmic correction to the entropy obtained in the large $\Delta$-expansion of the leading term in the Radamacher expansion of the microscopic result seen in~\eqref{RadexpC} (corresponding to the regular black hole geometry) was reproduced in~\cite{Sen:2011ba}. More recently however, the one-loop determinant on the four-dimensional~$\mathbb Z_c$ orbifold geometry was also analyzed using similar 
heat-kernel methods~\cite{Gupta:2013sva, Gupta:2014hxa} where it was found that for $\cN=8$ supergravity, the logarithmic correction in the area of the black hole is $c$-independent. This is once again consistent with the large-$\Delta$ expansion of the subleading terms in the Radamacher expansion \eqref{RadexpC} whose power of $\Delta$ is $c$-independent. This computation is however insufficient to determine the full $c$-dependence of the one-loop determinant and for that we will have to go in detail through the localization computation. 
Nevertheless, the results of \cite{Gupta:2014hxa} serve as a non-trivial check for our localization results for each individual $\cN=2$ Weyl, vector supermutiplet and hypermultiplet: 
they show that each one of these $\cN=2$ supermultiplets yields a logarithmic correction in the area that has a non-trivial $c$-dependence, which we reproduce from our index computations.

\subsection{One-loop determinant on AdS$_2\times S^2$ from an index theorem}
\label{sec:one-loop-determinant-AdS2-S2}

In this subsection we review the calculation of~\cite{Jeon:2018kec} of the superdeterminant of the operator~$\qeq \CV$ 
with~$\CV$ defined in~\eqref{Vact}. We first organize all the fluctuating fields of the theory into 
cohomological variables, i.e.~representations of the equivariant algebra~$\qeq^2= H$ with~$H=L_0-J_0$ 
as in~\eqref{specificQ}. The representations take the form~$(\Phi\,, \qeq \Phi\,,\Psi\,, \qeq \Psi)$, 
where~$\Phi$ and~$\Psi$ are a set of local bosonic and fermionic fields, respectively, called elementary 
fields and their~$\qeq$-partners are called descendants. In this basis one can see the one-loop determinant is elegantly obtained from the Atiyah-Bott fixed-point formula. 
Nevertheless, this formula is not complete and has corrections from boundary modes.

We denote the contribution to the one-loop determinant obtained from the index by~$Z_{\rm index}$, and 
use~$Z_\text{one-loop}$ for the full one-loop determinant that includes the full path integral over the moduli space of zero-modes. We shall explain the relation between  $Z_{\rm index}$ and 
$Z_\text{one-loop}$ in great detail in Section \ref{sec:effect-of-large-gauge-transf}, while in this section we will solely focus on computing  $Z_{\rm index}$ for all supermultiplets in $\cN=2$ supergravity.

\vskip 0.2cm

\subsubsection{The superdeterminant from the Atiyah-Bott fixed-point formula}

As the supercharge~$\qeq$ pairs up the fields algebraically at each point in space,  
all the contribution to the superdeterminant can be understood as a mismatch between the elementary bosons 
and fermions, which is kept track by an operator~$D_{10}: \Phi \to \Psi$. 
An algebraic analysis then shows that the ratio of determinants of the fermionic 
and bosonic kinetic operators in~$\qeq \CV$ then reduces to the ratio
\be \label{detratio}
Z_\text{index} \= \sqrt{\frac{\det_{\Psi} (H/\ell)}{\det_{\Phi}(H/\ell)}} \,,
\ee
where the scale $\ell$ gives the  overall scaling in the physical size of the spacetime with~$\ell^{2} = \rme^{-\CK (\phi^{\Lambda}+\ii p^\Lambda)}$.

Any mode which is not in the kernel or cokernel of~$D_{10}$ does not contribute to this ratio.
Thus the ratio of determinants on the right-hand side can thus be computed from the knowledge of the index,\footnote{Here we will use a convention in which $H$, $L_0$, and $J_0$ are all anti-Hermitian.}
\be \label{indD10}
\text{ind} (D_{10})(t) \; \coloneqq \; \Tr_\text{Ker$D_{10}$}  \, \rme^{\frac{tH}\ell} - \Tr_\text{Coker$D_{10}$}  \, \rme^{\frac{t H}\ell} \,.
\ee
which following from \eqref{detratio} is thus simply related to the difference between the heat-kernels of the bosonic and fermionic differential operators. Specifically, 
writing the index as a series,
\be \label{indexser}
\text{ind} (D_{10})(t) \= \sum_{n} \a(n) \, \rme^{\ii  {\lambda_{n} t}} \, , 
\ee 
we can read off the eigenvalues~$\l_{n}$ of~$H/\ell$, as well as their indexed degeneracies~$\a(n)$, and 
the ratio of determinants in \eqref{detratio} is
\be \label{detratio1}
Z_\text{index}\= \prod_{n} \, \left({\l_{n}}\right)^{-\half \a(n)} \, ,
\ee
where the infinite product is defined using zeta-function regularization.  

The calculation thus reduces to the calculation of the equivariant index~\eqref{indD10}, with respect 
to the action of~$H/\ell$. This can be done using the Atiyah-Bott fixed-point 
formula~\cite{Atiyah:1974}, which says that it reduces to the quantum-mechanical modes at the 
fixed points of the manifold under the action of~$H/\ell$. This can intuitively be explained as follows: in \eqref{indD10}, 
the index is the difference between the traces evaluated on the space of fields in $\Phi$ and $\Psi$, respectively, for the operator~$ \rme^{\frac{tH}\ell}$, that implements a finite diffeomorphism $x \mapsto \wt x = \rme^{\frac{tH}{\ell}} x$. Writing the trace as a sum over diagonal elements, the trace of $ \rme^{\frac{tH}{\ell}} $ can be written as an integral over $d^4 \,x$ with an insertion of $\delta^4(\tilde x - x)$. 
One therefore finds,
\be \label{ABFormula}
{\rm ind}(D_{10})(t) \= \sum_{\{x \mid \wt x = x\}}  
\frac{{\rm Tr}_{\Phi} \, \rme^{\frac{tH}\ell } \, - \, {\rm Tr}_{\Psi}\,e^{\frac{tH}\ell}}{{\rm det}(1-\partial \wt x/\partial x)} \,.
\ee
Writing the~AdS$_2 \times$~S$^2$ metric in complex coordinates as
\begin{equation}\label{metriccomplex}
    ds^2\=
    d\rho^2 + \sinh^2\rho \, d\tau^2 + d\theta^2 + \sin^2\theta \, d\psi^2 
     \= 
      \frac{4 dw d\bar w}{(1- w \bar w)^2} + \frac{4 dz d\bar z}{(1 + z \bar z)^2}  \,,
\end{equation}
the $U(1)$ action is
\be 
H \; \equiv \; L_{0} - J_{0} \= (\partial_\tau -\partial_\psi)\,.
\ee
Its fixed points are given by~$w=0$ ($\eta = 0$), and~$z=0$ or~$1/z=0$ (and~$\psi=0$ or~$\psi=\pi$) which are  the center of $AdS_2$, 
with the North Pole or South Pole of $S^2$ respectively.
The action of the operator~$e^{-\frac{\ii Ht}\ell}$ on the spacetime coordinate is~$(w\,,z) \to (e^{ \ii t/\ell} w\,,e^{ -\ii t/\ell}z)$
and therefore the determinant factor in the denominator of~\eqref{ABFormula} is, 
with~$q = \rme^{\ii t/\ell}$, 
\be
\label{eq:AB-numerator}
{\rm det}(1-\partial \wt x/\partial x) \= (1-q)^2(1-q^{-1})^2 \,,
\ee
regardless of which supermultiplet we are considering. 
Near the fixed points the space looks locally like~$\IR^{4}$, and we have an 
associated~$SO(4)=SU(2)_{+} \times SU(2)_{-}$ rotation symmetry.
At the North Pole and South pole, the chiral and anti-chiral part of the Killing spinor and the Hamiltonian reduce to
\be
NP: \quad \ve^i_{+\alpha}\=0\,, \qquad \ve^{\,i}_{-\dot\alpha} \= 
 \left(\sigma_3 \exp\left[{\ii \frac{(\tau+\psi)}{2}\,  \sigma_3}\right]\right)^i{}_{\dot\alpha}\,, \qquad 
 H\= L_0-J_0\= 2 J^{(+)}_{0} \,,
\ee
\be
SP: \quad \ve^{\,i}_{+\alpha}=  \left(-\ii \sigma_3 \exp\left[{\ii \frac{(\tau+\psi)}{2}\,  \sigma_3}\right]\right)^i{}_{\alpha}\,,
\qquad \ve^{\,i}_{-\dot\alpha}= 0\,, \qquad H\= L_0-J_0=2J^{(-)}_0 \,.
\ee
where $J_0^{(\pm)}$ are also defined to be anti-Hermitian.  
Therefore, a representation of~$SU(2)_+ \times SU(2)_{-}\times SU(2)_R$ is twisted to the representation 
of $SU(2)_{+}\times SU(2)_{-R}$ and $SU(2)_{+R}\times SU(2)_{-}$ (where $SU(2)_{\pm R}$ is in the diagonal of $ SU(2)_{\pm}\times SU(2)_R$), at the north pole and the south poles, respectively. 

Given a field in a representation~$(m,n)$, the trace in the numerator of~\eqref{ABFormula} 
is easily calculated at the north pole and south pole:
\be\label{TrNPSP}
\begin{split}
NP: & \quad {\rm Tr}_{(m,n)}e^{ \frac{tH}\ell} \= n (q^{-|m-1|}+q^{-|m-1|+2}\cdots + q^{|m-1|-2}+ q^{|m-1|}) \,, \\
SP: & \quad {\rm Tr}_{(m,n)}e^{ \frac{tH}\ell} \= m (q^{-|n-1|}+q^{-|n-1|+2}\cdots + q^{|n-1|-2}+ q^{|n-1|}) \,.
\end{split}
\ee
To obtain the determinant we finally have to take integral over the index, 
\be 
\log Z_{\rm index} \= \frac{1}{2} \int_{\epsilon}^\infty \frac{dt}{t} \, {\rm ind}(D_{10})(t).
\ee
where $\epsilon$ is a cut-off that is independent of the scale is the size of the spacetime $\ell$.

If all we care about the one-loop determinant is its dependence on the area of the black hole, i.e.~the scaling with $\ell$, then we to compute the coefficient $a_0$ such that $\log Z_\text{index} \sim a_0 \log \ell + O(1)$ as we scale the charges of the black hole to take $\ell \to \infty$. The coefficient~$a_{0}$, in turn, is obtained
from the constant term in the~$t\to 0$ expansion of the index,
\be\label{a0index}
\frac14 \, \text{ind} (D_{10})(t) \= \cdots + \frac{a_{-2}}{t^2} + a_{0} + a_{2} \, t^{2} + \cdots \,.
\ee
However, in this paper we are interested 
in the dependence of the one-loop determinant on the orbifold parameters,
and therefore we need to calculate~$\text{ind} (D_{10})(t)$ exactly.

To make things concrete, as a first example, we will compute the index of the vector supermultiplet in Section~\ref{sec:N=2-vector-supermultiplet-example}. This is known for the leading AdS$_2$ saddle \cite{Murthy:2015yfa} but the result is new for the orbifold geometry.

\vskip 0.2cm

\subsubsection{The construction of the~$\qeq$-complex}

Given the above formalism, what remains is to assemble all the fields in the cohomological variables
for a given supercharge and read off the charges of the various modes under this rotation. 
For a matter multiplet around a fixed background this is not a difficult task---one 
begins with the lowest component of the supermultiplet and follows the variation of a fixed supercharge 
to all the fields of the multiplet. When the fields of the theory have gauge invariance, the problem is a 
little bit more subtle as we need to include the Fadeev-Popov ghosts in the list of fields in order to maintain 
the covariance of the problem. The problem when doing this in supergravity, rather than in usual gauge theory, is 
\begin{enumerate}[label=(\alph*)]
    \item that the usual supersymmetry variations on the fields only holds up to gauge transformations,
    \item to work out the action of the supercharge on the ghosts. 
\end{enumerate}

Both these problems are solved consistently by demanding that the algebra~$Q^2 = H$ holds as such 
(without additional gauge variations) on all the fields of the theory (see e.g.~\cite{Pestun:2007rz,Hama:2012bg}). 
For the Weyl multiplet, the problem becomes more subtle because all transformations of the theory---including 
the supersymmetry variation---are realized a priori as gauge variations and one needs to first fix a background 
supersymmetry variation around a curved background. 
The technical issue boils down to essentially the same one as mentioned in Section~\ref{sec:locformalism},
namely the softness of the gauge algebra. The deformation of the BRST algebra presented in~\cite{deWit:2018dix,Jeon:2018kec} 
is thus also practically useful to calculate the one-loop determinant.

Now we move on to the second issue, the need to determine the action of the supercharge on the ghosts. To explain how this works its useful to provide more details on the Weyl supermultiplet in the off-shell superconformal formalism, which after coupling to an auxiliary vectormultiplet and a hypermultiplet can be gauge fixed to the Poincare $\mathcal{N}=2$ graviton multiplet. The Weyl multiplet consists on gauge fields for the following transformations\footnote{Where upper (lower) $i,j,\ldots$ denote indices in the (anti)-fundamental of $SU(2)_R$, $\mu,\nu,\ldots$ denote spacetime indices, and finally $a,b,\ldots$ denote frame indices directions for spacetime.}: translations generated by $P^a$, dilatations which we denote by $Dil$, special conformal transformations by $K$, Lorenz transformations by $M_{ab}$, $U(1)_R$ transformations by $A$; $SU(2)_R$ transformations by $V^i_{~j}$, supersymmetry transformations by $Q^i$ and superconformal transformations by $S^i$. Besides these gauge fields the Weyl multiplet includes auxiliary tensor fields $T_{ab}^\pm$ (where the superscript indicates the chirality) a scalar $D$ and fermions $\chi^i$. We refer to \cite{Jeon:2018kec} for more details, and in particular section 3 and 4 for the full development of the formalism, and restrict ourselves here to pointing out a couple of differences compared to rigid supersymmetry. 

While the action of the supercharge on the fields in the supermultiplet can be simply read off from the supersymmetric variation of each field, 
the action of the supercharge on the ghosts is determined by the structure constant of the symmetry algebra of supersymmetry variations. The supersymmetry algebra of two rigid supercharges typically has the form
\be 
\{Q^i\,, \overline{Q}_j\} \= (\gamma^a P_a +\ii Z) \delta^i _j\,,
\ee
where~$P^a$ is local translation and~$Z$ is a central charge. 
In conformal off-shell~$\CN=2$ supergravity, this has the following non-linear form,
\be\label{algebra1}
[\delta_{Q}(\varepsilon_{1}), \delta_{Q}(\varepsilon_{2})]=\delta_{\text{cgct}}(\xi)
+\delta_{M}(\varepsilon)+\delta_{K}(\Lambda_{K})+\delta_{S}(\eta) + \delta_{\text{gauge}}\,,
\ee
where~$\text{cgct}$ is the so-called covariant general coordinate transformation which
includes all the gauge symmetries of the theory, $M$, $K$, and~$S$ are, respectively, the 
local Lorentz, special conformal, and S-supersymmetry transformations,  
and~$\delta_{\text{gauge}}$ includes additional gauge transformations (including central charges) 
external to the minimal supergravity algebra.  
The composite parameters~$\varepsilon^{ab}$ are given by
\be  \label{parameters1}
\varepsilon^{ab}\=\ii \frac{1}{2}{\bar\varepsilon}_{2 i+}\varepsilon_{1+}^{i}T^{ab+ }
+\ii \frac{1}{2}{\bar\varepsilon}_{2 i-}\varepsilon_{1-}^{i}T^{ab- } \,,
\ee
and there are similar expressions for~$\xi^{\mu}$, $\Lambda_{K}^{a}$, and~$\eta^i$.

As a result, e.g.~there can be structure functions relating two supersymmetry transformations and the Lorentz 
transformation~$M^{ab}$ which is proportional to the field~$T^{ab}$. 
According to the deformed BRST formalism of~\cite{Jeon:2018kec}, this implies that the~$\qeq$-variation of 
the ghost~$c^{ab}$ for Lorentz transformations is proportional to~$T^{ab}$.
Similarly, the anticommutator
\be \label{cabghostvar}
\{Q^i \,, \overline{S}_j\}\=-\delta^i_j(\gamma^{ab}M_{ab}+Dil -\gamma_5 A) + 2V^{i}{}_{j} \,,
\ee
where~$Dil$ is the dilatation gauge symmetry of the conformal supergravity. By reading off the structure functions from this equation, we deduce that the~$\qeq$ variation of the dilatation 
ghost is the ghost for~$S$-supersymmetry. 

Having briefly reviewed how to obtain the transformation under $Q_{\rm eq}$ for the quantum fluctuation of all the fields in supergravity around a certain background,  
we  now describe the procedure~\cite{Jeon:2018kec}
of classification of all these fields either as cohomological primaries ($\Phi$ or $\Psi$) or descendants ($Q_{\rm eq} \Phi$ and $Q_{\rm eq} \Psi$). . 

\begin{itemize}
\item We will choose a certain twist of all the fields, either by considering the diagonal $SU(2)_{+R}$ in $SU(2)_+ \times SU(2)_R$ or the diagonal  $SU(2)_{-R}$ in $SU(2)_- \times SU(2)_R$ which both commute with the action of $\qeq$. This choice of twist is not unique and some choices do not lead to a valid classification of all the fields in a $Q_{\rm eq}$-cohomology complex.  

\item We start with a bosonic or fermionic field, $\phi_R$ or $\psi_R$ respectively, transforming in a given irrep $R$ of the twisted group. The variations $\qeq \phi_R$  and $\qeq \psi_R$ are both in the  irrep  $R$.   

\item We then try to find a twisted fermionic or bosonic field, which we denote by $\widetilde \psi_R$ or $\widetilde \phi_R$, that appears linearly, without any derivatives acting upon it, in $\qeq \phi_R$ or in $\qeq \psi_R$ respectively. If such a term exists, then we classify $\phi_R \in \Phi$ or $\psi_R \in \Psi$, respectively, and $\widetilde \psi_R \in \qeq \Phi$ or $\widetilde \phi_R \in \qeq \Psi$. If no such terms exists in $\qeq \phi_R$ or in $\qeq \psi_R$ then we try to find another fermionic or bosonic field whose variation under $\qeq$ has a term linear in the fields $\phi_R$ or $\psi_R$ respectively. In such a case, we would conclude that  $\phi_R \in \qeq \Psi$ or $\psi_R \in \qeq \Phi$ and the field whose variation we considered 

\item We repeat the process until it is no longer possible to pair fields into cohomological primaries and descendants.  If we cannot classify all the fields into $(\Phi\,, \qeq \Phi\,,\Psi\,, \qeq \Psi)$ by failing to preform the previous two steps, then we have to consider a different twist and repeat the previous steps until all the fields are classified.   
\end{itemize}

\subsection{One-loop determinant on the orbifold from index theorem}
\label{sec:one-loop-det-orbifold}

In the previous subsection we have abstractly explained how to compute the one-loop determinant around the leading black hole saddle using the Atyiah-Bott index theorem. We will now extend this method of computation for the orbifold geometries. As previously mentioned, from the $4d$ perspective the orbifold geometry can be understood as 
a~$\mathbb Z_c$ quotient, generated by~$\gamma = \rme^{\frac{2\pi H}c }$, which in terms of the coordinates in \eqref{metriccomplex} is given by the identification, 
\be 
(\tau, \psi) \sim \left(\tau+\frac{2\pi}c, \psi+\frac{2\pi}c\right)\sim (\tau, \psi+2\pi)\,.
\ee

Since we are working with ungauged supergravity, no field is charged under the gauge fields in the vector multiplets. 
This implies that the square of the supercharge does not involve any rigid gauge transformation
of these gauge fields.\footnote{This fact was implicitly used before in the one-loop calculations of~\cite{Murthy:2015yfa,Jeon:2018kec}. Note that the presence of such a rigid gauge transformation in~$\qeq^2$ would  affect  the index, even on the AdS$_2\times S^2$ geometry.}
Consequently, the one-loop determinant is independent of the orbifold parameter~$d$, which appears as a twisted boundary condition on fields with D0-charge. The same argument implies that the one-loop determinant is independent of~$n$.
This is consistent with the structure of the Kloosterman sum since the $d$ and $n$ dependence appear in an exponential reproduced by a term in the action, and not as a correction to the one-loop determinant. 

The traces in the index in \eqref{indD10} can then be related to those on the AdS$_2 \times S^2$ geometry by using the method of images. Instead of inserting $\delta^4(\widetilde x-x)$ in the integral over $d^4 x$ in the orbifolded geometry that computes the trace in~\eqref{indD10} on the orbifolds,
we can instead insert $\frac{1}{|\mathbb Z_c|}\sum_{\gamma \in \mathbb Z_c}  \delta^4(\widetilde x(\gamma, t)-x)$, with  $\widetilde x(\gamma, t) = (\gamma \cdot \rme^{\frac{tH}\ell}) x =  \rme^{\left(\frac{t}{\ell}+\frac{2\pi}c\right)H} x$,  within the  $d^4 x$ integral over an un-quotiented AdS$_2\times S^2$. Here, $|\mathbb Z_c|=c$ is the dimension of the quotient group which gives the total number of images. The index on the $\mathbb Z_c$ orbifold geometry, which we denote by $\text{ind}_{c}(D_{10})(t)$, can therefore be written as  
\be 
\text{ind}_{c}(D_{10})(t) \= \frac{1}{c} \sum_{\gamma \in \mathbb Z_c} \sum_{ \{x|\widetilde x(\gamma, t)=x\} } \frac{\Tr_\Phi \bigl( \, \gamma \cdot \rme^{\frac{tH}\ell}\, \bigr)-\Tr_\Psi \bigl(\, \gamma \cdot \rme^{\frac{tH}\ell}\, \bigr)}{\det\left(1 - {\partial \widetilde x }/{\partial x}\right) 
} \,,
\ee
which can now be evaluated explicitly using \eqref{TrNPSP} for the denominator and \eqref{eq:AB-numerator} for the numerator, now with $q = \rme^{\frac{t}{\ell}+\frac{2\pi}{c}}$.

\subsection{An example: the $\cN=2$ vector supermultiplet}
\label{sec:N=2-vector-supermultiplet-example}

To make things concrete, it is useful to review the computation of the one-loop determinant for the $\cN=2$ vector supermultiplets around the leading saddle using the results in Section~\ref{sec:one-loop-determinant-AdS2-S2}, after which the one-loop determinant around the orbifold geometries can be easily computed using the method of images presented in Section~\ref{sec:one-loop-det-orbifold}. In its off-shell formulation, the fields of the $\cN=2$ vector supermultiplets are 
\be 
(A_\mu, \, X, \bar X, \,\lambda^i, Y^{ij})
\ee 
where $A_\mu$ is the vector field, $X$ is a complex scalar whose complex conjugate in $\bar X$, $\lambda^i$ are two gaugini that form an $SU(2)$ doublet and $Y^{ij}$ are auxiliary scalars that form an $SU(2)$ triplet. In addition, in order for the supersymmetry transformations to close off-shell we have to include the ghosts $(b,c,B)$ associated to the $U(1)$ gauge transformations, where $b$ and $c$ have fermion statistics and $B$ has bosonic statistics. Within the BRST formalism described in the previous subsections, we can now extract the action of the equivariant supercharge on $(A_\mu, \, X, \bar X, \,\lambda^i, Y^{ij})$ by deforming the supersymmetric variation as described in Section~\ref{sec:one-loop-determinant-AdS2-S2}. Before doing so, we perform the following twist for the fields in the vector supermultiplet, 
{\begin{table}[h] 
\begin{center}
\begin{tabular}{|c|c|c|}
\hline
Field or ghost & $SU(2)_+ \times SU(2)_- \times SU(2)_R$ irrep &  $SU(2)_{\pm} \times SU(2)_{R\mp}$ irrep \\
\hline
$A_\mu$ & $(2,2,1)$ & $(2,2)$ \\
$X,\, \bar X$ & $(1,1,1)$ & $(1,1)$ \\
$\lambda_i$ & $(2,1,2)\oplus (1, 2,2)$ & $(2,2)\oplus (1, 1) \oplus (1, 3)$ \\
$Y^{ij}$ & $(1,1,3)$ & $(1,3)$ \\
$c,\,b,\,B$ & $(1,1,1)$ & $(1,1)$ \\
\hline
\end{tabular}
\label{twistedfieldsvecmultiplet}
\end{center}
\end{table}}

To make the calculation more explicit, it is useful to see how to perform the twist for the gaugini by projecting against the fixed Killing spinor $\varepsilon^i$ to obtain the three irreps listed in the table above:
\be 
\lambda_{(2,2)} \;\equiv \; \lambda_\mu \;\equiv \; \bar \varepsilon_i \gamma_\mu \lambda^i\,, \qquad \lambda_{(1,1)}\;\equiv \; \lambda \;\equiv \; \bar \varepsilon_i \gamma_5 \lambda^i\,, \qquad \lambda_{(1,3)}\;\equiv \; \lambda^{ij} \;\equiv \; -2 \varepsilon^{(i} C \lambda^{j)}\,,
\ee
where $C$ is the charge conjugation matrix.

We start with the transformation under $Q_\text{eq}$ of the quantum fluctuation for the gauge field $A_\mu$ which includes,
\be 
Q_\text{eq} A_\mu \= \lambda_\mu + \partial_\mu c\,.
\ee
Since the twisted variable $\lambda_\mu$ appears without derivatives, it is the cohomological descendant of $A_\mu $. Thus $A_\mu \in \Phi$ and $\lambda_\mu\in \Psi$. Due to the nature of the twisting, as previously mentioned $A_\mu$ and $\lambda_\mu$ lie within the same $(2,2)$ twisted representation. One can then easily check that $Q_{eq} \lambda_\mu$ contains no terms linear in the fields of the vector supermultiplet without any derivatives, confirming that $ \lambda_\mu$ is indeed a cohomological descendant. 

 Next, we see that the variation of the linear combination $\Im X = -i(X - \bar X)$ transforms as
\be 
Q_\text{eq} \, \Im X  \= \lambda \,,
\ee
from which we deduce that $\Im X\in \Phi$ and $ \lambda \in Q_\text{eq} \Phi$. The transformation under $Q_\text{eq}$ of the remaining twisted representation of the gaugino, $\lambda^{ij}$ which transforms as $(1,3)$ under $SU(2)_\mp \times SU(2)_{R\pm}$, involves
\be
Q_\text{eq} \lambda^{ij} \supset Y^{ij}\,,
\ee
which thus implies that $\lambda^{ij} \in \Psi$ and $Y_{ij} \in Q_\text{eq} \Psi$. What remains to be classified is the other linear combination of scalars, $\Re X = X+\bar X$ as well as all the ghost fields. We find $\Re X$ when looking at how the ghost $c$ transforms under $Q_\text{eq}$, 
\be
 Q_\text{eq} c \supset i\,\left(\bar \epsilon_i \epsilon^i\right)\Re X\,,
\ee
which thus means that the ghost $c\in \Psi$ (due to its spin statistics), and $\Re X\in Q_\text{eq} \Psi$, rather than being part of $\Phi$. Finally, the ghosts $b$ and $B$ are related through the transformation 
\be 
Q_{\rm eq} b \= B\,,
\ee
which finally means that $b \in \Psi$ and $B \in Q_{\rm eq} \Psi$.

To summarize, we thus have the following fundamental fields in the $Q_{\rm eq}$ cohomology complex, 
 \begin{table}[h!]
  \begin{center}
  \begin{tabular}{|c|c|}
\hline
$\Phi$ & $	\subseteq Q_{eq} \Phi$ \\
\hline 
\hline $A_{(2, 2)}$ & $\lambda_{(2,2)}$\\ 
\hline $\Im X_{1,1}$ & $\lambda_{(1,1)}$ \\
\hline
\end{tabular}
\quad
\begin{tabular}{|c|c|}
\hline
$\Psi$ & $	\subseteq Q_{eq} \Psi$\\
\hline
\hline $\lambda_{(1, 3)}$ & $Y_{(1,3)}$\\
\hline $(c)_{(1,1)}$ & $\Re X_{(1, 1)}$\\ 
\hline $(b)_{(1,1)}$ & $(B)_{(1,1)}$\\ 
\hline
\end{tabular}
 \end{center}
\end{table}
\\ 
where the subscript indicates the twisted representation of each cohomological primary/descendant. Thus, to obtain the index associate to this supermultiplet, we simply use the representations listed in the table to compute the bosonic and fermionic traces in \eqref{ABFormula} by using \eqref{TrNPSP}: 
\be 
\begin{split}
\Tr_{\Phi}^{NP}\left( \rme^{\frac{tH}\ell}\right) &\= \Tr_{\Phi}^{SP}\left( \rme^{\frac{tH}\ell}\right) \=  {\rm Tr}_{(2,2)}\left( \rme^{\frac{tH}\ell}\right) + {\rm Tr}_{(1,1)}\left( \rme^{\frac{tH}\ell}\right)  \,, \\
\Tr_{\Psi}^{NP}\left( \rme^{\frac{tH}\ell}\right) &\= \Tr_{\Psi}^{SP}\left( \rme^{\frac{tH}\ell}\right) \=  {\rm Tr}_{(1,3)}\left( \rme^{\frac{tH}\ell}\right) + 2\,{\rm Tr}_{(1,1)}\left( \rme^{\frac{tH}\ell}\right)\,.
\end{split}
\ee
Thus, in total we find that the index for the vector supermultiplet on the regular black hole geometry is given by 
\be 
\label{eq:index-vector-N=2-BH}
\text{ind}_{\text{vect.}}^{\cN=2} (D_{10})(t) \= \frac{4(q+q^{-1}) -8}{(1-q)^2 \, (1-q^{-1})^2} \=  \frac{4 \,q}{(1-q)^2}\,\qquad \text{ with }\,q=e^{\frac{it}\ell}\,.
\ee 
A useful check of \eqref{eq:index-vector-N=2-BH} can be made by computing its contribution to the logarithmic corrections of the black hole entropy. As described in Section~\ref{sec:one-loop-determinant-AdS2-S2}, we simply need to extract the coefficient $a_0^{\rm vect.} = -1/12$ in the small $t$ expansion of \eqref{eq:index-vector-N=2-BH}. This agrees with the result of \cite{Sen:2011ba}, which computed the heat-kernel associated to the differential operators acting on fields in an $\cN=2$ vector supermultiplet around the on-shell AdS$_2\times S^2$ saddle. 
The index on the orbifolded geometry is given by the sum over images and simplifies to  
\be 
\text{ind}_{\text{vect.},\,c}^{\cN=2 \text{ s.c.}} (D_{10})(t)\bigg|_{\rm NP/SP} \= \frac{2\,c \,q^c}{(1-q^c)^2}\bigg|_{\rm NP/SP} \,,\qquad \text{ with }\,q=e^{\frac{it}\ell}\,,
\ee
which leads to a one-loop determinant around each orbifold geometry that has non-trivial $c$-dependence. Extracting the coefficient $a_0^{{\rm vect.},\,c} = -c/12$ from the small $t$ expansion, we find that our result is consistent with the heat-kernel result obtained in \cite{Gupta:2013sva, Gupta:2014hxa}, that extracted the leading logarithmic correction for the entropy contribution of each orbifolded geometry. 

In the previous paragraphs we focused on how the one-loop determinant scales with the AdS radius in the semiclassical limit, but we can also see the full c-dependence. To do this we first need to expand the index as $\frac{2c q^c}{(1-q^c)^2} = \sum_{n\geq 1} 2 c n \rme^{- i t\frac{n}{\ell} c}$ for the north pole piece. The north pole contribution to the one-loop determinant is $e^{\partial_s \left(\sum_{n\geq 1} 2 c n (n c/\ell)^{-s}\right)_{s\to 0}}$. At the south pole we have a similar expression up to an overall factor of $(-1)^s$ since the expansion has to be done in powers of $q^{-1}$. After taking the derivatives and using zeta function regularization
\begin{eqnarray}
\log \left( Z_{1-loop}^{\rm vec} \right) \=  - \frac{\ii\pi}{12} \, c+ \frac{c}{6}\log c+2 c \,  \zeta'(-1) - \frac{c}{12} \log \ell^2 .
\end{eqnarray}
This calculation was done in \cite{Murthy:2015yfa} for $c=1$ and was pointed out that one can neglect overall numerical constants and focus on the power of $\ell^2 = \rme^{-\mathcal{K}(\phi)}$. When attempting to sum over orbifolds, it is important how the numerical overall constant depends on its order $c$. We will see below that to obtain the correct microscopic black hole index we need the one-loop determinant of bulk modes, and therefore the index computed above, to be independent of $c$. This will be the case when all fields of $\mathcal{N}=8$ supergravity are included in the one-loop determinant.

\subsection{The index of all $\cN=2$ supermultiplets }
\label{sec:one-loop-det-results}

In this section we discuss how to compute the index of the
the spin-$\frac32$, with the goal of 
treating the supersymmetry transformations of one complex supercharge in an off-shell manner.
In order to achieve this goal, we first discuss the~$\CN=4$ Weyl supermultiplet for which an off-shell
formalism is known \cite{Bergshoeff:1980is, Ciceri:2015qpa}. We decompose this multiplet into~$\CN=2$ multiplets, and identify the 
off-shell fields that reduce, upon gauge-fixing, to the on-shell spin-$\frac32$ multiplet in~$\CN=2$ supergravity.

\subsubsection{The $\CN=4$ Weyl supermultiplet}
\label{sec:N=4-Weyl-supermultiplet}

The~$\CN=4$ Weyl multiplet has an~$SU(4)_R$ symmetry. 
The fields of the $\CN=4$ Weyl multiplet are presented in Table~\ref{WeylFields}, along with the list of gauge transformations 
in the theory. The ghosts corresponding to these gauge transformations are presented in Table~\ref{WeylGhosts}. 
The fields and ghosts together comprise $252\, \text{B}+252 \, \text{F}$ off-shell degrees of freedom.
{\begin{table}[h] 
\begin{center}
\begin{tabular}{|c|c|c|c|}
\hline
Local symmetry &Gauge fields & $SU(4)$ rep &Degrees of freedom \\
\hline
g.c.t &  $e_{\mu}^{a}$ &1 & 16 B \\
Dilatation $D$&  $A^D_{\mu}$ &1 & 4 B \\
Sp. conf. $K^a$&  $f_{\mu}^{a}$ & 1& composite \\
Lorentz $M_{ab}$&  $\w_{\mu}^{ab}$ &1 & composite \\
$SU(4)_R$&  $\mathcal{V}_{\mu \, j}^{\, i}$ & 15 & 60 B \\
$U(1)_R$&  $A^R_{\mu} = a_\mu$  & 1& composite\\\hline
$Q$-susy &  $\psi_{\mu}^{\,i}{}$ & 4& 64 F\\
$S$-susy &  $\phi_{\mu}^{\,i}{}$ & 4& composite \\
\hline
\hline
~ & Auxiliary fields & & Degrees of freedom \\
\hline
&  $T^{ij}_{ab}$  & 6& 36 B \\
& $E_{ij}$  & 10& 20 B \\
&  $D^{ij}_{kl}$ &$20'$& 20 B \\
&  $\phi^{\a}$ & 1& 2 B \\
&  $\Lambda^{i}$ &4  & 16 F \\
&  $\chi^{ij}_{\,k}$  &20& 80 F \\
\hline
\end{tabular}
\caption{The 158 bosonic (B) and 160 fermionic (F) matter fields of the Weyl multiplet.
These fields have $30\,\text{B}+32\,\text{F}$ gauge degrees of freedom (see Table~\ref{WeylGhosts}).
Upon subtracting the gauge degrees of freedom we have~$128\,\text{B}+128\,\text{F}$ degrees of freedom as in~\cite{Bergshoeff:1980is}.} 
\label{WeylFields}
\end{center}
\end{table}}
{\begin{table}[h] 
\begin{center}
$\begin{tabular}{|c|c|c|}
\hline
Local symmetry &Ghosts & Degrees of freedom \\
\hline
g.c.t &  $(c^\mu,b_\mu,B_\mu)$  & 8 F + 4 B \\
Dilatation $D$&  $(c_D,b_D,B_D)$  & 2 F + 1 B \\
Sp. conf. $K^a$&  $(c_K^{~a}\,,b_K^{~a}\,,B_K^{~a})$ & 8 F + 4 B \\
Lorentz $M_{ab}$&  $(c_M^{ab}\,,b_M^{ab}\,,B_M^{ab})$ & 12 F + 6 B\\
$SU(4)_R$&  $(c_R^{\;i}{}_{\,j}\,,b_R^{\;i}{}_{\,j}\,,B_R^{\;i}{}_{\,j})$ & 30 F + 15 B \\
\hline
$Q$-susy &  $(c_Q^{\,i}{}\,,b_Q^{\,i}{}\,,B_Q^{\,i}{})$ & 32 B + 16 F\\
$S$-susy &  $(c_S^{\,i}{}\,,b_S^{\,i}{}\,, B_S^{\,i}{})$& 32 B + 16 F  \\
\hline
\end{tabular}$
\caption{The 94 bosonic (B) and~92 fermionic (F) ghosts of the Weyl multiplet.} 
\label{WeylGhosts}
\end{center}
\end{table}}

In order to reduce the~$\CN=4$ off-shell Weyl multiplet to~$\CN=2$ multiplets, 
we first decompose the various irreps of $SU(4)_R$ under the maximal $SU(2)_R \times SU(2)' \times U(1)$ 
subgroup, where the first~$SU(2)_R$ is the symmetry of~$\CN=2$ supergravity. 
Then we collect the fields in a given representations of~$SU(2)'$, and track the closure of 
supersymmetry transformations. The results are presented in Tables~\ref{Reduction4to2} 
and~\ref{Reduction4to2fields}.\footnote{We thank Dan Butter for 
useful conversations related to this reduction and, in particular, for 
pointing out the importance of the chiral multiplet in this reduction.}
{\begin{table}[h] 
\begin{center}
$\begin{tabular}{|c|c|c|c|c|}
\hline
$\CN=2$ rep & $SU(2)'$ rep &  d.o.f. & Gauge freedom & gauge-fixed d.o.f. \\
\hline
Weyl & 1  & 43 B + 40 F  & 19 B + 16 F & 24 B + 24 F \\
Spin-$\frac32$ & 2 & 2 (36 B + 40 F)  & 2 (4 B + 8 F) & 2 (32 B + 32 F) \\
Vector & 3 & 3 (9 B+ 8 F)  & 3 (1 B + 0 F) & 3 (8 B + 8 F) \\
Chiral & 1 & 16 B + 16 F  & 0 B + 0 F & 16 B + 16 F \\
\hline
\end{tabular}$
\caption{Reduction of the~$\CN=4$ off-shell Weyl multiplet to~$\CN=2$ multiplets with a total of 128B+128F degrees of freedom.} 
\label{Reduction4to2}
\end{center}
\end{table}}
{\begin{table}[h] 
\begin{center}
$\begin{tabular}{|c|c|c|c|c|c|}
\hline
Field &$SU(4)_R$ rep & $\cN=2$ Weyl & $\cN=2$ \,spin-$\frac32$ & $\cN=2$ vector & $\cN=2$ chiral \\
\hline
$e^a_\mu$ & 1 & $(1,1)$ & -- & -- & -- \\
$A_\mu^D$ & 1 & $(1,1)$ & -- & -- & -- \\
$V_\mu{}^i{}_j$ & 15  & $(3,1) \, \oplus \, (1,1)$ & $ 2 (2,2) $ & $(1,3)$ &-- \\
$\phi^{\alpha}$ & 1  & $(1,1)$ &  -- & -- & --  \\ 
$T_{ab}^{ij}$ & 6  & $(1,1)$ &  $(2,2)$ & -- & $(1,1)$   \\ 
$E_{ij}$& 10 & -- & $(2,2)$ &  $(1,3)$ & $(3,1)$ \\
$D_{ij}^{kl}$ & $20'$ & $ (1,1)$  & $2(2,2)$ & $(3,3)$ & $2(1,1) $ \\  \hline
$\psi_{\mu}^i$ & 4  & $(2, 1)$ & $(1, 2)$ & -- & -- \\ 
$\Lambda^i$ & 4  & --& $(1, 2)$ & -- & $(2,1)$ \\ 
$\chi^{ij}_{k}$ & 20  & (2,1) & $2(1, 2) + (3,2)$ & $(2,3)$ & $(2,1)$ \\ 
\hline
\end{tabular}$
\caption{Reduction of the~$\CN=4$ off-shell Weyl multiplet to~$\CN=2$ multiplets. Each~$SU(4)_R$ 
representation decomposes into a number of representations under~$SU(2)_R \times SU(2)'$, which 
falls into one of the four $\CN=2$ supermultiplets (Weyl/spin-$\frac32$/vector/chiral). 
The scalars~$\phi^\alpha$ and~$E_{ij}$
are complex and the other bosons are real. All the fermions are in a 4-dimensional spinor representation  
which is suppressed.} 
\label{Reduction4to2fields}
\end{center}
\end{table}}

{\begin{table}[h] 
\begin{center}
$\begin{tabular}{|c|c|c|c|c|}
\hline
Ghost &$SU(4)_R$ rep & $\cN=2$ Weyl & $\cN=2$ \,spin-$\frac32$ & $\cN=2$ vector \\
\hline
$(c^\mu,b_\mu,B_\mu)$ & 1 & $(1,1)$ & -- & --  \\
$(c_D,b_D,B_D)$ & 1 & $(1,1)$ & -- & --  \\
$(c_K^a,b_K^a,B_K^a)$ & 1 & $(1,1)$ & -- & --  \\
$(c_M^{ab},b_M^{ab},B_M^{ab})$ & 1 & $(1,1)$ & -- & --  \\
$(c_{R\,j}^{i},b_{R\,j}^{i},B_{R\,j}^{i})$ & 15 & $(3,1)\oplus (1,1)$ & 2(2,2) & (1, 3) \\
\hline
$(c_{Q}^{i},b_{Q}^{i},B_{Q}^{i})$ & 4 and $\bar 4$ & $(2,1)$ & -- & $(1, 2)$ \\
$(c_{S}^{i},b_{S}^{i},B_{S}^{i})$ & 4 and $\bar 4$ & $(2,1)$ & -- & $(1, 2)$ \\ \hline
\end{tabular}$
\caption{Reduction of the ghosts in the ~$\CN=4$ off-shell Weyl multiplet to~$\CN=2$ multiplets. Once again we list the representation of each ghost under~$SU(2)_R \times SU(2)'$.} 
\label{Reduction4to2ghosts}
\end{center}
\end{table}}

Next we discuss the reduction of the off-shell conformal supergravity to the on-shell Poincar\'e supergravity. 
We first remind ourselves of this reduction for~$\CN=2$ supergravity. Each off-shell multiplet has a number 
of non-propagating auxiliary fields which can be solved algebraically (i.e.~integrated out). 
A familiar example is the~$SU(2)_R$ triplet of auxiliary fields~$Y_{ij}$ in the~$\CN=2$ off-shell vector multiplet, 
which is integrated out by this procedure. 
In addition, the off-shell Weyl multiplet of superconformal gravity has additional gauge symmetries compared 
to the Poincar\'e theory. In order to reach the Poincar\'e theory, one adds additional matter multiplets which fix 
the extra gauge freedom of the superconformal theory. 
At the end of this procedure, we are left with a number of propagating fields which form the Poincar\'e gravity 
supermultiplet. 

It is useful to go through this procedure in a bit of detail. 

The gauge symmetries of the~$\CN=2$ Weyl multiplet of conformal supergravity 
are general coordinate invariance\footnote{In the superconformal formalism this appears after imposing 
some constraints which involve the local Lorentz transformations.} 
and $Q$-supersymmetry (both of which are present in the Poncar\'e theory), 
and dilatation~$Dil$, special conformal symmetry~$K$, $SU(2)_R$ and~$U(1)_R$, $S$-supersymmetry 
(which are not present in the Poincar\'e theory). 
The Poincar\'e theory contains a~$U(1)$ vector field called the graviphoton. 
In order to reduce to Poincar\'e supergravity, we couple the Weyl multiplet to one~$\CN=2$
vector and one~$\CN=2$ hypermultiplet.\footnote{Instead of the hypermultiplet, we could also 
choose a tensor multiplet or a non-linear multiplet \cite{deWit:1980lyi}.} 
Let us consider these matter multiplets to be on-shell, as in the off-shell formalism, the auxiliary fields 
of the matter multiplets can be integrated out.\footnote{Recall that there is no known Poincar\'e-invariant 
off-shell formalism with a finite number of fields for the hypermultiplet. Here we are only interested in 
the off-shell completion for one complex supercharge for which we do have an off-shell formalism~\cite{Hama:2012bg,Murthy:2015yfa}.}
So the vector multiplet contains~$(A_\mu, X, \overline{X}, \lambda^i)$, $i=1,2$, and the hyper 
contains~$(A^i_I, \zeta_I)$, $I=1,2$. First let us consider bosonic fields and their associated transformations. 
The special conformal symmetry~$K$ is fixed by setting the gauge field for dilatations~$A_\mu^D$ 
(often called~$b_\mu$) to zero. $Dil$ and~$U(1)_R$ are fixed by setting~$\sqrt{g} \rme^{-\CK}$ to its asymptotic value, and $SU(2)_R$
is fixed by setting~$A^i_I=A \delta^i_I$. 

Further, the Weyl multiplet has a non-propagating field~$D$ which couples linearly (as a Lagrange multiplier) 
to a linear combination of~$A$ and~$X$, and the scalar~$A$ is fixed by integrating out~$D$.
At the end of this procedure, the extra gauge invariances and the non-propagating fields in the Weyl multiplet 
are all fixed, and the only propagating field in the matter sector is the gauge field itself which is identified with 
the graviphoton of the Poincar\'e supergravity. 
In the fermionic sector, there is a similar story. The~$S$-supersymmetry transformations account for 8 degrees 
of freedom, and the non-propagating Lagrange multiplier~$\chi^i$ field has 8 degrees of freedom, which together 
fix the 8+8 propagating fermions of the matter multiplets.

Now we repeat the procedure for the off-shell~$\CN=4$ Weyl supermultiplet. The new gauge symmetries 
compared to the~$\CN=2$ case are: four instead of two~$Q$ and~$S$ supersymmetries, and~$SU(4)_R$ 
instead of~$SU(2)_R$. The procedure to go to the Poincar\'e theory is presented in \cite{Bergshoeff:1980is}. The Poincar\'e 
theory has 6 vector fields (graviphotons). 
So we add 6~$\CN=4$ vector multiplets,\footnote{In this case these multiplets are necessarily on-shell, as there 
are no known off-shell $\CN=4$ matter multiplets.} each of which has 1 vector field, 6 scalars, and 16 fermionic 
degrees of freedom.
The K-gauge is fixed as usual by setting~$A_\mu^D=0$. The $SU(4)_R$ gauge symmetry (15) and the 
dilatation symmetry (1) together fix 16 scalars, and the 20 Lagrange multipliers~$D^{ij}_{kl}$
fix 20 scalars, so that, together, all the 36 scalars are fixed. 
In the fermionic sector we have~$16$ S-supersymmetry transformations and~$80$ Lagrange 
multipliers~$\chi^{ij}_{k}$, which together fix the 96 degrees of freedom.

We now go through the above gauge-fixing procedure keeping track of~$\CN=2$ multiplets. 
As explained above, the~$\CN=4$ Weyl multiplet consists of 1~$\CN=2$ Weyl + 2~$\CN=2$ spin-$\frac32$ + 3~$\CN=2$ vector multiplets. 
We want to keep track of how the gauge symmetries plus Lagrange 
multipliers fix the propagating fields in the matter multiplets. We read off the fields from 
Table~\ref{Reduction4to2fields}, and present the results in Table~\ref{ParametersGaugeFixing}. 
{\begin{table}[h] 
\begin{center}
$\begin{tabular}{|c|c|c|c|c|c|c|}
\hline
$\CN=2$  & Gauge (B) & $D$ & Tot (B) & Gauge (F)  &  $\chi$  & Tot (F) \\
multiplet & symmetry  &  EOM & & symmetry  & EOM  &  \\
\hline
Weyl & $Dil$~(1)+ $SU(2)_R$~(3) & 1 & 6 & $S$-susy (8) &  $\chi^i$~(8) & 16 \\
 & + $U(1)_R$~(1) &  &  &  &   &  \\
Spin-$\frac32$ & $U(1)$~(4) &4 & 8 & $S$-susy (4) &  20 & 24 \\
Vector & $U(1)$~(1) &3 & 4 & -- &  8 & 8 \\
Chiral &-- & 2 & 2 & -- & 8 & 8 \\
\hline
\end{tabular}$
\caption{The number of parameters in a multiplet that need to be fixed by the addition of matter. This includes 
parameters labelling gauge freedom and parameters from the Lagrange multipliers $D^{ij}_{kl}$ and $\chi^{ij}_k$.}  
\label{ParametersGaugeFixing}
\end{center}
\end{table}}

\ndt The bosonic and fermionic totals in the above table should be matched with the scalars and fermions, respectively,
of the gauge-fixing matter. We assume that the only~$\CN=2$ matter can be vector multiplets (2~scalars + 8~fermions)
or hyper multiplets (4~scalars + 8~fermions).
It is now clear that, under an~$\CN=2$ decomposition, in order to achieve a 
consistent $\CN=2$ Poincar\'e theory,
we have the following unique pairing of multiplets with gauge-symmetry + gauge-fixing matter (all multiplets 
refer to~$\CN=2)$:
\begin{center}
Weyl  + (1~vector  + 1~hyper)$_\text{matter}$, \\
Spin-$\frac32$  + (2~vectors + 1~hyper)$_\text{matter}$, \\
Vector  + (1~hyper)$_\text{matter}$, \\
Chiral  + (1~vector)$_\text{matter}$. 
\end{center}
To obtain the original~$\CN=4$ gauge-fixing, we can only have~$\CN=4$ vectors as matter, which 
implies  (again, all multiplets refer to~$\CN=2)$:
\begin{center}
Weyl  + (vector  + hyper)$_\text{matter}$, \\
(Spin-$\frac32$ + Vector) + 2(vector + hyper)$_\text{matter}$, \\
Vector + Chiral  + (vector + hyper)$_\text{matter}$.
\end{center}

The final conclusion from this subsection is that in a formalism where we have off-shell supersymmetry with
one complex supercharge and arrange the fields in a manifestly~$\CN=2$ covariant manner,  
the spin-$\frac32$ multiplet should be gauge-fixed using~2~$\CN=2$ vectors + 1~$\CN=2$ hyper multiplet
in order to obtain an on-shell spin-$\frac32$ multiplet of Poincar\'e supergravity.

In the following subsection we perform a consistency check of this conclusion: we calculate the index of all $\cN=2$ supermultiplets in both the superconformal theory and in the super-Poincar\'e theory.
We will show that the resulting one-loop determinants indeed agree with past one-loop calculations, not around the localization locus, but rather around the on-shell geometries, in a limit where we can do a consistent comparison.

\subsubsection{The $\qeq$-complex and the index of the $\cN=2$ supermultiplets}

In this subsection we decompose the~$\CN=4$ Weyl multiplet into~$\qeq$-elementary and descendant fields
(the~$\qeq$-complex), using the procedure developed in~\cite{Jeon:2018kec}. Once we have achieved this we use the 
separation of the~$\CN=4$ Weyl multiplet into~$\CN=2$ multiplets as described in the previous subsection, 
in order to read off the elementary and descendant fields of the~$\CN=2$ spin-$\frac32$ multiplet. This then 
leads us to the calculation of the index of the spin-$\frac32$ multiplet. 

The main ideas of the construction of the~$\qeq$-complex were sketched in the previous subsection.
The practical steps are as follows: we begin by twisting all the fermionic fields using the Killing spinor~$\ve^i$ 
in the~AdS$_2 \times$~S$^2$ background. We then write down the supersymmetry variation of all the 
supergravity fields and keep track of the charges under~$SU(2)_+ \times SU(2)_- \times SU(2)_R \times SU(2)'$ 
as described above. We start with the top component (which is say bosonic) in a given representation 
and consider its~$\qeq$-variation, which is typically a linear combination of products of various twisted 
fields and Killing spinor bilinears. Among these terms we isolate terms which are purely linear in the fields 
(in this case fermionic) without derivatives (so that the change of variables to the twisted fields is invertible). 
If we find such a fermionic field, it is a descendant and the original bosonic field is elementary.
Then we move down the supermultiplet, considering the representations of fields which have not been 
classified and repeat the process, interchanging bosons and fermions if needed. 
The~$c$-ghosts are classified according to the discussion sketched near Equation~\eqref{cabghostvar}.

In the off-shell superconformal formalism, the relevant part of the supersymmetric transformation under $Q$ and $S$ of the fields in the $\CN=4$ Weyl multiplet are \cite{Bergshoeff:1980is, Ciceri:2015qpa}:
\be 
\label{eq:N=4-variations}
\delta e_{\mu}\,^a  &=\bar \epsilon^i \gamma^a \psi_{\mu i} + \text{h.c.} \,,\nn \\
\delta \psi_{\mu}\,^i &\supset -b_\mu \epsilon^i - 2 V_\mu\,^i\,_j\nn \epsilon^j -\frac{1}2 \gamma^{ab} T_{ab}{}^{ij} \gamma_\mu \epsilon_j - \gamma_\mu \eta^i \,,\\
\delta b_\mu &\supset 
- \frac{1}2 \bar \psi_\mu{}^i \eta_i +
 \text{ h.c. }\,,\nn \\
 \delta V_{\mu}{}^i{}_j &\supset 
 \bar \epsilon^k \gamma_\mu \chi^i{}_{kj} - \bar \psi_{\mu}{}^i\eta_j + \frac{1}4 \delta^i{}_j \bar \psi_\mu{}^k \eta_k\,,\nn \\
 \delta \Lambda_i &\supset E_{ij} \epsilon^j + \frac{1}2 \varepsilon_{ijkl} T_{bc}{}^{kl} \gamma^{bc} \epsilon^j\,,\nn \\ 
 \delta E_{ij} &\supset -2 \bar \epsilon^k \chi^{mn}_{(i} \varepsilon_{j)kmn} + 2 \bar \eta_{(i} \Lambda_{j)}\,,\nn \\
 \delta T_{ab}{}^{ij} &\supset -2 \bar \epsilon^{[i}\left(\gamma_{[a} \phi_{b]}{}^{j]} + \frac{1}2 \gamma \cdot T^{j]k} \gamma_{[a} \psi_{b]k}\right) + \frac{1}2 \bar \epsilon^k \gamma_{ab} \chi^{ij}{}_k - \frac{1}4 \varepsilon^{ijkl} \bar \eta_k \gamma_{ab} \Lambda_l\,,\nn \\ 
 \delta \chi^{ij}{}_k &\supset -\frac{1}2 \slashed{D} T_{ab}^{ij} \epsilon_k + D^{ij}{}_{kl} \epsilon^l + \frac{1}2 T_{ab}^{ij} \gamma^{ab} \eta_k + \frac{2}3 \delta_{k}^{[i} T_{ab}{}^{j]l} \gamma^{ab} \eta_l - \frac{1}2 \varepsilon^{ijlm} E_{kl} \eta_m\,,\nn \\ 
 \delta D^{ij}{}_{kl} &\supset \varepsilon^{ijmn} \bar \epsilon^p T^{ab}{}_{kl} \left[2 T_{abnp} \lambda_m + T_{abmn} \Lambda_p \right]\,,
\ee 
where we have only included the variations that are either linear in one field or, when expanding around the localization background described in Section~\ref{sec:localization-in-supergravity}, could be linear in the quantum fluctuations of a given field.  
One can therefore then use \eqref{eq:N=4-variations} to write down the supersymmetric variations for the quantum field fluctuations around the localization background described in Section~\ref{sec:locformalism}. Including ghosts one finds the BRST variations which then need to be modified, as described in Section~\ref{sec:one-loop-det-theoretical-prelims}, to find the equivariant variation. To exemplify how the cohomological partners are found in this case, we will simply list the first few steps of the algorithm presented in Section~\ref{sec:one-loop-determinant-AdS2-S2}.   The first few variations are as follows. The first variation tells us that 
 \be 
 e_\mu{}^a &\= e_{ (3, 3, 1) \oplus (1, 3, 1)\oplus (3, 1, 1)\oplus (1, 1, 1)}\in \Phi,\nn \\   \psi_\mu^a &\= \psi_{(2, 2, 1) \otimes (2, 2, 1)} = \psi_{ (3, 3, 1) \oplus (1, 3, 1)\oplus (3, 1, 1)\oplus (1, 1, 1)} \notin \Psi\,.
 \ee
where we in the subscript we shall list the twisted $SU(2)_\pm \times SU(2)_{R\mp} \times SU(2)'$ representation associated to each field. The next relations tell us that
 \be 
 \psi_{(2, 4, 1)} \in \Psi\,, \qquad 
\psi_{(2, 2, 1) \otimes (1, 1, 1)} \in \Psi\,. 
 \ee
 and consequently, 
 \be 
 V_{(2,2,1)}\not\in \Phi\,, \qquad V_{(2, 4, 1)}\not\in \Phi\,.
 \ee
 There is an additional component of the gravitino which is also in the $(2,2,1)$ twisted representation. When modifying \eqref{eq:N=4-variations} to obtain the BRST variation the term $\gamma_\mu \eta^i$ includes the fluctuation for the ghost field $c_S$ which after twisting is also in the $(2, 2,1)$ representation. 
 Thus, we conclude that 
 \be 
 \psi_{(2, 2, 1)'} \in \Psi\,, \qquad (c_S)_{(2,2,1)} \not \in \Phi\,.
 \ee
Continuing on in this manner, we obtain the full classification of the~$\CN=4$ Weyl supermultiplet.
We present the results split into the~$\CN=2$ multiplets following the decomposition in Table~\ref{Reduction4to2fields}: 
the fields that are part of the ~$\CN=2$ Weyl multiplet are shown in Table~\ref{WeylClasseq2}, 
those that are part of the spin-$\frac32$ multiplet in Table~\ref{spin32class},
those in the vector multiplet in Table~\ref{vectorclass}, and 
finally those in the chiral multiplet in Table~\ref{chiralclass}.

 \begin{table}[h!]
  \begin{center}
 \begin{tabular}{|c|c|}
\hline
$\Phi$ & $	\subseteq Q_{eq} \Phi$ \\
\hline 
\hline $e_{(3,3, 1)}$ & $\psi_{(3,3,1)}$ \\ 
$e_{(3,1, 1)}$ & $\psi_{(3,1,1)}$ \\ 
 $e_{(1,3, 1)}$ & $\psi_{(1,3,1)}$ \\
 $e_{(1,1, 1)}$ & $\psi_{(1,1,1)}$ \\
 $V_{(2,2,1)}$ & $\chi_{(2,2,1)}$ \\ 
$T_{(1,3,1)}$ & $\chi_{(1,3,1)}$ \\ 
$(c_Q)_{(1,3,1)}$ & $(c_{R})_{(1, 3, 1)}$ \\ 
$(c_Q)_{(1,1,1)}$ & $(c_{R})_{(1, 1, 1)}$ \\ 
$(A_\mu^D)_{(2,2,1)}$ & $(c_{K})_{(2, 2, 1)}$ \\ 
$b_*$ & $B_*$\\ 
\hline
 \end{tabular}
\quad
\begin{tabular}{|c|c|}
\hline
$\Psi$ & $	\subseteq Q_{eq} \Psi$\\
\hline \hline  
$\psi_{(2,2,1)}$ & $(c_{S})_{(2,2,1)}$\\
$\psi_{(2,2,1)}$ & $V_{(2,2,1)}$\\ 
$\psi_{(2,4,1)}$ & $V_{(2,4,1)}$\\ 
$\chi_{(1,1,1)}$ & $D_{(1,1,1)}$\\ 
$(c_M)_{(3,1,1)}$ & $T_{(3,1,1)}$\\ 
$(c_M)_{(1,3,1)}$ & $(c_S)_{(1,3,1)}$\\ 
$(c_D)_{(1,1,1)}$ & $(c_S)_{(1,1,1)}$\\ 
$(c_\mu)_{(2,2,1)}$ & $(c_Q)_{(2,2,1)}$\\ 
$b_*$ & $B_*$\\ 
\hline 
\end{tabular}
\caption{Cohomological primaries and descendants in the $\CN=2$ Weyl supermultiplet. 
The representation labels are the Cartans of $SU(2)_\mp \times SU(2)_{R\pm} \times SU(2)'$.
All fermionic/bosonic $b$-ghosts are cohomological primaries whose descendents are the bosonic/fermionic 
$B$-ghosts with the same $SU(2)_\mp \times SU(2)_{R\pm} \times SU(2)'$ representations. 
All the fields in this multiplet are singlets under~$SU(2)'$ and, if we suppress the notation~$1$, this table 
is the same as Table~6 of~\cite{Jeon:2018kec}.}
 \label{WeylClasseq2}
 \end{center}
\end{table}

 \begin{table}[h!]
  \begin{center}
  \begin{tabular}{|c|c|}
\hline
$\Psi$ & $	\subseteq Q_{eq} \Psi$\\
\hline \hline 
$\psi_{(2, 3, 2)}$ & $V_{(2, 3,2)}$ \\ 
$\psi_{(3, 2, 2)}$ & $T_{(3, 2,2)}$ \\ 
$\psi_{(2, 1, 2)}$ & $V_{(2, 1,2)}$ \\ 
$\psi_{(1,2,2)}$ & $(c_S)_{(1,2,2)}$ \\ 
$\Lambda_{(1,2,2)}$ & $E_{(1,2,2)}$ \\
$\chi_{(1,2,2)}$ & $D_{(1,2,2)}$ \\ 
$\chi_{(1,2,2)}$ & $D_{(1,2,2)}$ \\ 
$(c_R)_{(1,2,2)}$ & $E_{(1,2,2)}$ \\ 
$b_*$ & $B_*$\\ 
\hline
\end{tabular}
\quad
 \begin{tabular}{|c|c|}
\hline
$\Phi$ & $	\subseteq Q_{eq} \Phi$ \\
\hline \hline 
$V_{(2, 3, 2)}$ & $\chi_{(2, 3,2)}$ \\ 
$V_{(2, 1, 2)}$ & $\chi_{(2, 1,2)}$ \\ 
$T_{(1, 2, 2)}$ & $\chi_{(1, 2,2)}$ \\ 
$T_{(1, 4, 2)}$ & $\chi_{(1, 4,2)}$ \\ 
$(c_S)_{(2, 1, 2)}$ & $\chi_{(2, 1,2)}$ \\ 
$(c_Q)_{(2, 1, 2)}$ & $\Lambda_{(2, 1,2)}$ \\ 
$(c_Q)_{(1, 2, 2)}$ & $(c_R)_{(1, 2,2)}$ \\ 
$b_*$ & $B_*$\\ 
\hline
\end{tabular}
\caption{Cohomological primaries and descendants in the $\CN=2$ spin-$3/2$ supermultiplet. }
 \label{spin32class}
 \end{center}
\end{table}

  \begin{table}[h!]
  \begin{center}
  \begin{tabular}{|c|c|}
\hline
$\Phi$ & $	\subseteq Q_{eq} \Phi$ \\
\hline 
\hline $V_{(2, 2, 3)}$ & $\chi_{(2,2,3)}$\\ 
\hline $E_{(1,1,3)}$ & $\chi_{(1,1,3)}$ \\
\hline
\end{tabular}
\quad
\begin{tabular}{|c|c|}
\hline
$\Psi$ & $	\subseteq Q_{eq} \Psi$\\
\hline
\hline $\chi_{(1, 3, 3)}$ & $D_{(1,1,3)}$\\
\hline $(c_{R})_{(1,1,3)}$ & $\bar E_{(1, 1, 3)}$\\ 
\hline $(b_{R})_{(1,1,3)}$ & $(B_{R})_{(1,1,3)}$\\ 
\hline
\end{tabular}
\caption{Cohomological primaries and descendants in the $\CN=2$ vector supermultiplet.}
 \label{vectorclass}
 \end{center}
\end{table}

 \begin{table}[h!]
  \begin{center}
 \begin{tabular}{|c|c|}
\hline
$\Phi$ & $	\subseteq Q_{eq} \Phi$ \\
\hline \hline 
$\phi_{(1,1,1)}$ & $\Lambda_{(1,1,1)}$  \\
$\phi_{(1,1,1)}$ & $\Lambda_{(1,1,1)}$ \\
$E_{(3,1,1)}$ & 
$\chi_{(3,1,1)}$ \\ 
$T_{(1,3,1)}$ & 
$\chi_{(1,3,1)}$ \\ 
\hline 
\end{tabular}
\quad
\begin{tabular}{|c|c|}
\hline
$\Psi$ & $	\subseteq Q_{eq} \Psi$\\
\hline \hline 
$\Lambda_{(1, 3, 1)}$ & $\bar E_{(1, 3, 1)}$ \\ 
$\Lambda_{(3, 1, 1)}$ & $ T_{(3, 1, 1)}$ \\ 
$\chi_{(1, 1, 1)}$ & $ D_{(1, 1, 1)}$ \\ 
$\chi_{(1, 1, 1)}$ & $ D_{(1, 1, 1)}$ \\ 
\hline 
\end{tabular}
\caption{Cohomological primaries and descendants in the $\CN=2$ chiral supermultiplet.}
 \label{chiralclass}
 \end{center}
\end{table}

\newpage
 
\ndt {\bf Index calculation}

Using the results in the tables above we  can now easily calculate the index of the various multiplets using the fixed-point formula~\eqref{ABFormula},
and the trace formula~\eqref{TrNPSP}. In the super-conformal formalism, the results for the various~$\CN=2$ multiplets on the smooth AdS$_2 \times S^2$ are thus given as follows:
\bea
\text{Vector}: && \qquad \text{ind}_{\text{vect.}}^{\cN=2 \text{ s.c.}} (D_{10})(t) \big|_{\rm NP/SP}\= \frac{2(q+q^{-1}) -4}{(1-q)^2 \, (1-q^{-1})^2} \big|_{\rm NP/SP} \,, \\
\text{Hyper}: && \qquad  \text{ind}_{\text{hyper}}^{\cN=2 \text{ s.c.}}  (D_{10})(t)\big|_{\rm NP/SP} \= - \frac{2(q+q^{-1}) -4}{(1-q)^2 \, (1-q^{-1})^2} \big|_{\rm NP/SP} \,, \\
\text{Weyl}: && \qquad \text{ind}_{\text{Weyl}}^{\cN=2 \text{ s.c.}}  (D_{10})(t)\big|_{\rm NP/SP} \=  \frac{2(q^2+q^{-2}) - 6(q+q^{-1}) + 8}{(1-q)^2 \, (1-q^{-1})^2} \big|_{\rm NP/SP}\,, \\
\text{Spin-$\frac32$}: && \qquad \text{ind}_{\text{spin 3/2}}^{\cN=2 \text{ s.c.}}  (D_{10})(t) \big|_{\rm NP/SP} \=  \frac{-2(q^2+q^{-2}) + 4(q+q^{-1}) -4}{(1-q)^2 \, (1-q^{-1})^2} \big|_{\rm NP/SP} \,, \\
\text{Chiral}: && \qquad \text{ind}_{\text{chiral}}^{\cN=2 \text{ s.c.}}  (D_{10})(t)\big|_{\rm NP/SP} \= 0\big|_{\rm NP}\,,
\eea
where the ``s.c.'' upper-script indicates that we are still working in the off-shell formulation of supergravity in the super-conformal formalism. As in the example of the $\cN=2$ vector supermultiplet given in Section~\ref{sec:N=2-vector-supermultiplet-example}, we have separated the contribution of the north and south poles since one needs to carefully treat them in a small $q$, respectively $q^{-1}$, expansion.

Following the discussion in Section~\ref{sec:N=4-Weyl-supermultiplet}, we can now write the indices of $\cN=2$ after fixing to super-Poincar\'e. We will denote such results using the ``s.P.'' upperscript. Directly listing the the results for the orbifold geometries using the sum over images discussed in Section~\ref{sec:one-loop-det-orbifold}, we find
\bea
\text{Vector}: &&\qquad \text{ind}_{\text{vect.},\,c}^{\cN=2 \text{ s.P.}} (D_{10})(t)\big|_{\rm NP} \=\,  \frac{2 cq^c}{(1-q^c)^2} \big|_{\rm NP} , \label{eq:N=2-vector-sp-indexfirst}\\
\text{Hyper}: && \qquad  \text{ind}_{\text{hyper},\,c}^{\cN=2 \text{ s.P.}}  (D_{10})(t) \big|_{\rm NP} \= \frac{-2 cq^c}{(1-q^c)^2} \big|_{\rm NP}  \,, \\
\text{graviton}: && \qquad \text{ind}_{\text{Graviton},\,c}^{\cN=2 \text{ s.P.}}  (D_{10})(t) \big|_{\rm NP} \= \frac{2 q^{2c} + 2(c-2)q^{c}+2}{(1-q^c)^2} \big|_{\rm NP} \,, \\
\text{gravitino}: && \qquad \text{ind}_{\text{Gravitino},\,c}^{\cN=2 \text{ s.P.}}  (D_{10})(t) \big|_{\rm NP} \= -\frac{2 q^{2c} + 2(c-2)q^{c}+2}{(1-q^c)^2} \big|_{\rm NP}\,,\label{eq:N=2-vector-sp-indexlast} 
\eea
We can finally put together the $\cN=2$ supermultiplets gauged fixed to super-Poincar\'e, to form the $\cN=4$ vector and Weyl or $\cN=8$ Weyl multiplets. These multiplets are found by putting these together
\be
\label{eq:N=4-vector-index}
\CN=4 \; \text{vector} &\rightarrow  \cN=2\,\text{vector + hyper}: \nn \\  &\text{ind}_{\text{vector}}^{\cN=4 \text{ s.P.}} (D_{10})(t) \= 0 \,,
\ee
\be
\label{eq:N=4-Weyl-index}
\CN=4 \; \text{graviton} 
&\rightarrow \cN=2\,\text{ graviton + 2~gravitini  + vector in s.P.}: \nn \\ &\text{ind}^{\cN=4 \text{ s.P.}}_{\text{graviton}} (D_{10})(t) \= -2 \,,
\ee
\be
\label{eq:N=8-Weyl-index}
\CN=8 \; \text{graviton}& \rightarrow \cN=2\,\text{ graviton + 6~gravitini + 15 vectors + 10 hypers in s.P.}: \nn \\ & \text{ind}^{\cN=8 \text{ s.P.}}_{\text{graviton}} (D_{10})(t) \= -10 \,.
\ee
We emphasize that the results \eqref{eq:N=4-vector-index}-\eqref{eq:N=8-Weyl-index} are exact, and are not expansions at leading order in $t$. Importantly, the indices of the $\cN=4$ and $\cN=8$ supermultiplets are all independent of $t$. Because of this on the orbifold geometry, the sum over images discussed in Section~\ref{sec:one-loop-det-orbifold}  implies that the index of such super-multiplets is $c$-independent. Since the result is $t$-independent we have also summed the contribution of the north and south pole since we not longer have to distinguish the two in the small $q$, respectively small $q^{-1}$, expansion.

The results \eqref{eq:N=4-vector-index}--\eqref{eq:N=8-Weyl-index} also agree with the leading quantum correction found for the orbifold entropy \cite{Gupta:2014hxa}, who noticed that the coefficient in front of the logarithmic correction in the black hole area is independent of $c$. Nevertheless, \eqref{eq:N=4-vector-index}--\eqref{eq:N=8-Weyl-index} are much stronger results: as we shall see, it predicts that not only this coefficient in the quantum correction to the entropy is $c$-independent but also the entire one-loop determinant of all bulk modes is also $c$-independent.

To obtain the one-loop determinants from the results \eqref{eq:N=2-vector-sp-indexfirst}--\eqref{eq:N=2-vector-sp-indexlast} for the indices of the various supermultiplets that we have discussed we simply have to follow the same procedure as in Section~\ref{sec:N=2-vector-supermultiplet-example} and perform the integral over $t$, appropriately keeping track of the expansion in $q$ and $q^{-1}$ at the north and south poles, respectively. Below, we shall list the one-loop determinants obtained through this procedure.

\subsection{A summary of results}

\label{sec:summary-results-bulk-one-loop-det}

The one-loop determinants are given by a contribution captured by the index $Z_{\rm index}$, and a correction from a proper treatment of zero-modes which is explained in the next section. Since the results for the indices associated to each supermultiplet had an intricate derivation, below we summarize the contribution to the one-loop determinants from these indices for all $\cN=2$, $4$ or $8$ supergravity supermultiplets, 
when gauge fixed to the super-Poincar\'e theory.  For the AdS$_2 \times S^2$ geometries as well as for the orbifold geometries, the one-loop determinants of all $\cN=2$ super-multiplets are:
\be
\log \left( Z_{\rm index}^{\cN=2\,{\rm vector}} \right) &\=  - \frac{\ii\pi}{12} \, c+ \frac{c}{6}\log c+2 c \, \zeta'(-1) - \frac{c}{12} \log \ell^2 \,, \nn \\
\log \left( Z_{\rm index}^{\cN=2\,{\rm hyper}} \right) &\=   \frac{\ii\pi}{12} \, c- \frac{c}{6}\log c-2 c \, \zeta'(-1) + \frac{c}{12} \log \ell^2 \,, \nn \\
\log \left( Z_{\rm index}^{\cN=2\,{\rm graviton}} \right) &\= -\frac{\ii \pi}{12} \, c + \frac{c}{6}\log c + 2 c \, \zeta'(-1)  + \left(1 -\frac{c}{12}\right) \log \ell^2 \nn \\ 
\log \left( Z_{\rm index}^{\cN=2\,{\rm gravitino}} \right) &\= \frac{\ii \pi}{12} \, c  - \frac{c}{6}\log c - 2 c \, \zeta'(-1) -\left(1 -\frac{c}{12}\right) \log \ell^2 \,.
\ee
It is also useful to have the expression for the index ${\rm ind}(D_{10})(t)$ associated to each multiplet, which we gave above in equations \eqref{eq:N=2-vector-sp-indexfirst}-\eqref{eq:N=2-vector-sp-indexlast}. These expression present complicated dependence with the orbifold parameter. For $\cN=4$ super-multiplets we have a much simpler answer:
\be
Z_{\rm index}^{\cN=4\,{\rm vector}} &\= 1\,,\nn \\ 
 \log\left( Z_{\rm index}^{\cN=4\,{\rm graviton}} \right)&\= -\log \ell^2\,.
\ee
Finally, for $\cN=8$ super-multiplets we obtain:
\be 
 \log \left( Z_{\rm index}^{\cN=8\,{\rm graviton}} \right) &\= -5\log \ell^2\,.
\ee
Thus we see that
the answer is independent of the orbifold parameter~$c$ for the~$\mathcal{N}=4$ theory
as well as the $\CN=8$ theory.

\section{The role of large gauge transformations}
\label{sec:role-of-large-gauge-transformations}

Large (super)diffeomorphisms that are nearly zero-modes have been found to control the low-temperature limit of near-extremal (or near-BPS) black holes (see our companion paper \cite{Iliesiu:companionPaper} and references therein). In this context, their role within the localization framework presented in this paper should be re-analyzed. At the beginning of Section~\ref{sec:one-loop-det-theoretical-prelims}, we noted that in addition to the modes whose one-loop determinant we have considered in the previous section there are additional large diffeomorphisms, some of which are zero-modes of $\qeq \CV$ and some that are not. We denote the full one-loop determinant by:
\be 
\label{eq:separation-one-loop}
Z_\text{1-loop}^{\qeq \CV} \;\equiv \; Z_\text{zero-modes}^{\qeq \CV} Z_\text{non-zero}^{\qeq \CV}\,.
\ee
 $Z_\text{zero-modes}^{\qeq \CV}$ is given by the full integral over the space of large super-diffeomorphisms that are zero-modes. $Z_\text{non-zero}^{\qeq \CV}$ includes the contribution of all other modes, part of it is reproduced by the index calculation of the previous section but also has contributions from treating boundary non-zero-modes carefully. In Section~\ref{sec:effect-of-large-gauge-transf}, we follow~\cite{Jeon:2018kec} and clarify how the presence of such boundary non-zero modes affects the relation between the index contribution $Z_\text{index}^{\qeq \CV}$ computed in the previous sections and $Z_\text{non-zero}^{\qeq \CV}$. Nevertheless, in \cite{Jeon:2018kec} the path integral over the zero-modes and therefore the entire moduli space of large super-diffeomorphisms was not considered and thus the full $\phi^0$ and $c$-dependence of their contribution was left undetermined beyond leading order. 
 In Section~\ref{sec:integration-measure-large-gauge-transf} and \ref{sec:connection-to-N=4-super-Schw}, we compute this path integral explicitly to finally obtain the full one-loop determinant around the localization locus. A more general analysis about the role of these large diffeomorphisms in the gravitational path integral for extremal black holes is discussed in a companion paper \cite{Iliesiu:companionPaper}.

\subsection{Effect of large gauge transformations on the bulk one loop determinant}
\label{sec:effect-of-large-gauge-transf}

Large gauge transformations or large diffeomorphisms are pure gauge modes whose gauge parameters are not normalizable and thus, in our case, are non-vanishing at the boundary of AdS$_2$. Such modes are physical and should be integrated over since they have a non-trivial effect on physically observables quantities. 
For example, both large diffeomorphisms affect the extrisic curvature of the boundary while large gauge transformations affects the value of the electro-magnetic boundary term. This is true as long as such modes do not correspond to isometries of the spacetime or of the gauge bundle. In fact, as we shall see the exclusion of such isometry modes from the space of physical large diffeomorphisms will play an important role below. 

To analyze the role of all large diffeomorphisms modes in the relation between $Z_\text{index}^{\qeq \CV}$ and $Z_\text{non-zero}^{\qeq \CV}$, we will start by studying the illustrative  example of the simplest large-gauge transformation,  that for the $U(1)$ gauge fields $A_\mu^I$ that are part of the vector supermultiplets. They are normalizable modes with $A_\mu^\text{bdry} = \partial_\mu \Lambda$ obtained from non-normalizable scalar modes $\Lambda$ where for the purposes of this section we will drop the $I$ label associated to each vector supermultiplet\footnote{Non-normalizable gauge parameters generate zero-modes only for certain choice of the dependence on $S^2$. 
For example for vector zero-modes they involve a constant gauge transformation along $S^2$. Some zero-modes in the metric involve spin one harmonics, e.g.~the ones realizing the rotational $SU(2)$ isometries of $S^2$.}. Such modes transform under the equivariant variation as $\qeq A_\mu^\bdy = \lambda_\mu^{\bdy}$ where $\lambda_\mu^{\bdy}$ is a boundary mode associated to the gaugino. Boundary modes that are zero-modes also satisfy 
$\qeq^2 A_\mu^\text{bdry} = H  A_\mu^\text{bdry} = n^\nu F_{\nu \mu}^\text{bdry} =  0\,$. More generally, all large bosonic gauge transformations and diffeomorphisms (denoted by $\Phi^\text{bdry}$) that are zero-modes also satisfy
\be 
 \qeq^2 \Phi^\text{bdry} \= H  \Phi^\text{bdry} \=  0\,.
\ee
For all such modes $\cV$ is vanishing---this can be easily seen for $A_\mu^\text{bdry}$ since for such modes  $\cV$ solely contains composites of the vanishing boundary field strength. This consequently implies that 
$\qeq \CV|_{A_\mu^\text{bdry} } = 0\,$ and, more broadly, that
\be 
\qeq \CV|_{\Phi^\text{bdry} } \= 0\,.
\ee
Thus, 
all modes $\Phi^\text{bdry}$ are zero modes of the $\qeq$-exact deformation term in supergravity, confirming that they are part of the localization locus. For extremal black holes it has also long been known that such modes are zero modes of the original supergravity 
action making their path integral somewhat more subtle as we explain in Section~\ref{sec:connection-to-N=4-super-Schw}. The superpartners $\qeq \Phi^\text{bdry}$ are however not necessarily zero-modes of the deformation term, even though they are boundary modes. Explicitly, the deformation term for such superpartners can be consistently chosen to be
\be 
\label{eq:QV-for-QPhiBdy}
\qeq \CV|_{\qeq \Phi^\text{bdry}} =  -(\qeq \Phi^\bdy) {H} (\qeq \Phi^\bdy) = -\qeq\left[\Phi^\bdy {H} (\qeq \Phi^\bdy)\right]\,,
\ee
which, from the last equality, is also clearly  $\qeq$-exact and thus remains a valid deformation in the localization procedure. 
Thus, the contribution of the modes $\qeq \Phi^\text{bdry}$ for which the one-loop determinant can be explicitly obtained from \eqref{eq:QV-for-QPhiBdy}, should be distinguished from that of $\Phi^\text{bdry}$ for which a full path integral should be considered. To simplify notation we have rescaled $H$ by $\ell$ when discussing the quadratic fluctuations of $\qeq \CV$. 

Similarly, let us consider large fermionic gauge transformations, i.e.~the fermionic parts in the super-diffeomorphisms of supergravity.
Such fermionic modes are either cohomological primaries, which we denote by $\Psi^\bdy$, or they are cohomological descendants, in which case they can be written as $\qeq \Phi^\text{pre-bdry}$ for the appropriate bosonic mode  $\Phi^\text{pre-bdry}$.\footnote{In supergravity, it can be checked non-elementary bosons are not zero modes of either the supergravity action or of the deformation term and, for this reason, they have not been considered above.} In both cases,  for such fermionic modes $\qeq$ is nilpotent. Similarly, to the bosonic case,  $\cV$ vanishes for such modes, implying
\be 
\qeq \CV|_{\Psi^\text{bdry} }= \qeq \CV|_{\qeq \Phi^\text{pre-bdry} } = 0\,,
\ee
which means that both $\Psi^\text{bdry}$ and $\qeq \Phi^\text{pre-bdry}$ are part of the localization locus and should be fully integrated over to obtain their contribution to $Z_\text{zero-modes}^{\qeq \CV}$. The superpartners of such modes, $\qeq \Psi^\bdy$ and $\Phi^\text{pre-bdry}$ are however not zero-modes of the deformation term, given the definition \eqref{Vact} of $\CV$. Specifically, evaluating the deformation term for these superpartners yields
\be 
\label{eq:QV-for-QPsiBdy}
\qeq \CV|_{\qeq \Psi^\text{bdry}} =  -(\qeq \Psi^\bdy) 1 (\qeq \Psi^\bdy)\,,\qquad \qeq \CV|_{\Phi^\text{pre-bdry}} =  \Phi^\text{pre-bdry} H^2 \Phi^\text{pre-bdry}\,,
\ee
again $\qeq$-exact due to the nilpotency of $\Psi^\bdy$ and $\qeq \Phi^\text{pre-bdry}$. 

Thus, we find that $Z_\text{non-zero}^{\qeq \CV} = \sqrt{{\det^{*_f} K_f}/{\det^{*_b} K_b}}$, where ${*_b}$ includes all bosonic  modes with the exception of $\Phi^\bdy$, while ${*_f}$ includes all fermionic modes with the exception of $\Psi^\text{bdry}$ and $\qeq \Phi^\text{pre-bdry}$. Using \eqref{eq:QV-for-QPhiBdy} and \eqref{eq:QV-for-QPsiBdy} we can explicitly separate out the contribution of the superpartners of large super-diffeomophisms to the one-loop determinant:
\be 
 Z_\text{non-zero}^{\qeq \CV}  =  \sqrt{\frac{\det^{*_f} K_f}{\det^{*_b} K_b}} = \sqrt{\frac{\text{det}_{\qeq \Phi^\bdy} (H)}{\text{det}_{\Phi^\text{pre-bdy}}( H^2 )}\frac{\det' K_f}{\det' K_b}}  = \sqrt{\frac{\text{det}_{\Phi^\bdy} (H)}{\text{det}_{\Phi^\text{pre-bdy}}( H^2 )}\frac{\det'_\Psi H}{\det'_\Phi H}} 
\ee
where the $'$ indicates that we are only considering the determinant over the bulk modes, no longer including the contribution of $ \Phi^\bdy$,  $Q_{\rm eq} \Phi^\bdy$, $Q_{\rm eq} \Psi^\text{bdy}$, or $\Phi^\text{pre-bdy}$. After the second equality sign, 
we have dropped the one-loop determinant of the $\qeq \Psi^{\rm bdry}$ modes since \eqref{eq:QV-for-QPsiBdy} implies that this determinant equals $1$. 
Now, in the $'$-space, all fields have~$\qeq$superpartners. 
Consequently, the differential operators $K_f$ and $K_b$ are related and the third equality follows.

In total, writing the heat-kernel transformation associated to $ Z_\text{non-zero}^{\qeq \CV} $, we find
\be 
\log Z_\text{non-zero}^{\qeq \CV} 
&\=-\frac{1}2 \int \frac{dt}{t} \left(\Tr_{\Phi^\bdy}e^{\frac{tH}\ell} - 2\Tr_{\Phi^\text{pre-bdry}}e^{\frac{tH}\ell} + \Tr_{\Psi'} \rme^{\frac{tH}\ell} - \Tr_{\Phi'} \rme^{\frac{tH}\ell}\right)\nn \\ 
&\=\int \frac{dt}t \left[-\Tr_{\Phi^\bdy}e^{\frac{tH}\ell} +\frac{1}2 \left(\Tr_{\Psi^\bdy}e^{\frac{tH}\ell}+ \Tr_{\qeq \Phi^\text{pre-bdry}}e^{\frac{tH}\ell}\right)\right]  \nn \\ 
& \qquad - \frac{1}2 \int\frac{dt}{t} \left(\Tr_{\Psi} \rme^{\frac{tH}\ell} - \Tr_{\Phi} \rme^{\frac{tH}\ell}\right)\,,
\ee
where we have reintroduced the scale $\ell$ in order to make the comparison with the previous section easier. The spaces $\Psi$ and $\Phi$ is now the entire space of normalizable modes associated to fields that are cohomological primaries, including all boundary modes. The difference of traces for such modes is precisely what the index of the operator $D_{10}$ computes, 
\be 
\text{ind}_{\text{Weyl}}^{\cN=8\text{ s.P.}}(D_{10})(t) \= \Tr_{\Phi} \rme^{\frac{tH}\ell} - \Tr_{\Psi} \rme^{\frac{tH}\ell}\,.
\ee 
Additionally, since all modes $\Phi^\bdy$, $\Psi^\bdy$ and $\qeq \Phi^\text{pre-bdry}$ are nilpotent under the action of $\qeq$, the traces in the square paranthesis simply count the number of zero modes in each set.  Since the number of zero modes is technically divergent, it has to be regularized by considering the ratio of bulk one-loop determinants for different values of $\phi^0$. We will denote the number of regularized zero modes by 
\be 
n^\text{bos}_\text{bdry} \;\equiv\; \Tr_{\Phi^\bdy}e^{{\tilde t H}} \,, \qquad n^\text{ferm}_\text{bdry} \; \equiv\; \Tr_{\Psi^\bdy}e^{{\tilde t H}}\,, \qquad \text{ with} \qquad \tilde t = \frac{t}{\ell}\,.
\ee
The regularized number of zero modes can be concretely obtained by zeta function regularization, excluding the contribution of the modes that correspond to bosonic or fermionic isometries and thus should not be integrated-over. On the smooth AdS$_2\times S^2$ geometry the super-isometry group is $H_{\rm Disk} = PSU(1,1|2)$ which has $6$ bosonic generators and $8$ fermionic generators. Each generator has an associated large super-diffeomorphism mode that has to be subtracted in the regularization of the number of bosonic and fermionic zero-modes. One should similarly subtract the contribution of modes with $\Lambda = $ constant for each vector supermultiplet. Consequently, for the smooth AdS$_2\times S^2$ geometry 
\be
\label{eq:regularized-number-zero-modes-smooth}
n^\text{bos}_\text{bdry}  \;\equiv\; n_\text{bdry}^\text{1-form} + n_\text{bdry}^\text{grav}\,, \qquad   n_\text{bdry}^\text{grav} \=  -6\,, \qquad n_\text{bdry}^\text{1-form} \= -n_v\,, \qquad  n^\text{ferm}_\text{bdry}  \= -8\,, \qquad \text{ for }c \= 1.
\ee
For the orbifolded geometry the super-isometry group is $\orbiso$ which has $2$ bosonic generators and $4$ fermionic generators, which implies that 
\be 
\label{eq:regularized-number-zero-modes-orbifold}
  n_{\text{bdry},\,c}^\text{grav} \=  -2\,, \qquad n_{\text{bdry},\,c}^\text{1-form} \= -n_v\,,  \qquad  n^\text{ferm}_{\text{bdry},\,c}  \= -4\,, \qquad \text{ for }\, c>1\,.
\ee 
Thus, we find that $Z_\text{bulk}^{\qeq \CV}$ is related to the index computed in the previous section through 
\be 
\label{eq:Z-bulk-final-heat-kernel-formula}
\log \,Z_{\text{non-zero}, \,c}^{\qeq \CV} =  
\left(-n^\text{bos}_{\text{bdry},\,c} +\frac{1}2 n^\text{ferm}_{\text{bdry},\,c}\right) \log(\ell)+  \log \,Z_{\text{index}, \,c}^{\qeq \CV}\,,
\ee
where 
\begin{equation}
    \log \,Z_{\text{index}, \,c}^{\qeq \CV} \=  \frac{1}2 \int_{\epsilon}^{\infty} \frac{d t}{ t} \text{ind}_{\text{Weyl},\,c}^{\cN=8\text{ s.P.}}(D_{10})(t).
\end{equation}
The fact that $-n^\text{bos}_{\text{bdry}, c}  +\frac{1}2 n^\text{ferm}_{\text{bdry},\,c} = 2+\frac{n_v}2$ for the $c=1$ smooth AdS$_2 \times S^2$ geometry is different than $-n^\text{bos}_{\text{bdry}, c}  +\frac{1}2 n^\text{ferm}_{\text{bdry},\,c} =\frac{n_v}2$ contribution for the orbifolded geometries with $c>1$ might seem problematic at first sight since it seemingly suggests that the $\phi^0$-dependence (and not only $c$-dependence) of $Z_{\text{index}, \,c}^{\qeq \CV}$ on the two geometry types is different. However, as we shall see in the next subsection, the measure of the integral over all physical modes is such that $Z_\text{1-loop}^{\qeq \CV}$ has the same $\phi^0$-dependence on all geometries that we sum over in the gravitational path integral.

\subsection{Integration measure in the space of large gauge transformations}
\label{sec:integration-measure-large-gauge-transf}

Thus, what remains to be evaluated is $Z_{\text{zero-modes},\,c}^\text{bdy.}$, the path integral over the localization locus of the zero modes:
\be 
\label{eq:zero-mode-integral}
Z_{\text{zero-modes},\, {c}}^\text{bdy.} &= \int_{\frac{\text{Large-diffeomorphisms}}{\text{Super-isometries}}} [d(A_\mu)^{\text{Large gauge transf.}}][d{h}_{\mu\nu}^{\text{Large super-diffs.}} ] [d{\psi}_{\mu}^{\text{Large super-diffs.}} ] \,,
\ee
where $(A_\mu)^{\text{Large gauge transf.}}$ and $h_{\mu \nu}^{\text{Large gauge transf.}}$ are part of $\Phi^\text{bdry}$ and, depending on their twisted representation, twisted components of ${\psi}_{\mu}^{\text{Large super-diffs.}}$ are either part of $\Psi^\bdy$ or of $\qeq \Phi^\text{pre-bdry}$. 

The integration space for large super-diffeomorphisms is given by $\text{Diff}(S^{1|4})$. There  are two types of bosonic modes in $\text{Diff}(S^{1|4})$, which correspond to fluctuations of the metric 
that we integrate in \eqref{eq:zero-mode-integral} when travelling along the  AdS$_2$ boundary: the first 
captures fluctuations in the shape of the asymptotically AdS$_2$ boundary and the second captures rotational fluctuations around $S^2$. 
Similarly, the fermionic modes in $\text{Diff}(S^{1|4})$ correspond to fluctuations of the gravitino when travelling along the  AdS$_2$ boundary, that we also integrate in \eqref{eq:zero-mode-integral}. Finally, the integration space of the large gauge transformations in \eqref{eq:zero-mode-integral} is $\text{Loop}(U(1))$ for each gauge field $A_\mu^I$.  

From $\text{Diff}(S^{1|4})$ we have to quotient out the appropriate super-isometry group, corresponding to the modes that are not physical. Thus, on the smooth AdS$_2\times S^2$ geometry the integration space is $\text{Diff}(S^{1|4})/PSU(1,1|2)$ and on the orbifold geometry it is $\text{Diff}(S^{1|4})/\orbiso$. The integration space for the large gauge transformations is similarly $\left[\text{Loop}(U(1))/U(1)\right]^{n_v}$ for black holes in the grand canonical ensemble.
If we instead want to consider black holes in the canonical ensemble,\footnote{Recall that we are still studying an index in which angular momenta are not fixed. Here, by canonical ensemble we mean that all $U(1)^{n_v+1}$ charges are fixed. }
we need to perform an additional integral over the holonomy around the AdS$_2$ boundary in order to fix the charges. Therefore, the integration space for large gauge 
transformations in the canonical ensemble is~$\left[\text{Loop}(U(1))\right]^{n_v}$. 

We take the measure over all fields to be given by the ultralocal measure. 
For example, for the gauge fields in the vector supermultiplets, we require that  
\be 
\label{eq:ultralocal-measure-example}
\int [d(A_\mu)] \rme^{-\int d^4 x \sqrt{g} g^{\mu\nu} A_\mu\, A_\nu} = 1\,.
\ee
In \eqref{eq:ultralocal-measure-example} and in the analogous equations for the metric and gravitino, we can rescale the fields to fully extract the dependence in the integration measure on the overall scale $\ell$ entering in the metric $g_{\mu \nu}$. (We remind the reader that $\ell^2 \sim 1/(\phi^0)$ on the localization locus.) 
This is a correct rescaling of the fields close to the localization locus and, since we are only interested in computing the quadratic fluctuation 
determinant around this locus, we do not have to consider fluctuations in the metric within the definition of the integration measure \eqref{eq:ultralocal-measure-example}. 
This rescaling of the fields in   \eqref{eq:ultralocal-measure-example} 
and in the analogous equations for the metric and gravitino was discussed in \cite{Sen:2011ba, Sen:2012cj, Jeon:2018kec} for the smooth AdS$_2\times S^2$ geometry. A similar analysis can be performed on the orbifolded geometries taking the different super-isometry group into account. 
The resulting path integral~\eqref{eq:zero-mode-integral} over zero-modes is 
\be 
\label{eq:zero-mode-integral-smooth}
Z_{\text{zero-modes},\, {c=1}}^\text{bdy.} &=(\ell)^{c_{\text{zero-modes}}} \int_{\frac{\text{Diff}(S^{1|4})}{PSU(1,1|2)}}  D(\text{Large-diffs}) \int_{\left[{\text{Loop}U(1)}\right]^{n_v}} D(\text{Large gauge})\,,  \\ 
\label{eq:zero-mode-integral-orbifold}
Z_{\text{zero-modes},\, {c}}^\text{bdy.} &= (\ell)^{c_{\text{zero-modes},\,c}} \int_{\frac{\text{Diff}(S^{1|4})}{\orbiso}} D(\text{zero-modes})\int_{\left[{\text{Loop}U(1)}\right]^{n_v}} D(\text{Large gauge})  \,,
\ee 
where the coefficient $c_{\text{zero-modes},\,c}$ 
takes the form 
\be
c_{\text{zero-modes},\,c} =  \beta^\text{1-form} n^\text{1-form}_{{\rm bdry},\,c} +\beta^\text{grav} n^{\rm grav}_{{\rm bdry},\,c} - \frac{1}2   \beta^\text{ferm} n^{\rm ferm}_{{\rm bdry},\,c}\,. 
\ee
Here~$ \beta^\text{1-form}$, $\beta^\text{grav}$, and $\beta^\text{ferm}$ are coefficients that capture the number degrees of freedom of the 1-forms, graviton and gravitino and were found to be \cite{Sen:2011ba,Sen:2012cj}
\be 
\beta^\text{1-form} = 1\,,\qquad \beta^\text{grav} =2 \,, \qquad  \beta^\text{ferm} = 3\,,
\ee
while the coefficients $ n^\text{1-form}_{{\rm bdry},\,c}$, $n^{\rm grav}_{{\rm bdry},\,c}$, and $n^{\rm ferm}_{{\rm bdry},\,c}$ are given in \eqref{eq:regularized-number-zero-modes-smooth} for the smooth AdS$_2\times S^2$ geometry, and in \eqref{eq:regularized-number-zero-modes-orbifold}  for the orbifold geometry. 
The remaining integrals over zero-modes in~\eqref{eq:zero-mode-integral-smooth}--\eqref{eq:zero-mode-integral-orbifold} is independent of $\ell$ but can still depend on the topology of the geometry, and therefore on the parameter $c$ of the orbifold. 

We now combine the powers of $\ell$ appearing in \eqref{eq:zero-mode-integral-smooth}--\eqref{eq:zero-mode-integral-orbifold} with the powers of $\ell$ appearing from counting the total number of 
zero-modes in~\eqref{eq:Z-bulk-final-heat-kernel-formula}. In total we find a power $(\ell)^{c_\text{non-index}}$ coming from the zero-modes, with $c_\text{non-index}$ given by
\be 
\label{eq:c-non-index}
c_\text{non-index} \= (\beta^\text{1-form}-1)\, n^\text{1-form}_{{\rm bdry}, \, c} +(\beta^\text{grav}-1) \, n^{\rm grav}_{{\rm bdry},\,c} - \frac{1}2  ( \beta^\text{ferm}-1) \, n^{\rm ferm}_{{\rm bdry},\,c} \= 2 \,,~~\text{ for all } c \,.
\ee
Notice that due to the particular values of $n^{\rm grav}_{{\rm bdry},\,c} $ and $n^{\rm ferm}_{{\rm bdry},\,c} $ in \eqref{eq:regularized-number-zero-modes-smooth} for $c=1$ and \eqref{eq:regularized-number-zero-modes-orbifold} for $c>1$, we find that \eqref{eq:c-non-index} is the same for AdS$_2\times S^2$ and its orbifolds. 
Using the fact that the index of the $\cN=8$ graviton supermultiplet also has the same scaling with $\ell$ for all these geometries, this means that the overall scaling with $\ell$ of the full one-loop determinant is also precisely the same for all these geometries. 
As we shall see in Section~\ref{sec:putting-it-all-together} for the $\frac{1}{8}$-BPS black hole index, the fact
that~$c_\text{non-index}$ is independent of the geometry around which the one-loop determinant is computed is a crucial detail needed to obtain 
the correct form of the Rademacher expansion.

\subsection{The connection to the $\cN=4$ super-Schwarzian}
\label{sec:connection-to-N=4-super-Schw}

What remains to be done is to evaluate the path integrals in \eqref{eq:zero-mode-integral-smooth} and \eqref{eq:zero-mode-integral-orbifold} over the zero-modes, with their $\ell$-dependence stripped off. The integrals that we would like to compute are schematically given by
\be 
\label{eq:volumes-that-we-want-to-compute}
\Vol\left[\frac{\text{Diff}(S^{1|4})}{H_\text{Disk}}\right]&=\int_{\frac{\text{Diff}(S^{1|4})}{H_\text{Disk}}}  D(\text{zero-modes})\,,\qquad \Vol\left[\frac{\text{Diff}(S^{1|4})}{H_\text{Orb}}\right]= \int_{\frac{\text{Diff}(S^{1|4})}{H_\text{Orb}}}  D(\text{zero-modes})\,, \nn \\ & \hspace{0.5cm} \Vol\left[{\text{Loop}(U(1))}\right]=
\int_{\text{Loop}(U(1))}  D(\text{Large gauge})
\ee 
where the boundary conditions for the zero modes will be clarified momentarily. All integrals above are technically divergent and need to be regularized. This brings up two issues that we need to address:
\begin{itemize}
    \item All volumes in \eqref{eq:volumes-that-we-want-to-compute} should have a non-vanishing regularized value; otherwise, the gravitational path integral would incorrectly predict that BPS black holes do not have a degeneracy. This should be contrasted to the case of non-supersymmetric black holes discussed in the companion paper \cite{Iliesiu:companionPaper}. In that case a similar integral, over zero-modes appears but there we find that $\Vol\left[\text{Diff}(S^1)/SL(2,\mR)\right] = 0$ after a consistent regularization, suggesting that non-supersymmetric extremal black holes \emph{do not} have a degeneracy that scales as $\exp(\text{Area}/4G_N)$. 
    \item $\Vol\left[{\text{Diff}(S^{1|4})}/{H_\text{Orb}}\right]$ can have a $c$-dependence from that of the boundary conditions for the zero-modes that we integrate over in \eqref{eq:volumes-that-we-want-to-compute}. 
    To complete the computation of the one-loop determinant for all geometries we would thus like to determine this $c$-dependence. 
\end{itemize}

To regularize the quantities in \eqref{eq:volumes-that-we-want-to-compute} we would like to compare the partition functions of black holes at zero temperature to the partition function of black holes at some non-zero temperature. This is because in the latter case the zero-modes \eqref{eq:volumes-that-we-want-to-compute} of  the action and of the localizing deformation will now gain a ``mass'' in the supergravity action and will thus no longer be zero-modes \cite{Iliesiu:companionPaper}; 
instead, in the limit of low-temperatures we will refer to these modes as ``nearly zero-modes''. In this limit, these modes decouple from all other bulk modes whose one-loop determinant we have computed in Section~\ref{sec:one-loop-det-results} and the effective theory associated to these modes becomes strongly coupled. 
Nevertheless, as we explain in detail in the companion paper \cite{Iliesiu:companionPaper} the effective theory associated to these modes turns out to 
have a path integral that is exactly solvable even at strong coupling, at 
small temperatures.\footnote{For $\cN=0,1,2$ super-Schwarzian this has been understood in \cite{Stanford:2017thb} and \cite{Mertens:2017mtv,Lam:2018pvp}, while for the case of the $\cN=4$ super-Schwarzian its path integral was studied in \cite{Heydeman:2020hhw} using similar techniques.} 
Explicitly, the first two integrals in \eqref{eq:volumes-that-we-want-to-compute} are given by the zero-temperature limit of the  $\cN=4$ super-Schwarzian  path integral, with appropriate boundary conditions for the disk and orbifold geometries respectively. 
The final integral in \eqref{eq:volumes-that-we-want-to-compute} is given by the zero-temperature limit of a particle moving on a circle. For black holes in the canonical ensemble the path integral in \eqref{eq:volumes-that-we-want-to-compute}  should be taken in a sector with fixed $U(1)$ charge. 

\ndt {\bf The volume of the space of gauge zero modes}

We will first consider the path integral over the large gauge transformations, $A_\tau = \partial_\tau \Lambda$, for the gauge fields in the vector supermultiplets.  
At small temperatures the gauge transformations gain a non-zero mass due to the electromagnetic boundary term $S_\text{sugra, bdy} \supset \int_{\partial(\text{AdS}_2\times S^2)} d^3 x \sqrt{h} A^\mu F_{\mu \nu} n^\nu$. 
This term simply evaluates to \cite{Sachdev:2019bjn, Iliesiu:2019lfc, Iliesiu:2020qvm} 
\be 
\label{eq:particle-on-U(1)-action}
I_{\text{Particle on }U(1)} = -\frac{T}{E_{U(1)}}\int_0^{2\pi} d\tau \, (\partial_\tau \Lambda(\tau))^2\,,
\ee
where $T/E_{U(1)}$ is the coupling of the $U(1)$ nearly zero-mode where $E_{U(1)}$ is determined by the background charge of the black hole. As we shall see shortly, the exact value of $E_{U(1)}$ for the black holes that we study in this paper is unimportant in the extremal limit.  

While working in the grand canonical ensemble, with a chemical potential given by $\mu_{U(1)}/T =  \alpha_{U(1)}$, we need to impose the boundary condition  $\Lambda(\tau+2\pi) = \Lambda(\tau) + \alpha_{U(1)}$. The holonomy $\alpha_{U(1)}$ is conjugate not to the total charge of the higher dimensional $U(1)$ field, but its fluctuation $\tilde{q}$ around the background value. 
However, since we are interested in fixing the charges of the black hole for the gauge fields in all the supermultiplets we should fix $\tilde q=0$. 
This can be achieved by performing a Fourier transform in the holonomy associated to each gauge field, setting $\tilde q = 0$. Thus, the partition function for the theory of nearly zero-modes associated to each gauge field is given by
\be 
\label{eq:particle-on-U(1)-partition-function}
Z^{\text{Particle on }U(1)}(\beta, \tilde q=0) &= \int_0^{2\pi} d\alpha_{U(1)} \,Z(\beta, \alpha_{U(1)})\nn \\  &= \int_0^{2\pi} d\alpha_{U(1)}\int_{\frac{\text{Loop}(U(1))}{U(1)}} D\Lambda\, \,e^{-I_{\text{Particle on }U(1)}[\Lambda, \alpha_{U(1)}]}\,,
\ee
Since this is just the theory of a particle moving on a circle, \eqref{eq:particle-on-U(1)-partition-function} can be computed easily.  By taking the zero-temperature limit of the 
resulting expression we obtain the following regularized volume of the space of gauge zero modes
\be 
\Vol\left[{\text{Loop}(U(1))}\right] = \lim_{T\to 0} \left[Z^{\text{Particle on }U(1)}(\beta, \tilde q=0)\right]=1\,.
\ee

\ndt {\bf The volume of the space of disk zero modes}

We now consider the path integral over the large super-diffeomorphisms on the disk, which can be understood as fluctuations of the boundary curve inside rigid~AdS$_2$ or as specific metric fluctuations with a fixed boundary. These metric fluctuations are parametrized by a
reparametrization mode~$f(\tau)$, where we label the boundary time by~$\tau$.
Similarly metric fluctuations around $S^2$ by $g(\tau) \in SU(2)$ and fluctuations of the gravitino by $\eta(\tau)$ and $\bar \eta(\tau)$. Together, $(f, g, \eta, \bar \eta)$ describe the elements of $\text{Diff}(S^{1|4})$ that we have to integrate in the first path
integral in \eqref{eq:volumes-that-we-want-to-compute}. At small but non-zero temperature all these modes gain a non-zero mass just like in the case of large gauge transformations discussed above. In \cite{Iliesiu:companionPaper}, we show that the action associated to these modes is given by that 
of the $\cN=4$ super-Schwarzian theory,\footnote{
The appearance of the $\cN=4$ super-Schwarzian should not come as a surprise. The same effective theory was shown to arise when dimensionally reducing the 4D supergravity on $S^2$ in the near-horizon region of near-BPS black holes to a 2D theory of $\cN=4$ super-JT gravity \cite{Heydeman:2020hhw}. } 
\beq
\label{eq:N=4-Schwarzian-action}
I_{\cN=4 \text{ Schw.}} =- \frac{T}{\ESchw} \int_0^{2\pi} d\tau \left[\Sch(f,\tau) + \Tr(g^{-1} \partial_\tau g)^2 + ({\rm fermions})\right]\,.
\eeq
where $\ESchw \sim \frac{1}{\ell_\text{Pl}\Delta^{3/2}}$.
The ratio $E_\text{gap}/T$ is identified as the coupling of the theory. At low temperatures we indeed find that the effective theory is strongly coupled. The nomenclature ``$E_\text{gap}$''
arises from the fact that this energy scale was shown to give the gap scale between BPS states and the lightest non-BPS states in the super-Schwarzian theory~\cite{Heydeman:2020hhw}. 
Nevertheless, just as in the effective theory for large gauge transformations discussed above, the exact value of $E_\text{gap}$ is unimportant if we are solely interested in the degeneracy at extremality. Thus, to evaluate the first volume in \eqref{eq:volumes-that-we-want-to-compute} we will take 
\be 
\label{eq:super-Schwarzian-path-integral}
\Vol\left[\frac{\text{Diff}(S^{1|4})}{H_\text{Disk}}\right] = \lim_{T\to 0} \int_{\frac{\text{Diff}(S^{1|4})}{PSU(1,1|2)}} Df Dg D\eta D\bar \eta ~\rme^{-I_{\cN=4 \text{ super-Schw.}}}\,,
\ee

To compute \eqref{eq:super-Schwarzian-path-integral} we however need to specify the boundary conditions for the fields $f(\tau)$, $g(\tau)$ and $\eta(\tau)$. These are given by
\be
\label{eq:bdy-cond-disk}
f(\tau+2\pi)\=  f(\tau)\,, \qquad  g(\tau+2\pi) \= \rme^{ \ii 2\pi \alpha \sigma_3 } \,g(\tau)\,, \qquad 
\eta(\tau+2\pi) \= -\rme^{ \ii {2\pi \alpha\sigma_3}}\eta(\tau)\,,
\ee
with
\be 
\alpha\=\frac{{\beta \,\Omega_E}}{4\pi}\,,
\ee
where~$\Omega_E$ is the angular velocity on $S^2$.

The first boundary condition in \eqref{eq:bdy-cond-disk} imposes that the interior of the AdS$_2$ region is smooth. The second fixes that the black hole horizon rotates by an angle $ \beta \Omega_E$ when going around the boundary of AdS$_2$,\footnote{The rotation is around an axis whose rotation generator we identify as $\sigma_3$.}. The third boundary condition, for the fermionic modes, comes from identifying this angle of rotation as a chemical potential for the rotational $SU(2)$ symmetry under which the fermionic fields are charged. Since the gravitino and metric are related by $\qeq$ the boundary conditions should preserve supersymmetry which corresponds to computing an index. This corresponds to setting 
$\alpha =1/2$, in which both the fermionic boundary mode $\eta(u)$ in   \eqref{eq:bdy-cond-disk} is periodic. Nevertheless, in \cite{Iliesiu:2021are} it was shown that even with these boundary conditions, bulk fermions (such as the gravitino itself) are still anti-periodic around all
contractible bulk cycles which is necessary in order for the bulk to have a smooth spin structure. Later, in Section~\ref{sec:index-vs-partition-function}, we will use the fact the path integral of these zero-modes is computable for arbitrary $\alpha$ to  show that, in fact, the degeneracy and index of such black holes agree for all geometries that are locally  given by AdS$_2\times S^2$.

Next, we want to find the saddle-points of \eqref{eq:super-Schwarzian-path-integral}, to start expanding the fields $f$, $g$ $\eta$, and $\bar \eta$ around these saddles. Taking \eqref{eq:bdy-cond-disk} as our boundary conditions, we find an infinite number of solutions which we label by $n \in \mZ$,
\be 
\label{eq:solution-Schw-on-disk}
f(\tau) = \tan\left(\frac{\tau}{2} \right)\,, \qquad g(\tau) = \rme^{i \tau \left[n +\alpha\right] \sigma_3}\,, \qquad \eta(\tau) = \bar \eta(\tau) = 0\,.
\ee
Because the space of field configurations $\text{Diff}(S^{1|4})/H_\text{Disk}$ is symplectic, it turns out that fluctuations in the path integral around each saddle in \eqref{eq:solution-Schw-on-disk} are one-loop exact. This yields a partition function \cite{Heydeman:2020hhw}
\be 
\label{eq:N-4-Schw-disk}
Z^{\cN=4\text{ Schw.}}_\text{Disk}\left(\beta, \alpha\right) = \sum_{n \in \mathbb Z} \frac{\ESchw}{T} \,\frac{2\cot(\pi \a) (\a+n)}{\pi^3(1- 4(n+\a)^2)^2} \rme^{2\pi^2 \frac{T}{\ESchw} \left(1 -4  (n+\a) ^2 \right)}.
\ee
In the limit $\alpha \to 1/2$, all terms vanish with the exception of the $n=0$ and $n=-1$ saddles. In this case the sum of the two saddles gives
\be 
Z^{\cN=4\text{ Schw.}}_\text{Disk}\Big(\beta, \alpha = \frac{1}2\Big) = 1\,,
\ee
which is completely independent of~$\beta$,
as we would expect from the index. 
Since we are computing an index in a supersymmetric theory, this temperature independence should not come as a surprise. Thus, we find that the first volume in \eqref{eq:volumes-that-we-want-to-compute} gets regularized to 
\be 
\Vol\left[\frac{\text{Diff}(S^{1|4})}{H_\text{Disk}}\right]= \lim_{T\to 0} \left[Z^{\cN=4\text{ Schw.}}_\text{Disk}\left(\beta, \alpha = \frac{1}2\right)\right] = 1\,.
\ee

\ndt {\bf The volume of the space of orbifold zero modes}

We now consider the path integral over the zero modes on the orbifolded geometries. As explained above, on the orbifold we need to make the following identification within the coordinate system~\eqref{metriccomplex}, 
\beq
\label{eq:more-general-identification-main}
(\tau,\,\psi) \sim \Big(\tau + \frac{2\pi}{c},\,\psi+\frac{2\pi}c \Big)\sim \Big(\tau,\,\psi+2\pi\Big)\,.
\eeq
The identification \eqref{eq:more-general-identification-main} now dictates different boundary conditions for the nearly zero-modes in \eqref{eq:N=4-Schwarzian-action} than those considered on the disk. We now have:
\be
\label{eq:bdy-cond-orbifold}
f(\tau+2\pi)= \frac{ f(\tau) + \tan(\pi/c)}{1 - f(\tau) \tan(\pi/c)}\,, \qquad  g(\tau+2\pi) = \rme^{2\pi i \alpha \sigma_3} \,g(\tau)\, \rme^{2\pi i \frac{2\pi}c \sigma_3}\,, \qquad \eta(\tau+2\pi) = -e^{2\pi i \alpha \sigma_3}\eta(\tau)\,,
\ee
where $\alpha$ once again corresponds to the chemical potential associated to black hole rotation. The index is once again computed by taking the limit $\alpha \to 1/2$. The effective theory arising by evaluating the gravitational action at finite temperature remains the $\cN=4$ super-Schwarzian \eqref{eq:N=4-Schwarzian-action}, unchanged from the case of the smooth geometry. Just like on the smooth geometry, there are multiple saddles satisfying the equations of motion of the $\cN=4$ super-Schwarzian and satisfying the boundary conditions \eqref{eq:bdy-cond-orbifold}. These saddles, that we label by $n$, come in pairs and are given by
\be 
\label{eq:solution-Schw-on-orbifold}
f(\tau) = \tan\left(\frac{\tau}{2c} \right)\,, \qquad g(\tau) = \rme^{ i \tau \left[n +\alpha \pm \frac{1}{2} \left(1 -  \frac{1}{c} \right)\right] \sigma_3}\,, \qquad \eta(\tau) = \bar \eta(\tau) = 0\,,
\ee

Using the fact that the path integral of the $\cN=4$ super-Schwarzian is one-loop exact, we can use \eqref{eq:solution-Schw-on-orbifold} to find that the partition function with a supersymmetric defect is given by
\beq
\label{eq:super-defect-part-function}
Z_{\,(c)}^{\cN=4\text{ Schw.}}(\beta,\alpha)=  \sum_{n\in\mathbb{Z}} \frac{\ESchw}{T} \frac{\cot(\pi \alpha)}{4\pi^3} \left( \frac{e^{2\pi^2 \frac{T}{\ESchw}\left[\frac{1}{c^2}-4(n+\frac{1}2 -\alpha-\frac{1}{2c})^2\right]}}{(1-2\alpha+2n)^2} -\frac{e^{2\pi^2 \frac{T}{\ESchw}\left[\frac{1}{c^2}-4(n+\frac{1}2 +\alpha-\frac{1}{2c})^2\right]}}{(1+2\alpha+2n)^2}\right) \, .
\eeq
When the partition function is computed for $\alpha \to 1/2$, the partition function in \eqref{eq:super-defect-part-function} simplifies drastically since only the $n=0$ and $n=-1$ terms survive, giving
\be 
\label{eq:super-defect-part-function-final}
Z_{\,(c)}^{\cN=4\text{ Schw.}}\left(\beta, \alpha=\frac{1}2\right)= \frac{1}c
\ee
at any temperature.  Just like in the case of the disk, in Section~\ref{sec:index-vs-partition-function} we will confirm that the entire contribution to \eqref{eq:super-defect-part-function-final} comes from states with zero angular momentum, indicating that for such geometries the degeneracy and index agree.  In summary, the integral over the zero-modes in a geometry with a supersymmetric deficit at the origin is
\beq
\label{eq:nearly-zero-modes-with-a-defect-final}
\Vol\left[\frac{\text{Diff}(S^{1|4})}{H_\text{Orb}}\right]  = \lim_{T \to 0} \left[ Z_{\,(c)}^{\cN=4\text{ Schw.}}\left(\beta, \alpha=\frac{1}2\right) \right]= \frac{1}c\,.
\eeq
All of the above results confirm that all volumes are non-zero, which confirms that, in contrast to non-supersymmetric black holes, $\frac{1}{8}$-BPS black holes have a large degeneracy.

\section{Putting it all together: $\frac{1}{8}$-BPS black hole index}
\label{sec:putting-it-all-together}

Having analyzed the localization manifold, its action, and the one-loop determinants around them including a careful treatment of boundary modes, we can put everything together and reproduce the microscopic index of the~$\frac18$-BPS black hole from supergravity.

There are four improvements over the previous results of~\cite{Dabholkar:2011ec} and its follow-ups.
Firstly, we work with the full $\mathcal{N}=8$ supergravity keeping all vector multiplets, instead of using the truncation considered in \cite{Dabholkar:2011ec}, 
and we work with arbitrary R-R charges $(q_\Lambda,p^\Lambda)$ in the type IIA formulation.
Note that the evaluation of the one-loop determinant even for $c=1$ in \cite{Dabholkar:2011ec} is incomplete and,
as explained in~\cite{Murthy:2015yfa}, a more careful treatment within this truncation leads to an answer inconsistent with the large-charge limit.
It is therefore crucial to consider the full theory. 
Secondly, we work with arbitrary R-R charges $(q_\Lambda,p^\Lambda)$ in the type IIA formulation and obtain a final black hole index, including all charge-dependent prefactors. The resulting index is invariant under the U-duality subgroup that shift only the R-R charges and thus provide a non-trivial consistency check on the measure over the localization manifold proposed in \eqref{eq:measurelocN8}.  Thirdly, we perform the exact path integral over the moduli space of large super-diffeomorphisms.
Finally, we perform the calculation for the orbifolds~$c\neq 1$ 
and verify the one-loop determinant and the measure give the right $c$ dependence to reproduce the microscopic Rademacher expansion of the index.

Now, we move on to the calculation.
Firstly, we compute the action on the localization manifold in terms of the prepotential given in \eqref{Fcub} as explained 
in Section~\ref{sec:locformalism}. The result can be written in terms of the intersection matrix and charges as 
\begin{eqnarray}
I[\phi]&=& \frac{\pi  \, q_\Lambda  \, \phi^\Lambda }{c}
 - \frac{4 \pi}{c} \, \Im{F \Big(\frac{\phi^\Lambda+\ii p^\Lambda}{2} \Big)}\\
&=& \frac{\pi C_{IJK}p^Ip^Jp^K}{c\phi^0} - \frac{3\pi C_{IJK} \phi^I \phi^J p^K}{c \phi^0} + \frac{\pi q_0 \phi^0}{c} + \frac{\pi q_I \phi^I}{c}.
\end{eqnarray}
The fact that the action is quadratic in $\phi^I$ will be crucial later.

The second step is to compute the one-loop determinant, combining the contribution from bulk and boundary modes found in Section~\ref{sec:one-loop-det-results} and regulating the boundary zero-modes as explained in Section~\ref{sec:connection-to-N=4-super-Schw}. First of all, the index computed in Section~\ref{sec:summary-results-bulk-one-loop-det} gives for $\mathcal{N}=8$ supergravity the contribution $ Z_{\text{index}, \,c}^{\qeq \CV}=e^{5 \mathcal{K}}$. 
The full non-zero mode determinant $Z_{\text{non-zero}, \,c}^{\qeq \CV}$ (which combines the contribution of the index with that of the boundary non-zero modes as in \eqref{eq:Z-bulk-final-heat-kernel-formula})  as well as the contribution of the zero modes $Z_{\text{zero-mode}, \,c}^{\qeq \CV}$ in~\eqref{eq:separation-one-loop} 
are given by 
\begin{eqnarray}
Z_{\text{non-zero}, \,c}^{\qeq \CV} \= \begin{cases} \;
\rme^{5 \mathcal{K}} \, \rme^{-\frac{17}2 \mathcal{K}}\,, & \quad c\=1 \,,\\
\; \rme^{5 \mathcal{K}} \, \rme^{-\frac{15}2\mathcal{K}} \,, & \quad c\;>\;1 \,,
\end{cases}
,~~~Z_{\text{zero-mode}, \,c}^{\qeq \CV} \=  \begin{cases} \;
 \, \rme^{\frac{15}2 \mathcal{K}}\,, & \quad c\=1 \,,\\
\;  \, \frac{1}{c}\rme^{\frac{13}2\mathcal{K}} \,, & \quad c\;>\;1 \,.
\end{cases}
\end{eqnarray}
The full one-loop determinant is the product of these two contributions. The final answer simplifies and can be written compactly for any $c$:
\begin{eqnarray}
Z^{\qeq\CV}_{\text{1-loop},\,c}(\phi) &=&    \,Z_{\text{non-zero}, \,c}^{\qeq \CV}(\phi)  \,Z_{\text{zero-modes}, \,c}^{\qeq \CV}(\phi) =\frac{e^{4 \mathcal{K}}}{c} ,\\
&=& \frac{1}{c} \left( \frac{\phi^0}{-2C_{IJK} p^I p^J p^K}\right)^{4}.
\end{eqnarray}
Upon putting together the action on the localization manifold, the one-loop determinant, and the measure 
in~\eqref{eq:measurelocN8}, we obtain the full contribution of the~$\IZ_c$ orbifolds to the index,\footnote{Note that the measure and one-loop determinants in $W^{(c)}(q,p)$ are independent of $\phi^I$ which will allow us to easily perform the integral over these fields. }
\begin{eqnarray}
W^{(c)}(q,p) &=& \sqrt{c} K_c(\Delta) \int \dd^{16} \phi ~\sqrt{\frac{ -2C_{IJK}p^Ip^Jp^K ~{\rm det} \left(3C_{IJ} \right)}{c^{16}(\phi^0)^{18}}}\frac{1}{c} \left( \frac{\phi^0}{-2C_{IJK} p^I p^J p^K}\right)^{4}\nonumber\\
&& \times\exp{\left(-\frac{\pi C_{IJK}p^Ip^Jp^K}{c\phi^0} + \frac{3\pi C_{IJK} \phi^I \phi^J p^K}{c \phi^0} - \frac{\pi q_0 \phi^0}{c} - \frac{\pi q_I \phi^I}{c} \right)}\,.
\label{eq:Wc-integral-phiI}
\end{eqnarray}
Here we have pulled out the Chern-Simons partition function~$\sqrt{c}K_c(\Delta)$ from the integral,
as is independent of $\phi^\Lambda$.
We now integrate out the scalars $\phi^I$,  $I=1,\,\dots,\,15$ 
which have a Gaussian action in \eqref{eq:Wc-integral-phiI}:
\begin{eqnarray}
\int d^{15} \phi~\exp \left(\frac{3\pi C_{IJK} \phi^I \phi^J p^K}{c\phi^0}  - \frac{\pi q_I \phi^I}{c}\right) \=\sqrt{-\frac{c^{15}(\phi^0)^{15}}{ {\rm det}(3C_{IJ})}} \exp \left(-\frac{\pi}{12 c} \phi^0 q_I C^{IJ} q_J \right),
\end{eqnarray}
The eigenvalues of $C_{IJ}$ are numbers fixed in terms of the charges which can be positive of negative. We will see below the need to pick the contour for the $\phi^0$-integral to be along the imaginary axis, both to make the integral convergent and to match with the microscopic result. Therefore the Gaussian integrals are oscillatory and not divergent. 

Using this result we obtain the black hole index as a single integral over $\phi^0$:
\be 
\label{eq:Wc-integral-phi0}
W^{(c)}(q, p) \= K_c(\Delta)  \int d\phi^0 ~\frac{1}{c} \frac{(\phi^0)^{5/2}}{\ii(-2C_{IJK}p^Ip^Jp^K)^{7/2}}  \exp \left(- \frac{\pi C_{IJK}p^Ip^Jp^K}{c\phi^0}- \frac{\pi \phi^0}{ c}(q^0+\frac{1}{12}  q_I C^{IJ} q_J) \right),
\ee
where we have identified $\Delta$ as in \eqref{eq:DeltaRRcharges} $\Delta = 4C_{IJK}p^Ip^Jp^K \big(q^0+\frac{1}{12}  q_I C^{IJ} q_J\big)$. Performing the following change of variables $\sigma = \frac{-\pi C_{IJK}p^Ip^Jp^K}{ c \phi^0}$, we find that the integral \eqref{eq:Wc-integral-phi0} can be finally rewritten in the same way it appears on the Rademacher expansion
\be 
W^{(c)}(q, p) \= K_c(\Delta)~ \frac{2\pi}{c^{9/2}}\left(\frac{\pi}{2}\right)^{7/2}~\frac{1}{2\pi \ii}\int \frac{d\sigma}{\sigma^{9/2}} \exp \left(\sigma + \frac{\pi^2 \Delta}{4c^2 \sigma} \right) \,.
\ee
If we take a contour along the imaginary axis with ${\rm Re}(\sigma)>0$ then we reproduce the integral representation of the Bessel function $\wt{I}_{7/2}(z)$ appearing in the microscopic formula \eqref{RadexpC}. As explain in \cite{Dabholkar:2011ec} the issues related to the precise integration contour of $\phi^\Lambda$ can be traced to the indefiniteness of the Euclidean action in gravity. 

We see at this point a benefit of having kept all charges general. Since the prefactors depend on charges it is a non-trivial result that the measure proposed in \eqref{eq:measurelocN8} is such that all factors that would spoil the invariance under the subgroup that acts on the 15 of the electric and magnetic RR-charges cancel, resulting in a function only of $\Delta$.  

In total we thus find from the localization calculation in supergravity that the index of the $\frac{1}{8}$-BPS  black hole is given by 
\be 
\label{eq:partition-function-conclusion}
 W(q,p) = \sum_c W^{(c)}(q,p)=\sum_{c=1}^\infty \frac{2\pi}{c^{9/2}}\left(\frac{\pi}{2}\right)^{7/2}~ K_c(\Delta) \tilde I_{7/2}\Big(\frac{\pi \sqrt{\Delta}}{c}\Big) = W_\text{micro}(\Delta)\,,
\ee
and it precisely matches the  Radamacher expansion of the microscopic index obtained in \eqref{RadexpC}.

\section{The index vs.~partition function: \\ why only orbifolds contribute to both}
\label{sec:index-vs-partition-function}

So far we have shown that the index obtained from the gravitational integral using localization precisely matches the index obtained from the microstate count of state in the D1/D5 construction. We would now like to argue that the supergravity path integral implies that this index exactly agrees with the actual degeneracy of $\frac{1}{8}$-BPS black holes, 
i.e.~all extremal black hole states are in fact bosonic. 
This statement applies to all configurations that are asymptotically $AdS_2 \times S^2$,
and not necessarily to degrees of freedom that live outside of this region, since those are out of our computational control. In particular, a subtlety we will not address is how the microscopic degeneracy of the D-brane system, whose degrees of freedom are not clearly located in the $AdS_2 \times S^2$ region, is related to the black hole degrees of freedom. 

These issues were addressed in \cite{Dabholkar:2010rm}. The new ingredients in this section are to carefully incorporate the large super-diffeomorphisms into this analysis and show that no other geometry contributes to the degeneracy while having vanishing index. This leads to a refined version of the previous arguments made in the literature.

We first show the contribution of the gravitational path integrals in the grand canonical ensemble when fixing the angular velocity to that corresponding to the index (with $\Omega_E \beta = 2\pi$)  and in the canonical ensemble with fixed angular momentum (with $J=0$), agree at zero temperature
\be 
\label{eq:what-we-want-to-show-index=degeneracy}
Z_{\frac{1}{8}\text{-BPS BH},\, \Sigma}^{T = 0}\left(\Delta, \Omega_E=\frac{2\pi}{\beta} \right) =  Z_{\frac{1}{8}\text{-BPS BH},\,\Sigma}^{T = 0}\left(\Delta, J = 0 \right)  \,,
\ee
when expanding about a large class of geometries, with each geometry $\Sigma$ being locally given by AdS$_2 \times S^2$. \eqref{eq:what-we-want-to-show-index=degeneracy} would not only indicate that the index is equal to the partition function but also that all $\frac{1}{8}$-BPS black hole states carry only zero spin. We will start by first showing that \eqref{eq:what-we-want-to-show-index=degeneracy} holds for manifolds $\Sigma$ that preserve supersymmetry and that we have considered in our localization computation---the smooth AdS$_2$ geometry and its orbifolds. We will then show that for geometries $\Sigma$ that do not admit the existence of a globally defined killing spinor, such as manifolds of higher genus or with multiple defects, both terms in \eqref{eq:what-we-want-to-show-index=degeneracy} vanish due to path integral over large super-diffeomorphisms.

If studying the partition function instead of the index the only difference are the boundary conditions that we impose at the edge of AdS$_2 \times S^2$. In a co-rotating frame the circle at the boundary of AdS$_2$ is contractible, and therefore all fermionic fields are anti-periodic when going around it for the computation of the index as well as the partition function. 
The only difference in boundary conditions in such a case is for the large super-diffeomorphism modes which fix the angular velocity of the black hole. As we've seen above when computing the index we fixed $\Omega_E = 2\pi/\beta$ while to compute the partition function we need $\Omega_E = 0$. To extract more information about what angular momenta contribute to the degeneracy (whether states with $J \neq 0$ contribute), we should further keep $\Omega_E$ arbitrary.

At zero-temperature, both the smooth AdS$_2\times S^2$ geometry and the orbifolds considered above have globally well-defined killing spinors which we can use to define the supercharge $\qeq$ with respect to which we localize. Following the procedure discussed in Sections~\ref{sec:BPS-BHs-in-N=8} 
and~\ref{sec:one-loop-det-theoretical-prelims} and using the fact that the boundary conditions for the bulk modes remain unchanged, the path integral over such modes also yields the same result. 
Thus, what remains to be done in order to confirm \eqref{eq:what-we-want-to-show-index=degeneracy} is to perform the path integral over large   super-diffeomorphisms with 
arbitrary~$\Omega_E$. In the case of smooth AdS$_2\times S^2$ geometry, taking the $T\to 0$ limit of \eqref{eq:N-4-Schw-disk} we find (with~$\alpha=\frac{\Omega_E \beta}{4\pi}$), 
\be 
\lim_{T\to 0} Z^{\cN=4\text{ Schw.}}_\text{Disk}\left(\beta, \alpha\right) = 1\,, \qquad \forall \,\,\,\,\alpha\,.
\ee
Similarly, for the orbifold geometries, taking the $T\to 0$ limit of \eqref{eq:super-defect-part-function} we find that 
\be 
\lim_{T\to 0} Z^{\cN=4\text{ Schw.}}_\text{Orb}\left(\beta, \alpha\right) = \frac{1}c\,, \qquad \forall  \,\,\,\, \alpha\,.
\ee
From doing the Fourier transform in $\alpha$ for arbitrary $J$, this implies that in the canonical ensemble for angular momenta, for the two types of geometries that contribute to the index
\be 
\lim_{T\to 0} Z^{\cN=4\text{ Schw.}}_\text{Disk}\left(\beta, J\right) = \delta_{J,0}\,, \qquad \lim_{T\to 0} Z^{\cN=4\text{ Schw.}}_\text{Orb}\left(\beta, J\right) = \frac{\delta_{J,0}}c\,. 
\ee
Thus, because we are able to localize the bulk modes and because the integral over the boundary modes remains unchanged, we conclude that for all the geometries that preserve supersymmetry
\be 
Z_{\frac{1}{8}\text{-BPS BH}, \text{ disk/orb}}^{T = 0}\left(\Delta, \Omega_E=\frac{2\pi}{\beta} \right) = Z_{\frac{1}{8}\text{-BPS BH}, \text{disk/orb}}^{T = 0}\left(\Delta, J = 0 \right) \,.
\ee

We now consider the path integral over the zero modes on geometries that do not preserve supersymmetry. We will start with the case of orbifolds that do not preserve supersymmetry; for these, the angle of the defect is not identified with the angle of rotation around $S^2$ when going around the defect. Taking the coordinates of AdS$_2 \times S^2$ as in \eqref{metriccomplex}, we consider the more general identification
\beq
\label{eq:more-general-identification}
(\tau,\,\psi) \sim \Big(\tau + \frac{2\pi}{c},\,\psi-4\pi\varphi\Big)\sim \Big(\tau,\,\psi+2\pi\Big)\,,
\eeq
where  $\varphi$ parametrizes the angle of rotation of $S^2$ as one goes around the defect, and, unlike the supersymmetric case, we will take $2\varphi \neq 1 - 1/c$. The identification \eqref{eq:more-general-identification} now dictate new boundary conditions for the nearly zero-modes in \eqref{eq:N=4-Schwarzian-action}, with $g(\tau+\beta) = \rme^{ i 2\pi \alpha\sigma_3} \,g(\tau)\, \rme^{-i {2\pi \varphi} \sigma_3}$.

For generic values of $\varphi$, the partition function in the grand canonical ensemble is \cite{Iliesiu:2021are}
\beq
\label{eq:generic-defect-part-function}
Z^{\cN=4\text{ Schw.}}_{(c,\varphi)}(\beta ,\, \alpha) = \sum_{n\in \mathbb{Z}} \frac{T}{\ESchw} \frac{ \cot(\pi \alpha)}{2\pi}  \left( \rme^{2\pi^2 \frac{T}{\ESchw}\left(\frac{1}{c^2}-4\left(n-\alpha+\varphi \right)^2\right)}-e^{2\pi^2 \frac{T}{\ESchw}\left(\frac{1}{c^2}-4\left(n+\alpha+\varphi\right)^2\right)}\right) \,.
\eeq
Contributions of these defects to both the index and the degeneracy vanish. A straightforward way to observe it is that now the overal factor of $T$ is in the numerator and therefore $Z_{c,\varphi}(\beta\to\infty,\alpha) \to 0$ for all $\alpha$. We thus conclude that orbifolds that do not preserve supersymmetry 
\be 
\label{eq:arbitrary-orbifold-part-function}
\lim_{T\to 0}Z_{\frac{1}{8}\text{-BPS BH}, {(b, \varphi)}}^{T = 0}\left(\Delta, J  \right)  = 0\,, 
\ee
for all angular momenta $J$. 

Next we consider higher genus geometries that could also possibly have defects on them. If we want to be able to smooth out the geometry by considering its fibration over the additional~$S^1$ that we used to make the supersymmetric orbifolds smooth in 5D, then the defect angle of all such defects 
needs to be quantized with $2\pi \varphi_i = 2\pi(1-1/c_i)$ with $c_i\in \mathbb Z$, for each defect $i \in \{1,\dots,m\}$.
All such geometries have a closed geodesic separating the asymptotic boundary from the higher genus multi-defect component.\footnote{This is the case as long as $g\geq 1$ and $m\geq 0$ or when $g\geq 0$ and $m\geq 2$. We have already considered the case $g=0$ and $m=1$ in \eqref{eq:arbitrary-orbifold-part-function}.} This changes the boundary conditions for the large super-diffeomorphism modes to 
\be
\label{eq:bdy-cond-trumpet}
f(\tau+2\pi)= \rme^{\frac{b}{2}}f(\tau) \,, \qquad  g(\tau+2\pi) = \rme^{ 2\pi i \alpha \sigma_3 }\,g(\tau)\, \rme^{ 2\pi i \varphi \sigma_3}\,, \qquad \eta(\tau+2\pi) = -e^{i 2\pi i \alpha \sigma_3}\eta(\tau)\,.
\ee
where $b$ parametrizes the  length of the closed geodesic that is homotopic to the asymptotic boundary. The space between the asymptotic boundary and this geodesic is usually referred to as a trumpet. Above,  $\varphi$ parametrizes the angle of rotation of $S^2$ as one goes around the closed geodesic of the trumpet.
The path integral over these modes can now be computed to be \cite{Iliesiu:2021are}
\be 
\label{eq:Trumpet-part-function-low-T}
\lim_{T \to 0}Z_{\text{Trumpet} \,(b, \varphi)}^{\cN=4\text{ Schw.}}\left(\beta, \alpha\right) = 0\,,\qquad \forall\,\alpha \,,
\ee
in the limit $T \to 0$, and  upon taking the Laplace transform of the  partition function for general $T$ (which is the same as \eqref{eq:generic-defect-part-function} upon replacing $1/c \to -ib$) with respect to $\beta$  we find a continuous density of states that starts at the energy $\ESchw$ above the extremal energy
$E_{\frac{1}{8}\text{-BPS BH}} \sim \Delta^{1/4}$,
\be 
\label{eq:density-of-states-cont-trumpet}
 \rho_{\text{trumpet},\,(b, \varphi)} \sim \Theta(E -E_{\frac{1}{8}\text{-BPS BH}} -\ESchw)\,.
\ee

We can now glue back the trumpet to a higher genus surface. To do this we in principle need to include the contribution of matter fields (which we denote by $ Z^\text{matter}_{(b, \varphi)}$) as well as the integral over the moduli space for a surface of genus $g$ with $m$ defects that is glued to the trumpet (which we we denote by $\Vol_{g,m}$). While in principle the contribution of matter fields is divergent for small values of the trumpet geodesic length \cite{Saad:2019lba} and the volume of the moduli space for such surfaces is unknown for the supersymmetry amount that we are interested in, \eqref{eq:density-of-states-cont-trumpet} implies  that in the strict $T\to 0$ limit of partition function we have\footnote{\eqref{eq:Trumpet-part-function-low-T} is insufficient to show that the partition function vanishes. For example, in the non-supersymmetric theory, the partition function on the trumpet still vanishes as $T\to 0$. Nevertheless, in the absence of matter, due to the fact that $\Vol_{g,m}(b)$ are polynomial in $b$, the partition function for surfaces whose genus is higher than $1$ is divergent as $T \to 0$. On the other hand, \eqref{eq:density-of-states-cont-trumpet} is a feature, albeit not obviously, of supersymmetry and is sufficient to show that the partition function is indeed vanishing at zero temperature. } 
\be 
\label{eq:overall-part-function-genus-g-with-m-defects}
\lim_{T \to 0} \int d\mu[b] \,d\mu[\varphi]\, Z_{\text{Trumpet} \,(b, \varphi)}^{\cN=4\text{ Schw.}}\left(\beta,\,J\right) Z^\text{matter}_{(b, \varphi)}\,\Vol_{g,m}= 0\,,\qquad 
\ee
for all angular momenta $J$.
Now that we have seen that the partition function of arbitrary non-supersymmetric orbifolds \eqref{eq:arbitrary-orbifold-part-function} and the partition function of surfaces with arbitrary genus and an arbitrary number of defects  \eqref{eq:overall-part-function-genus-g-with-m-defects} all vanish, we conclude that no other geometries that are locally AdS$_2\times S^2$ contribute to the degeneracy. For this reason, we find that the gravitational index computed in Section~\ref{sec:putting-it-all-together} (which we have denoted by $W(p, q)$) is the same as the actual degeneracy of a single black hole (which we denote by $d_{\frac{1}{8}\text{-BPS BH}}$):
\be 
\label{eq:degeneracy=index-final}
W(p, q) = Z_{\frac{1}{8}\text{-BPS BH}}^{T = 0}\left(\Delta ,\,\, \Omega_E \beta =2\pi \right) = Z_{\frac{1}{8}\text{-BPS BH}}^{T = 0}\left(\Delta , \,\, \Omega_E=0\right) = d_{\frac{1}{8}\text{-BPS BH}}\,.
\ee
Since $d_{\frac{1}{8}\text{-BPS BH}}$ is not a protected quantity on the string theory side, \eqref{eq:degeneracy=index-final} is not necessarily valid along the entire moduli space of the D1D5 CFT that is used to describe the black hole microstates. In fact, at the orbifold point it is known that the degeneracy of BPS states is much larger than the index, implying a large cancellation between bosonic and fermionic degeneracies. This degeneracy between bosonic and fermionic states is expected to be lifted when going away from the orbifold point in order for \eqref{eq:degeneracy=index-final} to be valid, 
and was explicitly verified in various  examples in~\cite{Dabholkar:2010rm}.
We discuss how this expectation is consistent with the path integral description of near-BPS states in more detail in the following discussion.

\section{Discussion and future directions}

\label{sec:discussion}

We have found that the the index of $\frac{1}{8}$-BPS states obtained in string theory is the same as the index obtained from localizing the gravitational path integral. On a technical level, this match is a culmination of two directions of research. 
The first is the 
development of localization in supergravity \cite{Dabholkar:2010rm, Dabholkar:2010uh, Dabholkar:2011ec, Gupta:2012cy, Dabholkar:2014ema, Gupta:2015gga, Murthy:2015yfa}, and in particular the formalism for BRST quantization of supergravity initiated in \cite{deWit:2018dix, Jeon:2018kec} which we extensively used to set up and calculate the functional integral.
The second direction of research which we have heavily relied upon has been that of the (super)Schwarzian \cite{Sachdev:2015efa, Almheiri:2016fws, Nayak:2018qej, Moitra:2018jqs, Castro:2018ffi, Larsen:2018cts,  Moitra:2019bub,Maldacena:2019cbz, Charles:2019tiu,Ghosh:2019rcj, Larsen:2020lhg,David:2020ems, Heydeman:2020hhw, Iliesiu:2020qvm, David:2020jhp, Castro:2021wzn, Boruch:2022tno}. 
Above, we have emphasized the role of this quantum mechanical system in the proper quantization of the boundary modes.
Without the exact contribution of both the bulk modes and of the $\cN=4$ super-Schwarzian around all the different geometries and field configurations that we localized to, it would have been impossible to obtain the exact match between the microscopic string theory computation and the macroscopic path integral calculation. 

Below we shall discuss how the above match, together with recent results in the literature \cite{Heydeman:2020hhw, Iliesiu:2021are}, offer a comprehensive view for the quantum mechanical properties of BPS
black holes and, to a lesser (but still meaningful) extent, near-BPS black holes. 
We also discuss a number of different future directions of research. 

\subsection{ BPS states vs.~non-BPS states}

In the introduction we motivated our work by asking whether the gravitational path integral is sufficiently powerful to reproduce the exact number of black hole microstates within an energy window. The results in Sections~\ref{sec:localization-in-supergravity} to \ref{sec:putting-it-all-together} 
show that by localizing gravitational path integral the difference in bosonic and fermionic degeneracies can be obtained exactly using localization. The fact that we have managed to also compute the degeneracy (which is not obviously protected 
by supersmmetry from the perspective of a putative boundary theory) is a non-trivial fact. Below, we shall explain how this calculation can be further extended to compute the total number of states within an energy window that starts at extremality.

As discussed in Section~\ref{sec:index-vs-partition-function}, what is perhaps more surprising is that the gravitational path integral allows us to probe observables (at strong string coupling, away from the orbifold point of the D1D5 system) that are unprotected, i.e.~the exact degeneracy of states at the BPS energy level in a sector of fixed charge and angular momentum. This is because from the perspective of the gravitational path integral, the same supersymmetry preserving geometries were shown to contribute to both the index and to the degeneracy, while geometries preserving no supersymmetries contributed to neither. The path integral over the nearly zero-modes of the super-Schwarzian reveals even more if studying the density of states of such a system rather than its partition function.  For the geometries that preserve some amount of supersymmetries the integral over the nearly zero modes gives, in addition to the degenerate number of states at the BPS energy, a continous density of states that starts at energies $E = E_{\frac{1}{8}\text{-BPS BH}} +\ESchw$~\cite{Heydeman:2020hhw, Iliesiu:2021are}:
\be 
\label{eq:density-of-states-cont-disk-and-orbifold}
\rho_{\text{Disk},\,(c, \varphi)}^{\text{cont.}} \,\sim \, \Theta(E -E_{\frac{1}{8}\text{-BPS BH}} -\ESchw)\,,  \qquad \rho_{\text{Orbifold},\,c}^{\text{cont.}} \, \sim \, \Theta(E -E_{\frac{1}{8}\text{-BPS BH}} -\ESchw)\,,
\ee
where for the black holes discussed above, we remind the reader that \cite{Heydeman:2020hhw} 
\be 
\ESchw \, \sim \, \frac{1}{\Delta^{3/2}} \,, \quad \Delta \to \infty  
\ee 
up to stringy corrections which are subleading in $1/\Delta$. Similarly, on geometries preserving no supersymmetries the integral over nearly zero modes yields \cite{Iliesiu:2021are} 
\be 
\label{eq:density-of-states-cont-trumpet-and-orbifold}
\rho_{\text{orbifold},\,(c, \varphi)} \, \sim \, \Theta(E -E_{\frac{1}{8}\text{-BPS BH}} -\ESchw)\,,  \qquad \rho_{\text{trumpet},\,(b, \varphi)} \, \sim \, \Theta(E -E_{\frac{1}{8}\text{-BPS BH}} -\ESchw)\,, 
\ee
In other words \eqref{eq:density-of-states-cont-disk-and-orbifold} and \eqref{eq:density-of-states-cont-trumpet-and-orbifold} indicate that the spectrum of near-BPS microstates only starts at $E_{\frac{1}{8}\text{-BPS BH}} + \ESchw \Bigl(1+\text{O}\bigl(1/\Delta\bigr) \Bigr)$, where one could imagine the correction to $\ESchw$ arises, via a mechanism similar to \cite{Maxfield:2020ale}, from summing geometries with arbitrary topologies obtained by gluing the answer in \eqref{eq:density-of-states-cont-trumpet-and-orbifold} as explained in Section~\ref{sec:index-vs-partition-function}. Going back to the question that we posed in the introduction, we can expect that not only is $d_{\frac{1}{8}\text{-BPS BH}}$ the degeneracy of BPS black holes, but additionally\footnote{Here, we assume that doubly non-perturbative effects in $\Delta$, which contribute beyond the topological expansion which we have considered when summing geometries that contain a ``trumpet'' geometry, might yield a non-zero number of states between $E_{\frac{1}{8}\text{-BPS BH}}$ and $E_{\frac{1}{8}\text{-BPS BH}} + \#\ESchw$. It would nevertheless be valuable to understand whether this is indeed the case.   } 
\be
\label{eq:total-number-of-states-energy-interval}
\int_{E_{\frac{1}{8}\text{-BPS BH}}}^{E_{\frac{1}{8}\text{-BPS BH}}+ \# E_\text{gap}} dE \rho(E) \= d_{\frac{1}{8}\text{-BPS BH}}\,.
\ee
for any number $\#$ that is smaller than $1$ by an amount that does not scale downwards with $1/\Delta$ as $\Delta$ is made large. As $\#$ becomes larger than 1 the path integral around any given saddle as well as the topological expansion becomes out of control and we find can only estimate the total number of states rather than recover an exact integer as in \eqref{eq:total-number-of-states-energy-interval}. Nevertheless, for $E> E_{\frac{1}{8}\text{-BPS BH}}+  E_\text{gap}$ the continuum density of states found in \cite{Heydeman:2020hhw} obtained from the expansion around the leading AdS$_2\times S^2$ offers a good course-grained approximation to the densely spaced discrete spectrum of black hole mirostates that we expect to exist for such energy values. Understanding how to resolve this discretum requires full control over both stringy corrections and over the sum of all geometries, a daunting task since the geometries in question  no longer preserve any supersymmetries.    

The contrast in the cases with $\#<1$ and $\#>1$ is also captured by looking at multi-boundary contributions to the gravitational path integral for the total number of states 
between $E_{\frac{1}{8}\text{-BPS BH}}$ and $E_{\frac{1}{8}\text{-BPS BH}}+\# E_\text{gap}$. There are no connected multi-boundary contributions to the number of states when $\#<1$ and thus the quantity \eqref{eq:total-number-of-states-energy-interval} does not suffer from the 
factorization puzzle that affects other observables---this is of course consistent with the fact 
that~$d_{\frac{1}{8}\text{-BPS BH}} \in \mZ$ which points towards a conventional Hilbert space interpretation. Counting states for~$\#>1$ \eqref{eq:density-of-states-cont-trumpet-and-orbifold} shows that wormhole contributions exist, 
pointing towards the fact that individual microstates cannot be distinguished but their chaotic energy statistics might be computable through the contribution of such connected geometries. 

Finally let us comment on the role that stringy corrections can play in the statements made above. In \cite{Boruch:2022tno}, it was shown that $E_\text{gap}$ receives stringy corrections where higher derivative terms are include in the supergravity action.\footnote{In \cite{Boruch:2022tno} the stringy corrections to $E_\text{gap}$ were determined 
in the context of near-BPS black holes in AdS$_5 \times S^5$ with a broken $SU(1,1|1)$ near-horizon isometry. While it would be informative to redo that analysis for the black holes discussed in this paper which have a $PSU(1,1|2)$ at extremality, 
we expect that the role of stringy correction on  $E_\text{gap}$ in the two cases to be qualitatively similar.} 
Stringy corrections are not of concern however for the gravitational index---we expect the value of the supergravity action along the localization locus to be independent of the higher-derivative correction. 
Furthermore, we saw that the integral over boundary zero-modes ended up being completely independent of $E_\text{gap}$ for all geometries. 
Therefore we only need to be careful about 
the upper bound of the integral in \eqref{eq:total-number-of-states-energy-interval} which does receive stringy corrections. 
At weak string coupling, we expect that $E_\text{gap}\to 0$ 
in order to find agreement with the D1-D5 system at the orbifold point for which there is a large difference between the degeneracy of states at the BPS energy and the index of such states. A related issue is that at weak string coupling we do not really know how to distill the would-be black hole degrees of freedom from the rest of the D-brane quantum mechanics.

\subsection{ BPS black holes have a Type I von Neumann algebra }

Understanding the algebra of observables in quantum gravity has been a problem of recent interest \cite{Leutheusser:2021qhd,Leutheusser:2021frk, Witten:2021unn, Chandrasekaran:2022cip}. Recent results show that semiclassical gravity can promote an algebra from type III to type II, and the results in this paper indicate that observables for BPS black holes in fact can be promoted to a type I algebra. While the fact that the gravitational path integral reproduces an exact integer as $T\to 0 $ meet seem sufficient to claim a type I algebra, it is useful to verify that this integer indeed has the interpretation of the dimension of the Hilbert space of BPS black holes.   To show this, we need to be able to be able to define density matrices in this space and show that the rank of all matrices is finite (and, importantly, integral). For this, we will check that the maximally mixed density matrix indeed has a size equal to the degeneracy $d_{\frac{1}{8}\text{-BPS BH}}$ discussed in the previous section. We construct the thermofield double state at zero temperature $ |{\text{TFD}(T = 0)}\>$ in a sector with fixed charges, define its associated pure density matrix, and the reduced density matrix on one of the sides (say, the left side),
\be 
\rho_\text{TFD} \= |{\text{TFD}(T = 0)}\>\<{\text{TFD}(T = 0)}|\,, \quad \rho_L \= \Tr_R (\rho_\text{TFD})\,.
\ee
The Renyi entropy $S_\alpha$ associated to $\rho_L$ is obtained by gluing $\alpha$ copies of a boundary segments, each with an infinite length to form a closed boundary that also has $\beta \to \infty$. Filling in the bulk with the geometries that contribute to the fixed charge partition function in the limit of $\beta \to \infty$,
\be
S_\alpha &= \frac{1}{1-\alpha} \frac{\Tr\, (\rho_L)^\alpha}{(\Tr\, \rho_L)^\alpha}\= \frac{1}{1-\alpha}\,\, \Biggl[{ \overbrace{\begin{tikzpicture}[baseline={([yshift=-.5ex]current bounding box.center)}, scale=0.45]
 \pgftext{\includegraphics[scale=0.35]{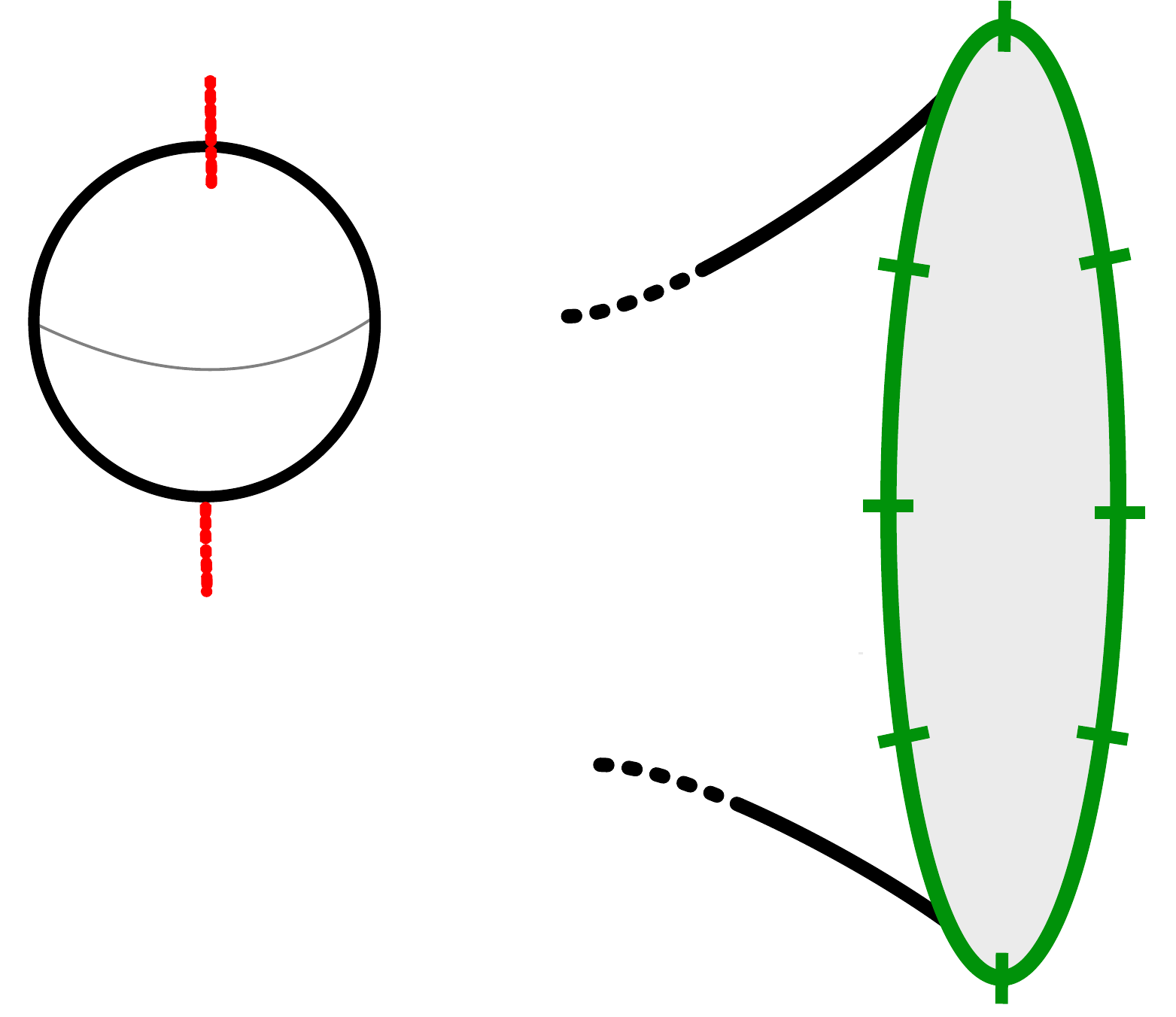}} at (0,0);
  \end{tikzpicture}}^{\alpha \text{ copies of segments of proper length } \beta}}\bigg{/}\,\,\,\,{
  \left(\begin{tikzpicture}[baseline={([yshift=-.5ex]current bounding box.center)}, scale=0.45]
 \pgftext{\includegraphics[scale=0.35]{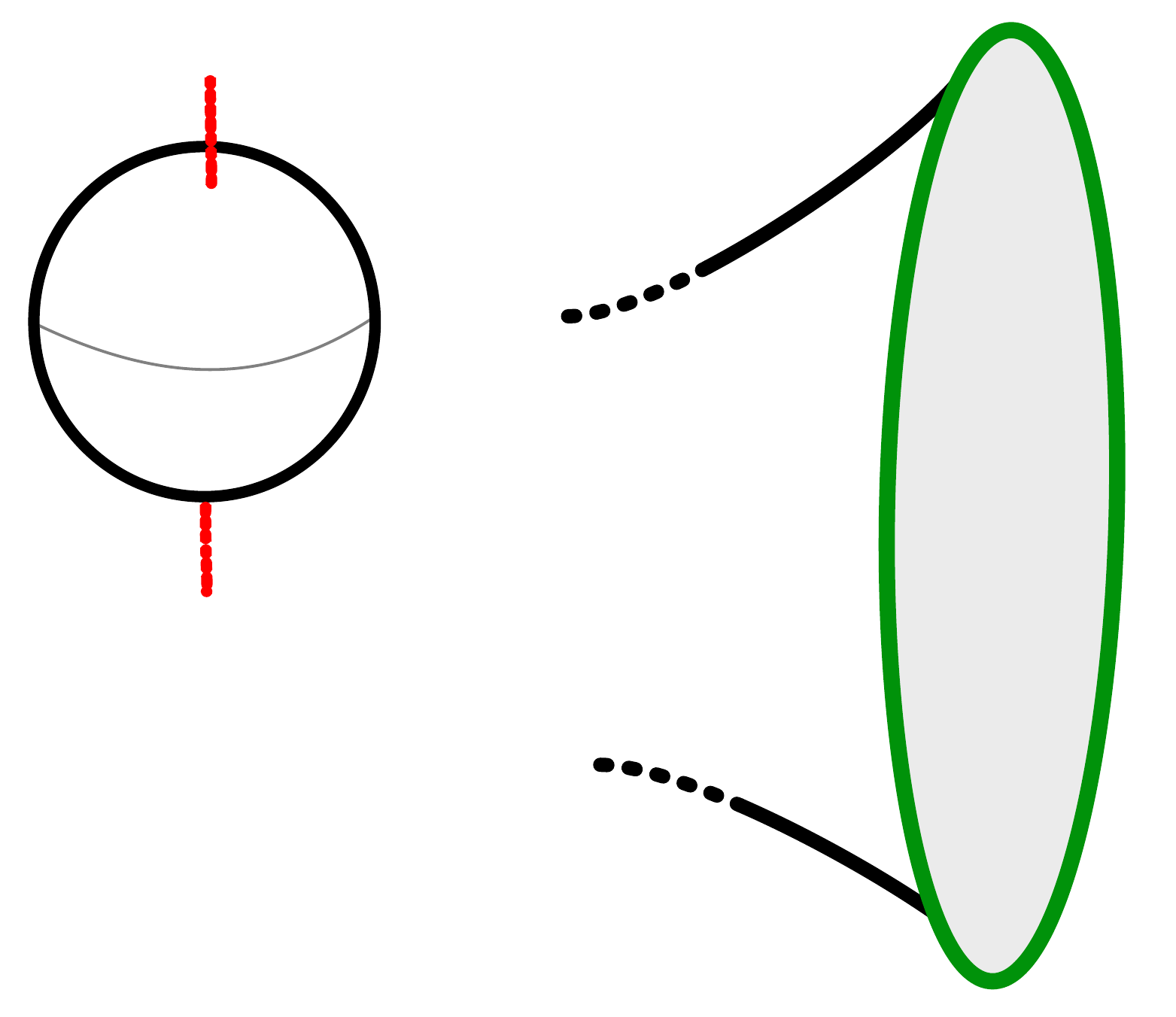}} at (0,0);
  \end{tikzpicture} \right)^\alpha
  }\Biggr] \nn \\ &\= \frac{\left(\begin{tikzpicture}[baseline={([yshift=-.5ex]current bounding box.center)}, scale=0.45]
 \pgftext{\includegraphics[scale=0.35]{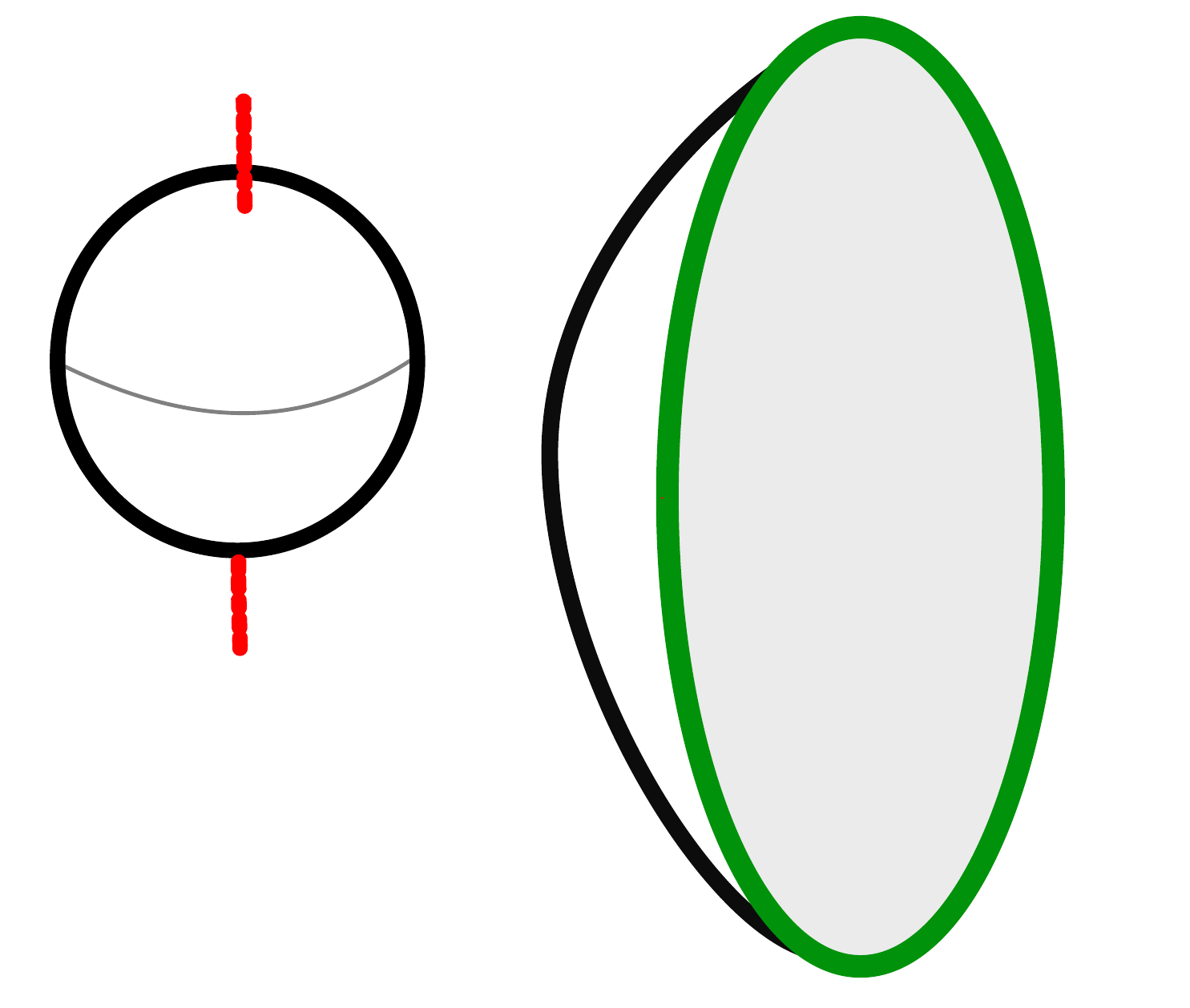}} at (0,0);
  \end{tikzpicture} \,\,\,+\,\,\, \begin{tikzpicture}[baseline={([yshift=1.0ex]current bounding box.center)}, scale=0.45]
 \pgftext{\includegraphics[scale=0.35]{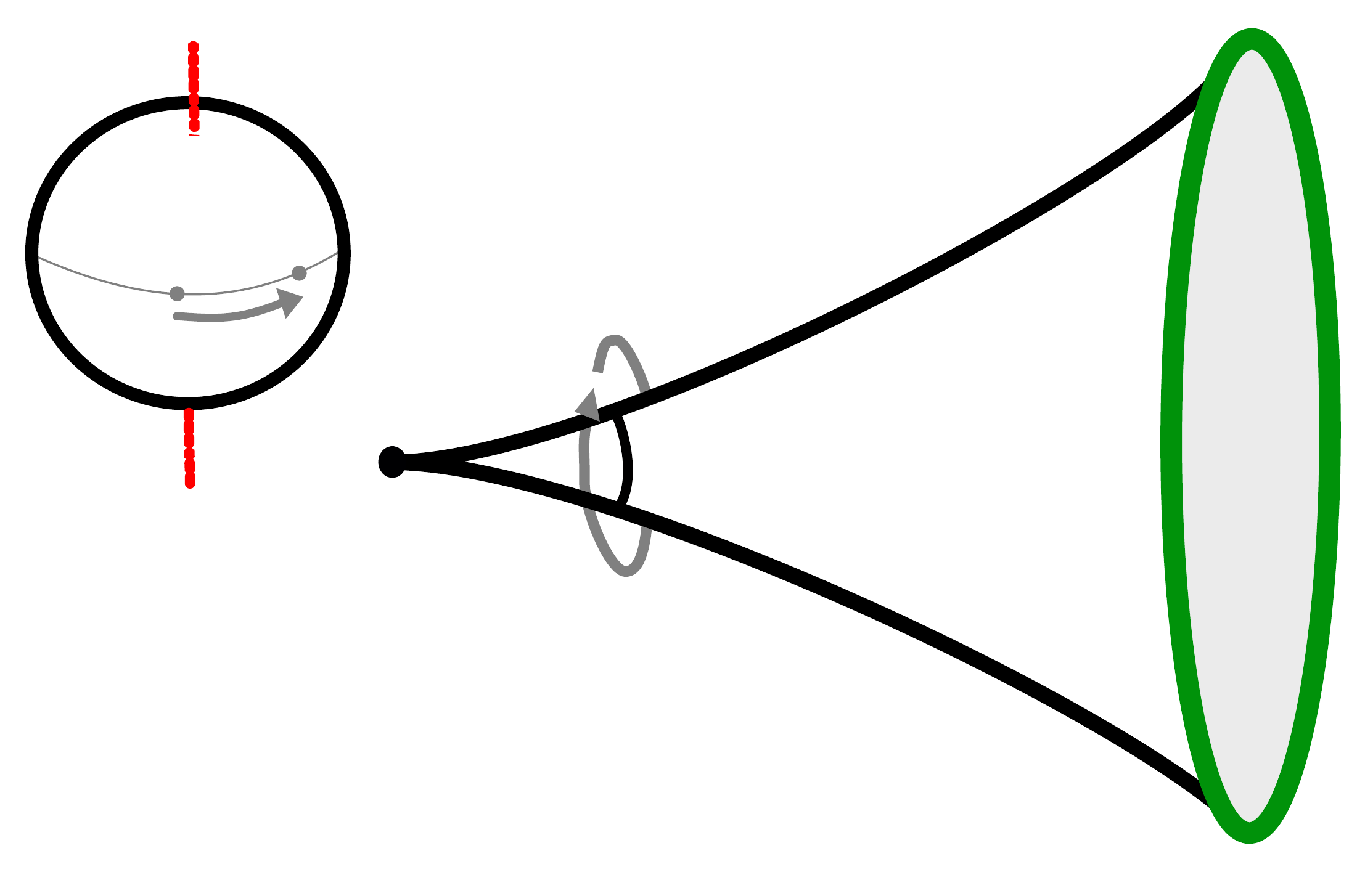}} at (0,0);
  \end{tikzpicture} \right)^{1-\alpha}}{1-\alpha}\=  \frac{(d_{\frac{1}{8}\text{-BPS BH}})^{1-\alpha}}{1-\alpha}  \,.
\ee
Here the third equality follows from the fact that the same geometries contribute to both the numerator and denominator in the limit $\beta \to \infty$. 
This confirms that $\rho_L$ is the maximally mixed density matrix for the finite $d_{\frac{1}{8}\text{-BPS BH}}$-dimensional Hilbert space of BPS black holes.\footnote{
For a discussion involving matter correlators see~\cite{Lin:2022rzw,Lin:2022zxd}.}

An important point to emphasize is that in the path integral preparation of the zero-temperature thermofield double state, the inclusion of orbifold geometries in the bulk is crucial in order to recover a Renyi entropy that is consistent with a maximally mixed state in a finite dimensional. One can therefore wonder whether the presence of the orbifolds in the state preparation plays an important role for other observables in the theory and, in particular, whether one can construct observables for which the orbifold geometries become dominant.

\subsection{ Future directions}

Before jumping into possible future directions we would like to make a list of technical questions regarding the calculations in this paper which would be good to address in the future: 
\begin{itemize}
    \item Perhaps the most important one is to find a principle that determines uniquely the correct measure over the localization manifold.
    Our choice of ultra-local measure turned out to be consistent with the U-duality invariance that is known to exist in the UV string theory completion but is not in any way explicit at the level of the AdS$_2\times S^2$ supergravity path integral. While we believe that this provides strong evidence for the validity of our choice of ultra-local measure, it would be highly satisfying to give a first-principles derivation of~\eqref{eq:measure-over-MQ}.
    
    \item It is possible to give a 4d interpretation of the terms in the Kloosterman sum as describing the topology of certain gauge fields. The evaluation of the action and in particular the phases involved in the Kloosterman sum was done by lifting from 4d to 5d and then reducing down to 3d Chern-Simons \cite{Dabholkar:2014ema}. It would be nice to compute the action in a fully 4d formalism to make the whole calculation self-consistent.
    \item We have done the calculation in a type IIA frame with only D0-D2-D4 charge. It would be nice to extend the calculation to allow for D6 charge. 
    \item We have used an $\mathcal{N}=2$ formalism to treat $\mathcal{N}=8$ supergravity. This works as long as we study states with vanishing charge under the NS-NS gauge fields in the gravitino multiplets, coming from the isometries of $T^6$. It would be nice to extend this calculation to non-zero charge and verify the full U-duality invariance.
\end{itemize}

The technical steps necessary for performing the localization computation rely on several steps, none of which require working solely with the $\cN=8$ supergravity theory, and should be applicable in theories with less supersymmetry. For instance, in theories of $\cN=4$ supergravity obtained by a dimensional reduction on $T^2\times K3$, one might hope to reproduce the exact degeneracy of $\frac14$-BPS and $\frac12$-BPS black holes. 
In the former case one has to remove the contribution of multi-center black holes from the index. The resulting index for single-center black holes is the Fourier coefficient of a mock Jacobi form~\cite{Dabholkar:2012nd}, whose Rademacher expansion is more complicated than the one for true Jacobi forms that we considered in this paper~\cite{Ferrari:2017msn, Cardoso:2021gfg}. 
In the latter case, while the Rademacher expansion of the associated modular form is, in fact, simpler than the one studied in this paper, the difficulty is that $1/2$-BPS black holes have a small horizon area and it is therefore difficult to define the gravitational path integral to start with. For theories of $\cN=2$ supergravity, finding the exact index of black holes is even more difficult~\cite{Denef:2007vg}---in such cases the prepotential can receive corrections at arbitrary loop order and an exact formula  is unknown. 

In this paper we have shown that the  gravitational functional integral in~AdS$_2$ takes the form of the infinite series~\eqref{eq:Rademacher-intro} which reproduces the 
convergent microscopic expansion~\eqref{RadexpC}. 
The elements of the sum in the gravitational picture are interpreted as contributions from a series of orbifolds. 
For~AdS$_3$/CFT$_1$ a very similar series of orbifolds are also known on both sides~\cite{Dijkgraaf:2000fq}. 
In fact, such a series of orbifolds is also known to contribute in higher dimensional~AdS$_d+1$/CFT$_d$ 
situations  for~$d=3,4$~\cite{Cabo-Bizet:2019eaf,ArabiArdehali:2021nsx,Aharony:2021zkr,BenCaboMur2021}.  
It would be interesting to make the link to the microscopic indices more precise in all these cases.

The fact that we have managed to fully localize the gravitational path integral for BPS black holes makes us hopeful that we might be able to perform this procedure to match a variety of boundary quantities that are protected by supersymmetry and do not rely on the asymptotics of the spacetime being AdS$_2\times S^2$. Two examples come immediately to mind. Firstly, as mentioned above, it is known that multi-center center solutions can contribute to the  $1/4$-BPS index in $\cN=4$ supergravity. 
It would be interesting to understand whether the contribution of such configurations to the gravitational path integral can also be computed from asymptotic flat space and matched to the string theory answer. Secondly, one could also hope to replicate the computation for spacetimes that are asymptotically AdS$_d\times \cM$ with the hope of matching quantities that are protected by supersymmetry in a boundary SCFT$_{d-1}$. Progress in this direction has been made in \cite{Dabholkar:2014wpa, Hristov:2019xku}, where the $S^3$ partition function of a boundary theory has been matched by performing the localization procedure in an asymptotically AdS$_4$ bulk. Now, armed with a better understanding of how to compute one-loop determinants  around the localization saddles, we hope to address a much greater variety of examples in near future.

\subsection*{Acknowledgements} 
We thank Dan Butter, Alejandra Castro, Clay Cordova, Atish Dabholkar, Rajesh Gupta, 
Zohar Komargodski,  Juan Maldacena, Boris Pioline, 
Ashoke Sen, Stephen Shenker, and Douglas Stanford for valuable discussions. GJT is supported by the Institute for Advanced Study and the National Science Foundation under Grant No. PHY-1911298, and by the Dipal and Rupal Patel funds. LVI was supported by the Simons Collaboration on Ultra-Quantum Matter, a Simons Foundation Grant with No. 651440. 
This work is supported by the ERC Consolidator Grant N.~681908, ``Quantum black holes: A macroscopic 
window into the microstructure of gravity'', and by the STFC grant ST/P000258/1. This work was performed in part at the Aspen Center for Physics, which is supported by National Science Foundation grant PHY-1607611.

\appendix

\section{A brief summary of Jacobi forms and the Rademacher expansion} \label{app:Jacobi}

We set~$q=\rme^{2 \pi \ii \tau}$ and~$\z=\rme^{2 \pi \ii z}$ with~$\t \in \IH$, $z \in \IC$. 
The odd Jacobi theta function is given by 
\be \label{theta1def}
\vth_1(\t, z) \= \sum_{n\in \IZ + \half} (-1)^n \,q^{n^2/2} \, \z^n \,.
\ee
The Dedekind eta function is 
\be \label{etadef}
\eta(\tau) \= q^{\frac{1}{24}} \prod_{n=1}^{\infty}(1 - q^{n}) \,.
\ee

We recall below a few relevant facts about Jacobi forms \cite{Eichler:1985ja}. 
A Jacobi form of  weight $k$ and index $m$  is a  holomorphic function $\varphi(\tau, z)$ 
from $\mathbb{H} \times\C$ to $\C$ which 
transforms under the modular group as
\be\label{modtransform}  
  \varphi \Bigl(\frac{a\t+b}{c\t+d}, \frac{z}{c\t+d} \Bigr) \= 
   (c\t+d)^k \, \rme^{\frac{2\pi i m c z^2}{c\t +d}} \, \varphi(\t,z)  \qquad \forall \quad
   \begin{pmatrix} a&b\\ c&d \end{pmatrix}  \in SL(2; \mathbb{Z}) 
\ee
and under the translations of $z$ by $\mathbb{Z} \tau + \mathbb{Z}$ as
\be\label{elliptic}  
 \varphi(\t, z+\lambda\tau+\mu)\= \rme^{-2\pi i m(\lambda^2 \t + 2 \lambda z)} \varphi(\t, z)
  \qquad \forall \quad \l,\,\mu \in \mathbb{Z} \, .
\ee
Here the integers~$k,m$ are called the weight and index, respectively, of the Jacobi form.
The equations~\eqref{modtransform}, \eqref{elliptic}  imply the 
periodicities $\varphi(\t+1,z) = \varphi(\t,z)$ and $\varphi(\t,z+1) = \varphi(\t,z)$, 
so that $\varphi$ has a double Fourier expansion
\be\label{fourierjacobi} 
\varphi(\t,z) \= \sum_{n, r \in \IZ} c(n, r)\,q^n\,\z^r\,. 
\ee
The function  is called a \emph{weak} Jacobi form if~$c(n, r)= 0$ unless~$n \geq 0$.   
Equation~\eqref{elliptic} is then equivalent to the periodicity property
\be\label{cnrprop}  
c(n, r) \= C_{r}(4 n m - r^2) \ ,
  \qquad \mbox{where} \; C_{r}(\Delta) \; \mbox{depends only on} \; r \, \text{mod}\, 2m \, . 
\ee

We have, for a Jacobi form~$\v$ of weight~$k$ and index~$m$, 
the double Fourier expansion 
\be \label{phikm}
\v(\t,z) \= \sum_{n, \ell \, \in \, \IZ} c(n, \ell)\,q^n\,\z^\ell\,.
\ee
The transformation property of a Jacobi form of index~$m$ under elliptic transformations implies that 
its Fourier coefficients obey
\be 
c(n, \ell) \= C_{\nu}(\Delta) \,, \qquad \Delta \= 4mn-\ell^2\,, \qquad \text{$\nu \= \ell$ mod~$2m$}\,,
\ee
for some functions~$C_\nu$. 
In effect, the elliptic symmetry reduces a function of two variables to a $2m$-dimensional vector of functions
of one variable~$\Delta = 4mn-\ell^2$, which is invariant under elliptic transformations.
In the case of our interest, the Jacobi form~\eqref{phi21} has 
index~$m=1$, and~$\ell = 4 n - \ell^2$~mod~2, so that the two functions~$C_\nu(\Delta)$, $\nu = 0,1$~mod~2
can be captured by one function,~$C=C_0$ (or $C_1$) for~$\Delta$ even (or odd), 
for all Fourier coefficients.

The Rademacher expansion is the following formula for the 
coefficients~$C_{\nu}(\Delta)$ of a Jacobi form of weight~$k<0$ and index~$m$ as in~\eqref{phikm}
for~$\Delta>0$,
\be\label{RademacherGeneral}
 C_{\nu} (\Delta) \= 
 \sum_{c=1}^\infty \, \Bigl(\frac{c}{2\pi} \Bigr)^{k-\frac{5}{2}} \, 
 \sum_{\wt \Delta < 0} \, C_{\mu}(\wt \Delta)  \, \text{Kl}(\Delta, \nu, \wt \Delta, \wt \nu;c) \,  
 \biggl\lvert\frac{ \wt \Delta}{4 k} \biggr\rvert^{\frac{3}{2}-k} I_{\frac{3}{2}-k}
 \biggl( \frac{\pi}{c} \sqrt{| \wt \Delta|  \Delta}\biggr)\,, 
\ee
where~$\wt{I}_{\rho}$, the \emph{modified Bessel function of index $\rho$}, is 
given by the following integral formula, 
\begin{equation}\label{intrep}
 \wt{I}_{\rho}(z)\=\frac{1}{2\pi \ii}\int_{\epsilon-\ii\infty}^{\epsilon+\ii\infty} \, 
 \frac{d\s}{\s^{\r +1}}\exp \Bigl(\s+\frac{z^2}{4\s} \Bigr) \, .
\end{equation}
It is related to the standard~$I$-Bessel function by
\be
\wt I_{\rho}(z) \= \Bigl(\frac{z}{2} \Bigr)^{-\rho} I_{\rho}(z) \,,
\ee
which admits the asymptotic expansion, with $\mu = 4\rho^{2}$, 
\be\label{Besselasymp}
I_\rho(z) \; \sim \;  
\frac{\rme^{z}}{\sqrt{2\pi z}} \biggl( 1- \frac{(\mu -1)}{8z} + \frac{(\mu -1)(\mu -3^{2})}{2! (8z)^{3} }- \frac{(\mu -1)(\mu -3^{2}) (\mu -5^{2})}{3! (8z)^{5} } + \ldots \biggr) \,.
\ee
The coefficients $ \text{Kl}(\Delta,\nu,\wt \Delta,\wt\nu ;c)$ are
generalized Kloosterman sums, equal to 1 for~$c=1$ and defined for~$c>1$ as
\be
\label{kloos0}
\text{Kl}(\Delta,\nu;\wt \Delta,\wt\nu ;c) \; \coloneqq \;
\rme^{\frac{\pi \ii}{2}(\frac12-k)}
\sum_{{-c \leq d< 0}  \atop { (d,c)=1}}
\rme^{2\pi \ii \frac{d}{c} (\Delta/4m)} \; M(\gamma_{c,d})^{-1}_{\nu\wt\nu} \; 
\rme^{2\pi \ii \frac{a}{c} (\wt \Delta/4m)} \,,
\ee
where~$\gamma=\bigl(\begin{smallmatrix} a & b \\ c & d \end{smallmatrix} \bigr) \in SL_2(\IZ)$ and 
$M$ is a~$2m$-dimensional representation of~$SL_2(\IZ)$ which transforms the 
standard theta functions of weight~$\frac12$ and index~$m \in \IZ^+$, 
\begin{equation}
 \vartheta_{m,\mu}(\tau,z) \=\sum_{\ell =\mu \, \text{mod} \, 2m} q^{\ell^2/{4m}} \, \z^\ell \, , 
\end{equation} 
as 
\begin{equation}\label{thetadef}
 \vartheta_{m,\mu}\left(\frac{a \tau +b}{c\tau +d}, \frac{z}{c\tau+d}\right) \=  
(c\tau+d)^{\frac12} \, \rme^{2\pi i m \frac{cz^2}{c\tau+d}} \, M(\gamma)_{\nu \mu}^{-1} \, \vartheta_{\mu,\ell}(\tau,z) \,,
\end{equation} 
where the square root is defined by choosing the principal branch of the logarithm.
The matrix $M(\gamma)$ is called the  multiplier system. 

An explicit formula for the value of~$M^{-1}$ on an arbitrary 
element~$\gamma=\bigl(\begin{smallmatrix} a & b \\ c & d \end{smallmatrix} \bigr) \in SL_2(\IZ)$ 
is given in~\cite{Jeffrey:1992tk} (with~$r=m+2$). In the above conventions we have~\cite{Dabholkar:2014ema}
\be \label{multiplier}
\begin{split}
M^{-1}(\gamma)_{\nu\mu} &\= 
\frac{\rme^{\frac{\pi \ii}{4}}\text{sign}(c)}{\sqrt{2r |c|}} \,
  \exp \Bigl( -\frac{\ii\pi}{6}\Phi(\gamma) \Bigr) \cr
& \quad   \sum_{\varepsilon\=\pm}\sum_{n=0}^{c-1}\varepsilon \, 
  \exp \Bigl(\, \frac{\ii\pi}{2rc} \bigl( d (\nu +1)^2-2(\nu +1)(2rn+\varepsilon (\mu +1))+a(2rn+\varepsilon (\mu +1))^2 \bigr) \Bigr) \,.
\end{split}
\ee
Here the Rademacher~$\Phi$-function is defined as
\begin{equation}\label{RadPhi}
 \Phi (\gamma) \= \frac{a+d}{c}-12\,\text{sign}(c) \, s(a,|c|) \,,
\end{equation}
where the Dedekind sum 
is, for~$c>0$,
\begin{equation}
 s(a,c)\=\frac{1}{4c}\sum_{j=1}^{c-1}\cot \bigl(\frac{\pi j}{c} \bigr) \,\cot \bigl(\frac{\pi ja}{c}\bigr) \,.
\end{equation}
It can be shown from these expressions that each term in the the generalized Kloosterman sum~\eqref{kloos0} 
depends only on the equivalence class~$\Gamma_{\infty}\backslash SL(2,\IZ)/\Gamma_{\infty}$.

\section{From BRST variation to equivariant variation}

\label{sec:BRST-variation-equiv-variation}

The main purpose of this appendix is to review the developments of the BRST quantization in the off-shell formulation of supergravity, discussed in detail in \cite{deWit:2018dix}. We will describe how to obtain the BRST variations of fields and ghosts as well as how to deform these variations to obtain the equivariant transformations with respect to the $\qeq$ which we used in our localization procedure. We start with the infinitesimal gauge transformation of some  bosonic or fermionic field $\phi^i$, whose gauge parameter is $\xi^\alpha$
\be 
\delta_{\xi} \phi^i = R(\phi)^{i}_{\,\alpha} \xi^\alpha\,,
\ee
where $R(\phi)^{i}_{\,\alpha}$ can include derivatives acting on $\xi^\alpha$, who itself might be bosonic or fermionic. Commuting or anti-commuting two such transformations we need to obtain a third, 
\be 
\delta_{\xi_1} \delta_{\xi_2}- \delta_{\xi_2} \delta_{\xi_1} = \delta_{\xi_3}
\ee
with $
\xi_3^\alpha = f(\phi)_{\beta \gamma}{}^\alpha \,\xi_1^\beta\,\xi_2^\gamma$. In usual gauge theories, $f(\phi)_{\beta \gamma}{}^\alpha$ are the structure constants of the gauge group and are thus independent of $\phi$. In the superconformal formulation of supergravity $f(\phi)_{\beta \gamma}{}^\alpha$ indeed generically depend on some of the supergravity fields $\phi^i$ and therefore we refer to them as ``structure functions'' and the gauge algebra is called ``soft''.

To quantize such theories, we need to add a set of ghost fields $c$, and write down the BRST transformations acting on all fields 
\be 
\label{eq:brst-transf-def}
\delta_{\rm brst} \phi^i = R(\phi)^i_{\,\alpha}\Lambda c^\alpha\,, \qquad \delta_{\rm brst} c^\alpha = \frac{1}2 f(\phi)_{\beta \gamma}{}^\alpha c^\beta \Lambda c^{\alpha}\,,
\ee
where we can see that the BRST transformation $\delta_{\rm brst}$ is nilpotent by acting on $\phi^i$ or $b$ repeatedly:
\be 
\delta_{\rm brst}^2 = 0\,.
\ee
Similar transformations apply to the anti-ghosts $b_\alpha$ and to the Lagrange multipliers $B_\a$. We shall come back to the variations of these fields after explaining how to deform the BRST transformations for $\phi^i$ and $c^\a$.

To determine the one-loop determinant around each point on the localization locus we need to define the action of the equivariant variation on all quantum fluctuations of fields and ghosts in the theory. To do this, we separate the fields and ghosts into their background values $\bck{\phi_i}$ and $\bck{c^\a}$, and their quantum fluctuations $\tilde \phi^i$ and $\tilde c^\alpha$. The BRST fluctuation acting on the variation is therefore,
\be 
\delta_{\rm brst}
\tilde \phi^i = \delta_{\rm brst}{{\phi^i}} - \delta_{\rm brst}{\bck{\phi^i}} \,, \qquad  \delta_{\rm brst} \tilde{c}^\alpha = \delta_{\rm brst} {c}^\alpha - \delta_{\rm brst} \bck{c}^\alpha\,.
\ee

What is this background for the fields $\phi^i$ and $c^\a$ along the localization locus that we are interested in? Close to the boundary where the spacetime is locally AdS$_2\times S^2$, there exists a set of locally defined Killing spinors associated to the $PSU(1,1|2)$ superisometry. We then choose one of the Killing spinors associated and require that it can be globally defined everywhere in the bulk. Field and metric configurations for which a Killing spinor cannot be globally defined do not contribute to the path integral when the boundary observable under study is protected. We then fix the ghost associated to the supersymmetry transformation with respect to which the Killing spinor to be equal to the Killing spinor itself. All other ghosts will be set to zero. Due to the Killing spinor equation or due to the vanishing of the other ghosts, we have that  
\be 
\label{eq:bck-brst-transf}
\delta_{\rm brst}{\bck{\phi^i}} = \Lambda  \bck{c}^\alpha\, R(\bck{\phi})_\alpha{}^i  = 0\,,
\ee
for all points on the localization locus. To properly define $\qeq$ we however need to deform the BRSY transformation to an equivariant transformation
\be 
\delta_\text{brst}\,\,\Rightarrow \,\,\delta_\text{eq}
\ee
in such a way that $\delta_\text{eq}$ is no longer nilpotent. This can be done by taking the BRST transformation on the background \eqref{eq:bck-brst-transf} and on the quantum fluctuations (which can be derived from \eqref{eq:brst-transf-def}) to remain unchanged. However, in contrast to the BRST transformation, we will fix the transformations on the non-zero background  ghost to also vanish. In other words, we thus define 
\be 
\label{eq:equiv-transf-def}
\delta_{\rm eq}\bck{\phi}^i&= 0 \,, \qquad\delta_{\rm eq} \tilde\phi^i = R(\phi)^i_{\,\alpha}\Lambda (\bck{c}+\tilde c)^\alpha\,,\nn \\ 
\delta_{\rm eq} \bck{c}^\alpha &= 0\,, \qquad \delta_{\rm eq} \tilde{c}^\alpha = \frac{1}2 f(\phi)_{\beta \gamma}{}^\alpha (\bck{c}+\tilde c)^\beta \Lambda (\bck{c}+\tilde c)^{\alpha}\,-\,\frac{1}2 f(\bck{\phi})_{\beta \gamma}{}^\alpha \,\bck{c}^\beta \Lambda \bck{c}^{\alpha}\,,
\ee

Squaring the newly defined transformation, one finds
\be 
\delta_{\rm eq}^2=\delta_{\bck\xi}\neq 0\,,\qquad  [\delta_{\rm eq}\,,\,\,\,\delta_{\bck\xi}]=0\,, 
\ee
where the transformation $\delta_{\bck\xi} $ is given by 
\be 
\label{eq:transf-background-xi}
\delta_{\bck\xi} \bck \phi = 0\,, \qquad  \delta_{\bck\xi} \tilde \phi^i =  \bck\xi^\alpha R(\phi^i)_\alpha{}^i\,, \qquad \delta_{\bck\xi} \bck c^\alpha = 0\,, \qquad \delta_{\bck\xi} \tilde c^\alpha = f(\phi)_{\beta \gamma}{}^\alpha (\bck c + \tilde c)^\beta \bck \xi^\gamma\,,
\ee
with 
\be
\tilde \xi = \Lambda_{[2} f(\bck \phi)_{\beta \gamma}{}^\alpha\, \bck c^\beta \Lambda_{1]} \bck c^\gamma \,,
\ee
Since the only non-vanishing ghosts are fixed to be given by one of the Killing spinors of the spacetime, the transformations \eqref{eq:transf-background-xi} are exactly what we expect under the bosonic isometry obtained by constructing the Killing vector from the bilinear of the Killing spinor that we used as our ghost background. In our case, this is the $U(1)$ generator $L_0-J_0$ in the AdS$_2\times S^2$ or orbifold geometries. To check that $\delta_{\rm eq}$ is a proper equivariant transformation we finally need to define its action on the anti-ghosts $b_\a$ and on the Lagrange multipliers $B_\a$. To start, the background of these fields is set to zero, $\bck{b}_\a = 0$ and $\bck{B}_\a = 0$. The equivariant variation on their quantum fluctuations is then defined to be
\be 
\delta_{\rm eq} \tilde b_\alpha = \Lambda B_\alpha\,, \qquad \delta_{\rm eq} B^\alpha = \frac{1}2 f(\bck{\phi})_{\delta \epsilon}{}^\beta\, \bck{c}^\delta\, \Lambda\, \bck{c}^\epsilon\,f(\bck{\phi})_{\a \b}{}^\g\,.
\ee
Acting once again on the anti-ghosts and on the Lagrange multipliers we find 
$\delta_{\rm eq}^2 b_\alpha = \delta_{\bck{\xi}} b_\alpha = \bck{\xi} f(\phi)_{\beta \alpha}{}^\gamma b_\gamma  
$, $\delta_{\rm eq}^2 B_\alpha = \delta_{\bck{\xi}} B_\alpha = \bck{\xi} f(\phi)_{\beta \alpha}{}^\g 
B_\g 
$ which is once again the expected transformation under the bosonic $U(1)$ transformation generated by $\delta_{\bck{\xi}}$. This indeed confirms that  $
\delta_{\rm eq}^2 = \delta_{\bck \xi}$ for all fields and ghosts and therefore $\delta_{\rm eq}$ is a valid equivariant transformation.

\section{On possible truncations of $\mathcal{N}=8$ supergravity}\label{app:truncation}
The first calculation of the Bessel function appearing in the Rademacher expansion of the $\mathcal{N}=8$ black hole index was outlined in \cite{Dabholkar:2011ec}. The calculation in that reference had some drawbacks which we addressed in this paper. In this appendix we clarify the relation between the two approaches for readers familiar with that work.

Firstly, instead of working with the full 4d $\mathcal{N}=8$ supergravity theory obtained from a toroidal compactification of type IIA theory, the calculation was done in a truncated theory. From the point of view of the low-energy effective theory, the truncation is obtained as follows. 
All the fields of the low-energy theory fall into a single representation of~$\CN=8$ supersymmetry 
containing the graviton, i.e.~the~$\CN=8$ graviton multiplet. We reduce this representation in terms 
of representations of a particular~$\CN=4$ sub-algebra. This gives  
an~$\CN=4$ graviton multiplet (= $\CN=2$ graviton + 2 spin-$\frac32$ + 1 vector), 
four~$\CN=4$ spin-$\frac32$ multiplets (each = $\CN=2$ spin-$\frac32$ + 2 vectors + 1 hyper), 
and 6 vector multiplets (each = $\CN=2$ vector + hyper). 
Next we throw away the spin-$\frac32$ multiplets, and then reduce the 
into~$\CN=2$ representations. 
Now we have an~$\CN=2$ graviton multiplet, 2 spin-$\frac32$ multiplets, 
7 vector multiplets, and 6 hyper multiplets. Finally, we throw away the 
the hyper multiplets. The remaining~$\CN=2$ theory is precisely~$\CN=2$ 
to~$\nv=7$ vector multiplets for which we write a prepotential below.
The~$T^6$ compactification has~$SO(6,6;\IZ)$ T-duality which is a subgroup of the full~$E_{7,7}$  U-duality. The 28 gauge fields of the theory decompose as~$\mathbf{28} = \mathbf{12} + \mathbf{16}$. The vector~$\mathbf{12}$ is in the~NS-NS sector while the spinor~$\mathbf{16}$ is in the~R-R sector. Now we can identify the~$16$ gauge fields in the R-R sector with the $4 \times 4 =16$ gauge fields in the four~$\CN=4$ spin-$\frac32$ multiplets that we drop in the truncation. The NS-NS fields behave as in usual~CY$_3$ compactifications, while the R-R fields decouple classically.  

The truncation consists on $\mathcal{N}=2$ supergravity coupled to $n_v=7$ vector multiplets and to perform the localization calculation we need to specify its holomorphic prepotential. In the M-theory 
description, each vector multiplet is associated with a 2-cycle (or its Hodge-dual 4-cycle) inside the~CY$_3$. For concreteness we label them in the following way:
\begin{eqnarray}
&& \hspace{-0.8cm}\Sigma^{45}\to X^1,~\Sigma^{67}\to X^2,~\Sigma^{68}\to X^3,~\Sigma^{69}\to X^4,~\Sigma^{78}\to X^5,~\Sigma^{79}\to X^6,~\Sigma^{89}\to X^7,\nn\\
&& \hspace{-0.8cm}\Sigma^{46}\to X^8,~\Sigma^{47}\to X^9,~\Sigma^{48}\to X^{10},~\Sigma^{49}\to X^{11},~\Sigma^{56}\to X^{12},\nn \\
&&\hspace{-0.8cm}\Sigma^{57}\to X^{13},~\Sigma^{58}\to X^{14},~ \Sigma^{59}\to X^{15}.\label{eqn:XvsSigma}
\end{eqnarray}
With these conventions, in the truncation described above, the vector multiplet~$I=1$ is associated to the~$T^2$ and the other 6 vector 
multiplets~$I=2,\dots, 7$ are associated to the~${4 \choose 2} = 6$ 2-cycles of the~$T^4$. 
Now we can read off the holomorphic prepotential by considering the intersection numbers of the 4-cycles 
on the~$T^6$. We obtain the following simple formula,
\be \label{Fcubtrunc}
F_{\rm trunc.}(X) \= -\frac12 \, \frac{X^1 \, C_{ab} \, X^a X^b}{X^0} \,, \qquad a = 2, \dots, 7,
\ee
where~$C_{ab}$ is the intersection matrix of the~2-cycles of the~$T^4$. We can use the conventions of Section~\ref{sec:N8prepot} to compute since its a particular case of $C_{IJK}$ when $I=1$ and $J,K=2,\ldots,7$.

In this appendix we want to use the fact that we did the calculation in the full $\mathcal{N}=8$ supergravity theory to integrate out the extra fields and reevaluate the validity of the truncation considered in \cite{Dabholkar:2011ec}. Obtaining the truncated theory is easy and involves turning off the hypermultiplets and turning off the vector multiplets $X^8$ to $X^{15}$. This implies we should also restrict to states with vanishing charges associated to these vectors $p^8=\ldots=p^{15}=0$ and $q_8=\ldots=q_{15}=0$. It is easy to see that the prepotential \eqref{Fcub} reduces to \eqref{Fcubtrunc} after identifying $X^a=\{ X^1,\ldots, X^7\}$, and similarly for the charges.

Now we want to compare the result from applying localization to this truncation to the result obtained from $\mathcal{N}=8$ supergravity after setting to zero the extra charges. This is not trivial since we still need to integrate out the extra localization manifold coordinates that the full theory has compared to the truncation. In our parametrization we need to integrate out $\phi^8,\ldots,\phi^{15}$. Since the measure over the $\phi$'s and the one-loop determinant both depend only on $\phi^0$ the integration is simply Gaussian over the exponential of the action on the localization manifold. This gives 
\begin{eqnarray}
\int \prod_{I=8}^{15} \dd \phi^I ~ \exp\left(\sum_{I,J=8}^{15}\frac{3\pi C_{IJK}\phi^I \phi^J p^K}{c \phi^0} \right) \Big|_{p^{8}=\ldots=p^{15}=0} = \frac{4c^4(\phi^0)^{4}}{(P^2)^2},~~~P^2 = p^aC_{ab}p^b. \label{eqn:intoutphi}
\end{eqnarray}
This seems to indicate that integrating out the extra vector multiplets involved in the truncation results in a non-trivial contribution to the remaining integrals. Nevertheless this factor disappears if we correct the measure over the $n_v=7$ vector multiplets to be over 
\begin{eqnarray}
[\dd^{8}\phi]_c \equiv  \dd^{8}\phi~ \sqrt{ 2~{\rm det}'\Big( \frac{1}{2c} ~{\rm Im} ~F_{\Lambda\Sigma} \Big) },
\end{eqnarray}
where the prime denotes that the determinant is taken only over the directions $\Lambda=0,\ldots,7$ and only $\phi^0,\ldots,\phi^7$ are turned on. Its easy to check that the mismatch between this measure and the measure over all the vector multiplets precisely cancels the integral over the remaining vectors \eqref{eqn:intoutphi}. The final answer is 
\begin{eqnarray}
W^{(c)}(q,p) = \sqrt{c} K_c(\Delta) \int [\dd^{8} \phi]_c ~c^{-1} \rme^{4\mathcal{K}(\phi)} ~e^{\frac{4\pi}{c} {\rm Im} F_{\rm trunc}\big(\frac{\phi^\Lambda + ip^\Lambda}{2} \big) - \frac{\pi}{c} q_\Lambda \phi^\Lambda}, \label{eq:truncW}
\end{eqnarray}
where we only turn on the charges and moduli in the truncation. Since this is equivalent to the original integral over 15 moduli we already know this gives the right answer to match the microscopic entropy. 

The integral in \eqref{eq:truncW} is precisely the starting point of \cite{Dabholkar:2011ec} for $c=1$ (and we extended it to the case $c\neq 1$ in this paper). This formula almost shows that the truncation is consistent off-shell except for the one-loop determinant. The action and measure is indeed the one corresponding to the truncation (as well as the boundary-mode contribution to the c-dependence $1/c$) but the bulk one-loop determinant $e^{4\mathcal{K}}$ corresponds only to the full $\mathcal{N}=8$ supergravity theory.  

One approach to justify the calculation in \cite{Dabholkar:2011ec} can be to postulate we include off-shell contributions from the $\mathcal{N}=8$ supergravity theory in the one-loop determinant, but not include the extra coordinates on the localization manifold. This is inconsistent since we would be treating two sets of off-shell modes differently without a good reason. Another approach is to see whether we can find an $\mathcal{N}=2$ theory (without gravitino multiplet) that reproduces the bulk one-loop determinant $e^{4\mathcal{K}}$. Imposing that the bulk one-loop determinant be c-independent restricts $n_H-n_V = 1$. Since in the truncation $n_V=7$ we can set $n_H=8$ without affecting the localization integral, other than the bulk one-loop determinant. The problem is that this unique choice set by looking at the c-dependence gives $e^{-2\mathcal{K}}$ for the bulk one-loop determinant. Therefore we conclude there is no $\mathcal{N}=2$ supergravity theory able to reproduce \eqref{eq:truncW}.

\bibliographystyle{utphys2}
{\small \bibliography{Biblio}{}}

\end{document}